\documentclass{aa}

\usepackage{graphicx}
\usepackage{gensymb}
\usepackage{float}
\usepackage{hyperref}
\usepackage{caption}
\usepackage{subcaption}
\hypersetup{colorlinks=true,linkcolor=blue,citecolor=blue,filecolor=blue,urlcolor=blue,}
\usepackage{siunitx}
\usepackage{placeins}
\newcommand{\orcid}[1]{\protect\href{https://orcid.org/#1}{\protect\includegraphics[width=10pt]{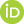}}}
\usepackage{txfonts,textcomp}

\newcommand{\A}{\r{A}\,} 
\newcommand{\ha}{H$\alpha$\,} 
 
\newcommand{\hb}{H$\beta$\,} 
\newcommand{\SBha}{$\Sigma_{H\alpha}$\,} 

\newcommand{\hii}{H\,{\footnotesize II}~\,}

\newcommand{\nii}{[N\,{\footnotesize II}]\,}
\newcommand{\sii}{[S\,{\footnotesize II}]\,}
\newcommand{\siii}{[S\,{\footnotesize III}]\,}

\newcommand{\oi}{[O\,{\footnotesize I}]\,}

\newcommand{\oii}{[O\,{\footnotesize II}]\,}
\newcommand{\oiii}{[O\,{\footnotesize III}]\,}

\newcommand{\hcm}{{\sc HCm}\,}

\defcitealias{BETISI}{BETIS I}
\defcitealias{BETISII}{BETIS II}
\usepackage{color}

\begin{document}

   \title{Bidimensional Exploration of the warm-Temperature Ionised gaS (BETIS)}

   \subtitle{III. The impact of eDIG on strong-line abundance determinations in galactic halos}

   \author{R. Gonz\'alez-D\'iaz
          \inst{1}\orcid{0000-0002-0911-5141}
          \and
          E. Pérez Montero
          \inst{1}\orcid{0000-0001-9737-4559}
          \and
          J. M. Vílchez
          \inst{1}\orcid{0000-0001-7299-8373}
          \and
           F. F. Rosales-Ortega
          \inst{2}\orcid{0000-0002-3642-9146}
          \and
          L. Galbany
          \inst{3,4}\orcid{0000-0002-1296-6887}
          }

   \institute{Instituto de Astrofísica de Andalucía (IAA-CSIC), Glorieta de la Astronomía s/n, E-18008 Granada, Spain\\ 
              \email{ragonzalez@iaa.es}
         \and
         Instituto Nacional de Astrof\'isica, \'Optica y Electr\'onica (INAOE-CONAHCyT),
              Luis E. Erro 1, 72840 Tonantzintla, Puebla, M\'exico.
        \and
        Institute of Space Sciences (ICE-CSIC), Campus UAB, Carrer de Can Magrans, s/n, E-08193 Barcelona, Spain.
         \and
        Institut d'Estudis Espacials de Catalunya (IEEC), 08860 Castelldefels, Barcelona, Spain.
            }

   \date{Received ... ; accepted ... }

% \abstract{}{}{}{}{} 
% 5 {} token are mandatory
 
\abstract{

The extraplanar diffuse ionised gas (eDIG) is a key tracer of the feedback processes that connect galactic discs and halos. We present the first study of spatially resolved gas-phase chemical abundances in the eDIG of edge-on galaxies, based on integral field spectroscopy obtained with MUSE. Using a sample of four galaxies from the BETIS sample, we derived maps of the oxygen abundance, 12+log(O/H), the nitrogen-to-oxygen ratio, log(N/O), and the ionisation parameter, log(U), with the HII-CHI-Mistry v5.5 code, and we tested the reliability of this method when different ionisation sources are present. We separated the eDIG into star-formation-dominated and shocks-contaminated regimes according to the relative contribution of photoionisation and shocks, and investigate their impact on the derived chemical abundances.

We find that HII-CHI-Mistry returns systematically lower O/H and N/O values in shock-contaminated regions, especially for N/O, with typical offsets of $\sim$0.1-0.15 dex relative to star-formation-dominated regions, together with a more homogeneous and lower ionisation state (log(U) $\approx -3.1$). In contrast, star-formation-dominated regions span a wider range of ionisation parameters (log(U) $\approx -3.3$ to $-2.8$) while maintaining approximately solar oxygen abundances. Despite these differences, the abundance offsets are comparable to the sampling step of the HII-CHI-Mistry model grid.%, indicating that the code provides robust metallicity estimates even in the presence of these mixed ionisation conditions.

The inferred oxygen abundance shows little variation between discs and halos. This could reflect a scenario in which star formation feedback and galactic fountains efficiently recycle oxygen-rich gas between both components, in agreement with hydrodynamical simulations. While O/H remains nearly constant between the disc and halo, compatible with the circulation of metal-rich gas driven by galactic fountains and feedback, the N/O ratio shows significant radial and vertical variations particularly in environments with weaker star formation feedback.

%In particular, galaxies with weaker feedback display enhanced halo N/O ratios, whereas the strong outflow activity in IC1553 is associated with lower halo oxygen abundances and higher N/O ratios, consistent with the redistribution of oxygen-enriched material by feedback-driven outflows. %These results demonstrate that combining chemical abundances with ionisation diagnostics provides a powerful tool for investigating the interplay between star formation, feedback, and the chemical evolution of galactic halos.

}


   \maketitle
%
%-------------------------------------------------------------------

\section{Introduction}

Studying the ionised gas emission in the halos in edge-on galaxies enables the detection of the warm and diffuse component of the ionised interstellar medium separately from the emission of the disc. Analogous to the diffuse ionised gas (DIG; \citealt{1963AuJPh..16....1H,1968Natur.217..709H,1971PhDT.........1R, 1974IAUS...60...51G,1985ApJ...294..256R, 1989ApJ...345..811R, 1989ApJ...337..761K}) in the galactic plane, the extraplanar diffuse ionised gas (eDIG) is a warm ($0.6-1\cdot10^4$ K) and low-density ($\sim10^{-1}$cm$^{-3}$) component that constitutes one of the most prominent phases of the interstellar medium (ISM). It is estimated to account for approximately 20\% of the ISM volume and up to 90\% of the ionised hydrogen mass \citep[e.g.][]{1996AJ....111.2265F,2014MNRAS.445..581H, 2016SAAS...43...85K, 2017PhDT........59G,2023ApJ...944L..22B, BETISI}.

The eDIG, through its ionisation budget and chemical properties, plays a crucial role in the physical processes that regulate the exchange of gas, metals, and energy between the galactic disc and halo, thereby influencing the evolution of disc galaxies \citep[e.g.][]{1988ARA&A..26..145T, 1990ApJ...352L...1R, 1990A&A...232L..15D, 2003ApJS..148..383M}. This role supports a scenario in which the disc and halo are connected through outflows of hot gas from the disc into the halo, occurring via filaments, superbubbles, or “chimneys” (also detected in X-rays and radio-emission; \citealt{2000A&A...364L..36T,2006A&A...448...43T}), a phenomenon typically associated with starburst galaxies, presenting eDIG layers that can reach distances from the midplane of 1-2 kpc, and exceptionally 5 kpc also in typical star-forming galaxies \citep[e.g.][]{2002ApJ...565L..79C,Veilleux2005,BETISII,ElliottGECKOS26}, thereby helping to constrain the physical processes in the disc from their observable effects in the halo. Nevertheless, in normal, non-starburst galaxies, there is only circumstantial evidence of the presence of bubbles or chimney structures in their halos, which is often attributed to observational limitations \citep[e.g. ][]{1989ApJ...345..372N,2000AJ....119..644H,2004AJ....128..674R,2009RvMP...81..969H}. These halo structures are formed as a consequence of this disc–halo connection, which drives a cycle of gas inflow and outflow. In this process, gas is transported from the disc into the halo and can subsequently fall back onto the disc, potentially lowering the gas-phase and stellar metallicities of the halo through mixing with less enriched material, a process that was observed on both Milky Way and local galaxies \citep[e.g. ][]{FontMetalHalo09,FontMetalHalo06,BregmanMetalHalo13,BrusadinMetalHalo13,QuMetalHalo19,VasilievMetalHalo23}. In addition, the enhancement of collisionally excited line ratios such as \sii/\ha or \nii/\ha has been observed to increase with distance from the plane of the disc (e.g. \citealt{2019ApJ...882...84L,2019MNRAS.482.4032L,2023A&A...678A..84D,Lu24,BETISII}).  These trends, which largely trace variations in electron density and ionisation conditions within the DIG, have also been suggested to arise from variations in metallicity \citep[e.g. ][and references therein]{2009RvMP...81..969H,BETISII}. For instance, \nii, \sii and \siii have been used for deriving both N/O \citep[e.g. ][]{PerezMontero09} and O/H relative abundances \citep[e.g. ][]{DiazS232000}, building empirical calibrations to measure these values (e.g. N2S2 and S23 calibrators; \citealt{1994ApJ...429..572S,2002MNRAS.330...69D,2015MNRAS.448.2030H}).

\noindent
This effect has been well documented for the DIG in galactic planes. Studies employing strong-line calibrations derived from long-slit spectroscopy and integral field spectroscopy (IFS) of local-universe galaxies have shown that the low excitation conditions characteristic of the DIG are associated with biases in the inferred O/H and N/O abundances \citep[e.g. ][]{SandersDIGabundance17,Asari2021}. Besides, the DIG contamination can lead to a flattening of the low-mass slope of the mass–metallicity relation (MZR), and can also induce a strong vertical negative metallicity gradient towards the galactic plane \citep[e.g.][]{AbundancesPerez11,SandersDIGabundance17,KumariDIGabundance19,PoetrodjojoDIGabundance19,ValeDIGabundance19}. However, although the extraplanar eDIG has recently been investigated using IFS such as Sydney-AAO Multi-object Integral
field spectrograph (SAMI; \citealt{2012MNRAS.421..872C,2015MNRAS.447.2857B}) Mapping Nearby Galaxies at Apache Point (MaNGA; \citealt{2015ApJ...798....7B}), Calar Alto Legacy Integral Field Area Survey (CALIFA; \citealt{2016RMxAA..52..171S,2019ApJ...882...84L}) or Multi Unit Spectroscopic Explorer (MUSE; \citealt{2022A&A...659A.153R,2023A&A...678A..84D,Li2023,BETISII,ElliottGECKOS26}), the relationship between the eDIG and chemical abundances in galactic halos, and the impact of the ionisation conditions and the physical properties of the eDIG on metallicity determinations has not yet been explored.

The bidimensional exploration of the warm-temperature ionised gas (BETIS) is a project aimed at characterising the emission from the DIG and providing new insights into its ionisation mechanisms and physical nature using MUSE IFS. The project consists of two main parts. The first, presented by \citet{BETISI} (hereafter \citetalias{BETISI}), characterises the MUSE galaxy sample, develops the analysis methodology and data-processing pipeline, and presents the first results for a sample of face-on star-forming galaxies. The second part, presented by \citet{BETISII} (hereafter \citetalias{BETISII}), applies the methodology and analysis pipeline developed in \citetalias{BETISI} to a sample of edge-on galaxies in order to investigate the eDIG. Using hybrid ionisation models, the work shows that the eDIG exhibits at least two main ionisation regimes resulting from different combinations of photoionisation and fast shocks given by the models of \citet{2010AJ....139..712L} and \citet{2008ApJS..178...20A}, with the relative contribution of each ionising mechanism varying across the galaxy. The shock-contaminated regime presents clear physical differences, including higher electron densities and temperatures, together with an enhanced contribution from low ionisation species to the emission-line spectrum. These differences, which may result from star formation feedback processes connecting the disc and halo, could also have a significant impact on metallicity determinations of the extraplanar gas.

\noindent
For these reasons, in this work we present the first spatially resolved investigation of the impact of extraplanar eDIG on the determination of the O/H and N/O abundance ratios in galactic halos, further exploring the properties of the two eDIG ionisation regimes identified by \citetalias{BETISII}. We use the sample of edge-on galaxies presented in \citetalias{BETISII}, and the code {\sc HII-CHI-Mistry} v5.5 \citep{PerezMonteroHCm14} to derive binned maps of the gas-phase metallicity and ionisation parameter, adopting the \citetalias{BETISII} data as input, and testing the reliability of using model-based codes to determine chemical abundances in galactic halos. In Sect. \ref{sec:sample} we present the sample. In Sect. \ref{sec:2_mapas} we present the code and the obtained maps. In Sect. \ref{sec:3_abundancias}, we examine how these results are affected when considering eDIG regions in which the ionisation is dominated either by star-formation feedback from the galactic plane or by fast shocks. Finally, in Section~\ref{sec:discuss}, we discuss the results on a case-by-case basis, comparing them with previous findings on the impact of the DIG in galactic planes, and assessing the potential biases introduced by different assumptions adopted when running {\sc HII-CHI-Mistry} with the available data.

The following notation is used throughout the paper: \nii$\equiv$\nii$\lambda6584$; \sii$\equiv$\sii$\lambda6717 +$\sii$\lambda6731$; \oiii$\equiv$\oiii$\lambda5007$; and \oi$\equiv$\oi$\lambda6300$.
We adopt the standard $\Lambda$CDM cosmology with H$_0$ = 70 km/s/Mpc,\, $\Omega_{\Lambda}$ = 0.7, and\,$\Omega_M$ = 0.3.

\begin{figure*}[ht!]
\centering
    \includegraphics[width=0.331\textwidth]{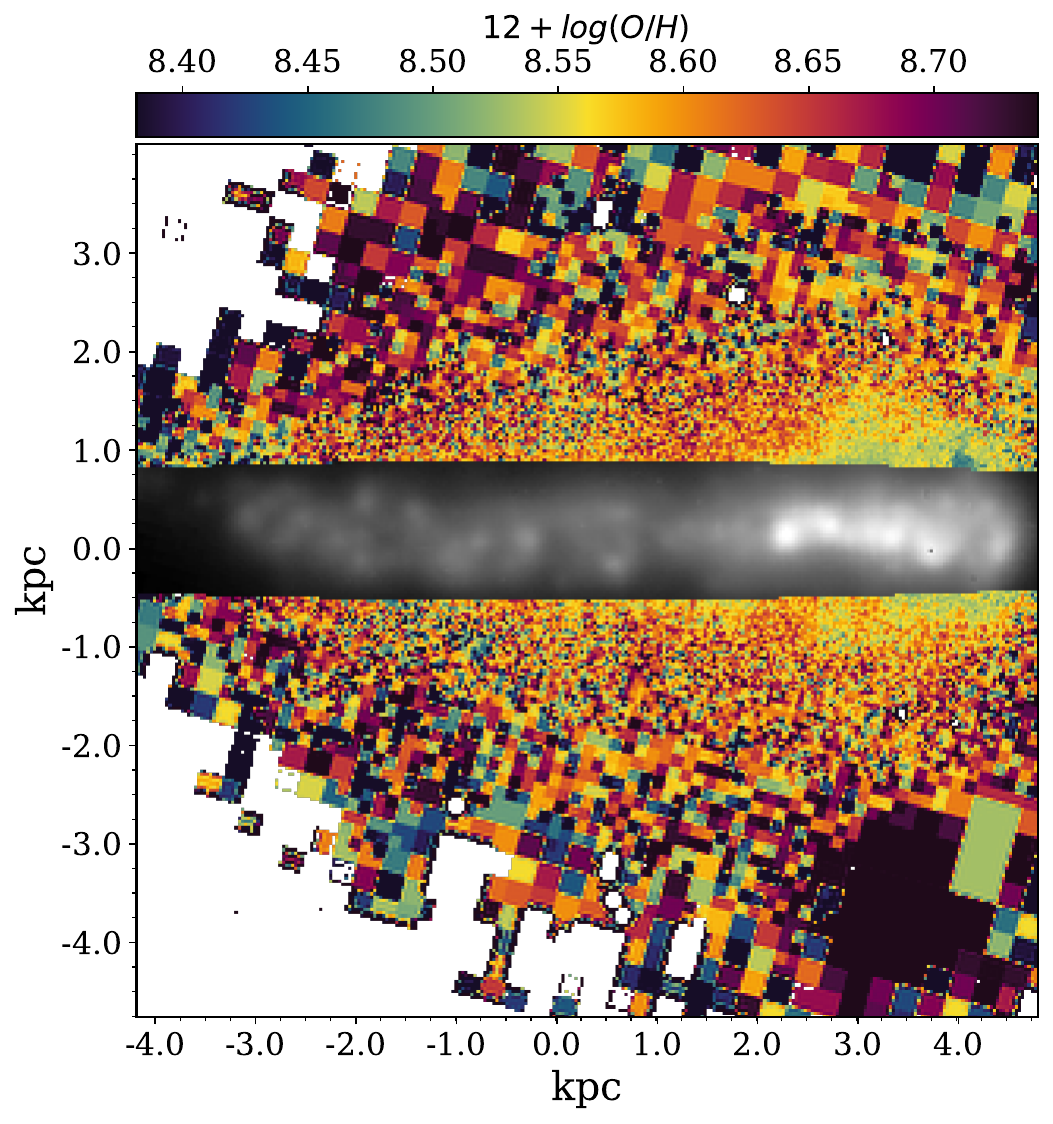}
    \includegraphics[width=0.33\textwidth]{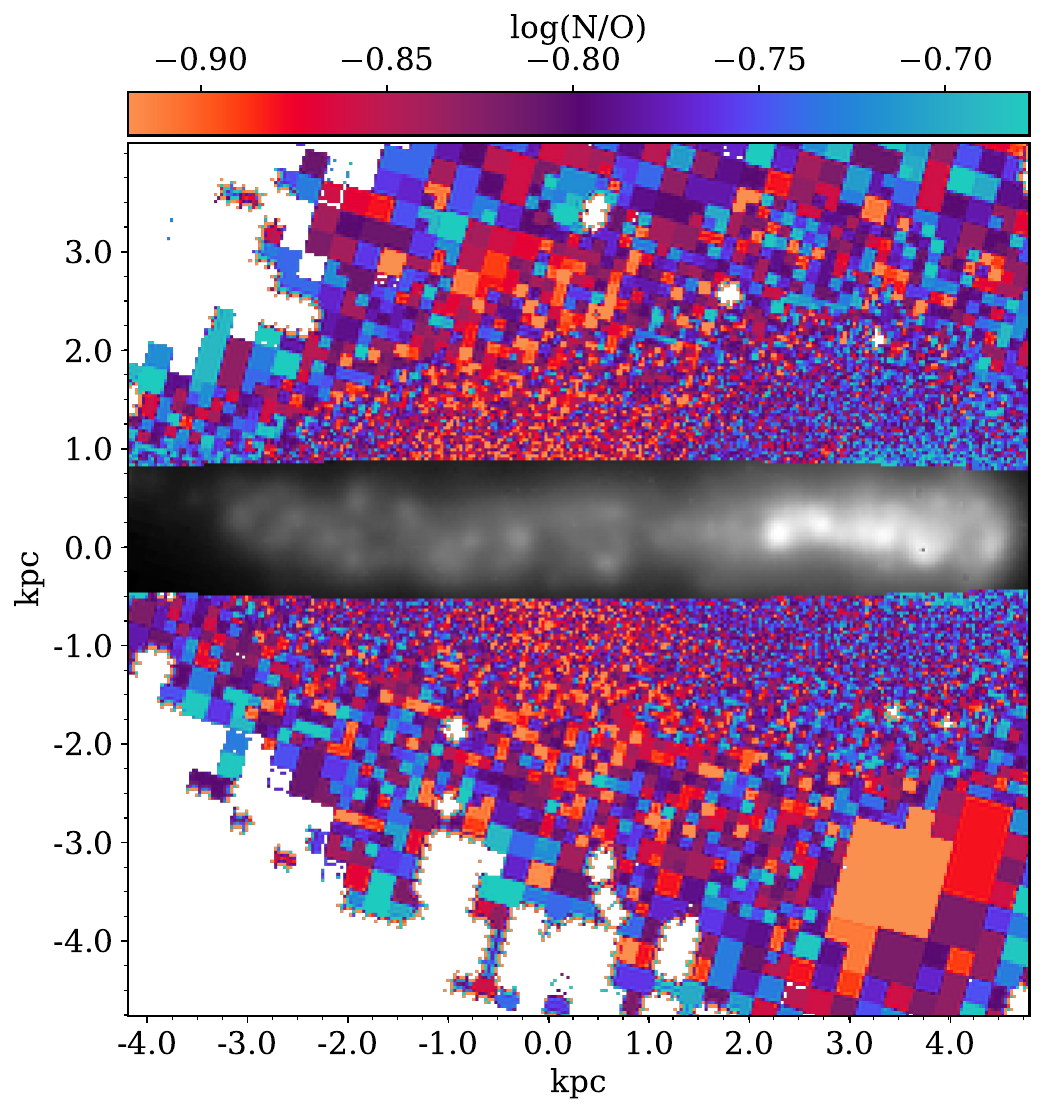}
    \includegraphics[width=0.33\textwidth]{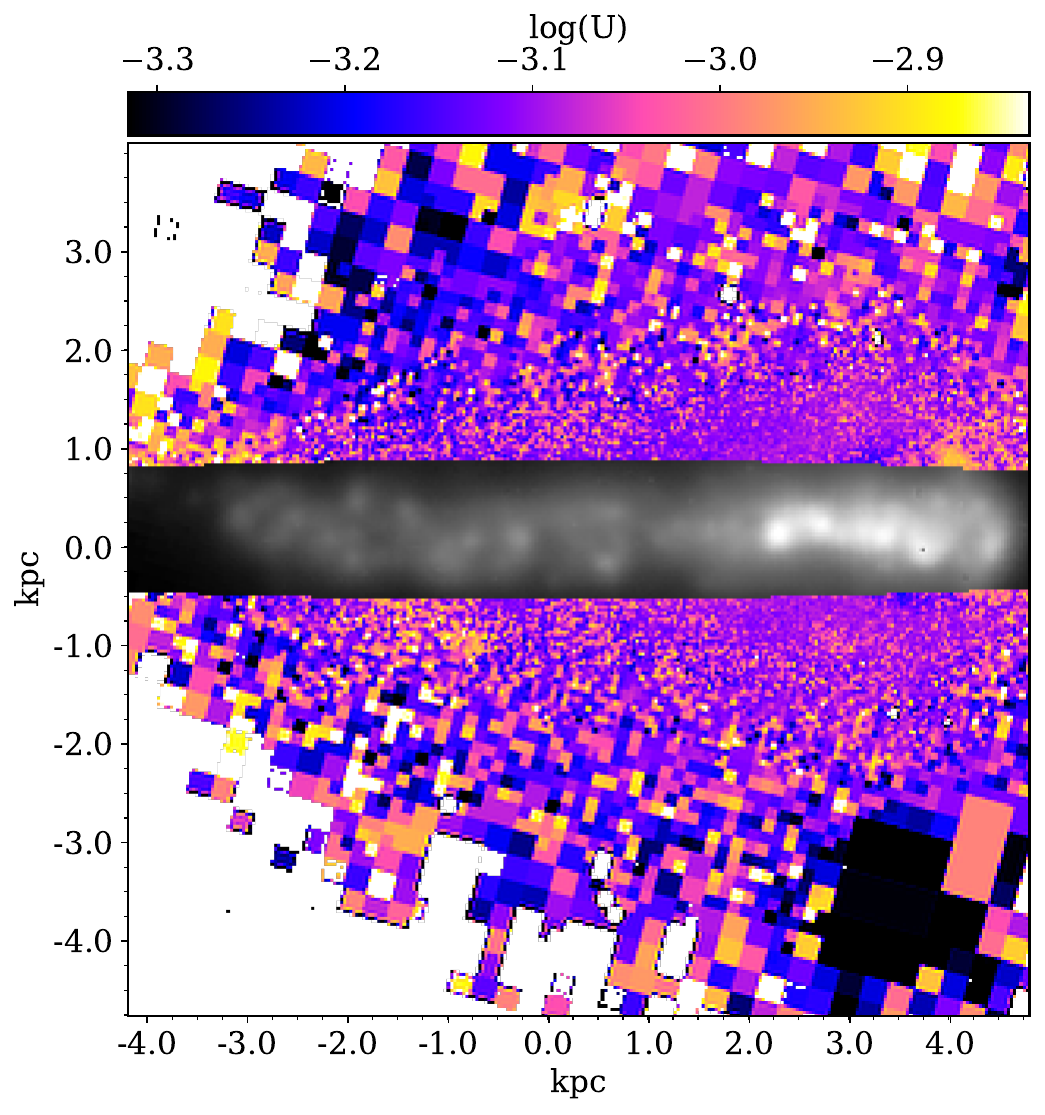}

    \includegraphics[width=0.331\textwidth]{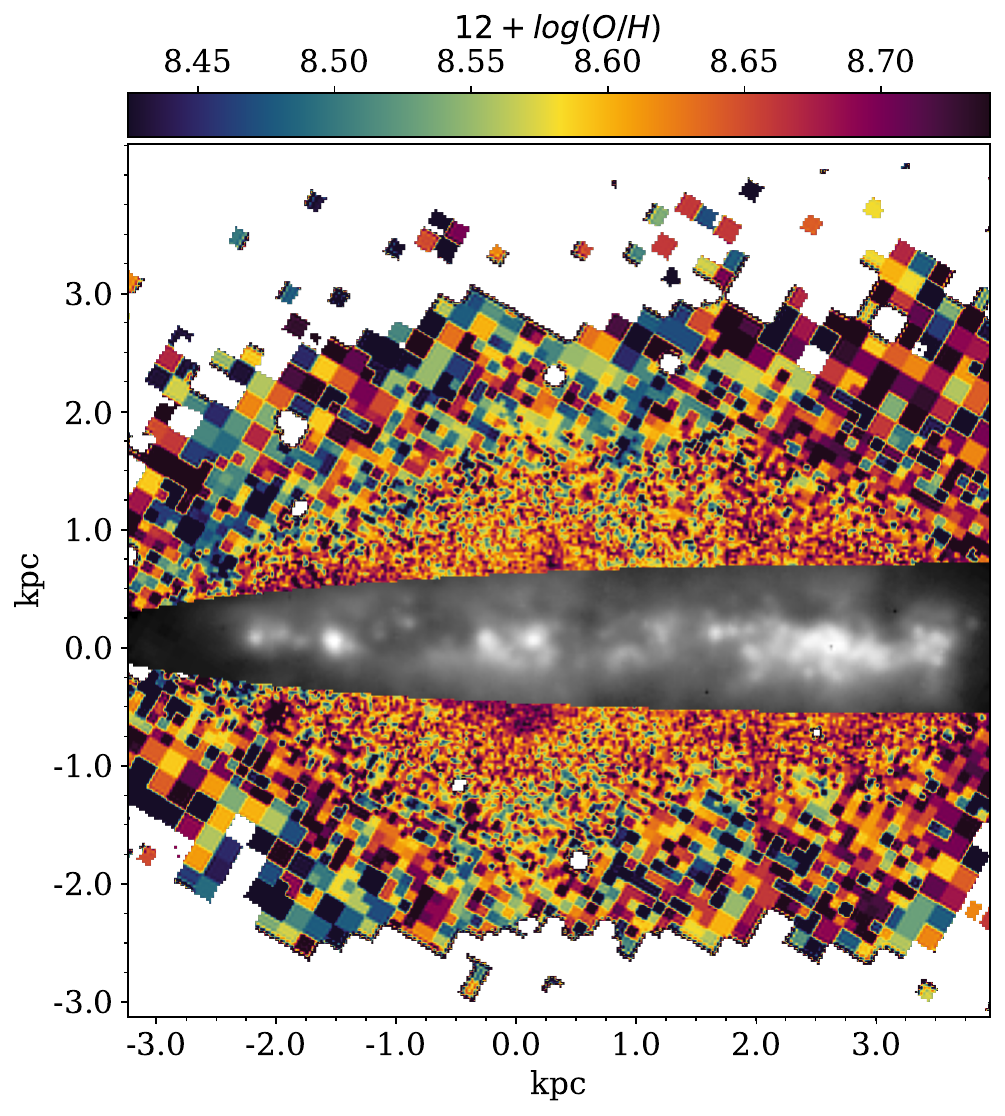}
    \includegraphics[width=0.33\textwidth]{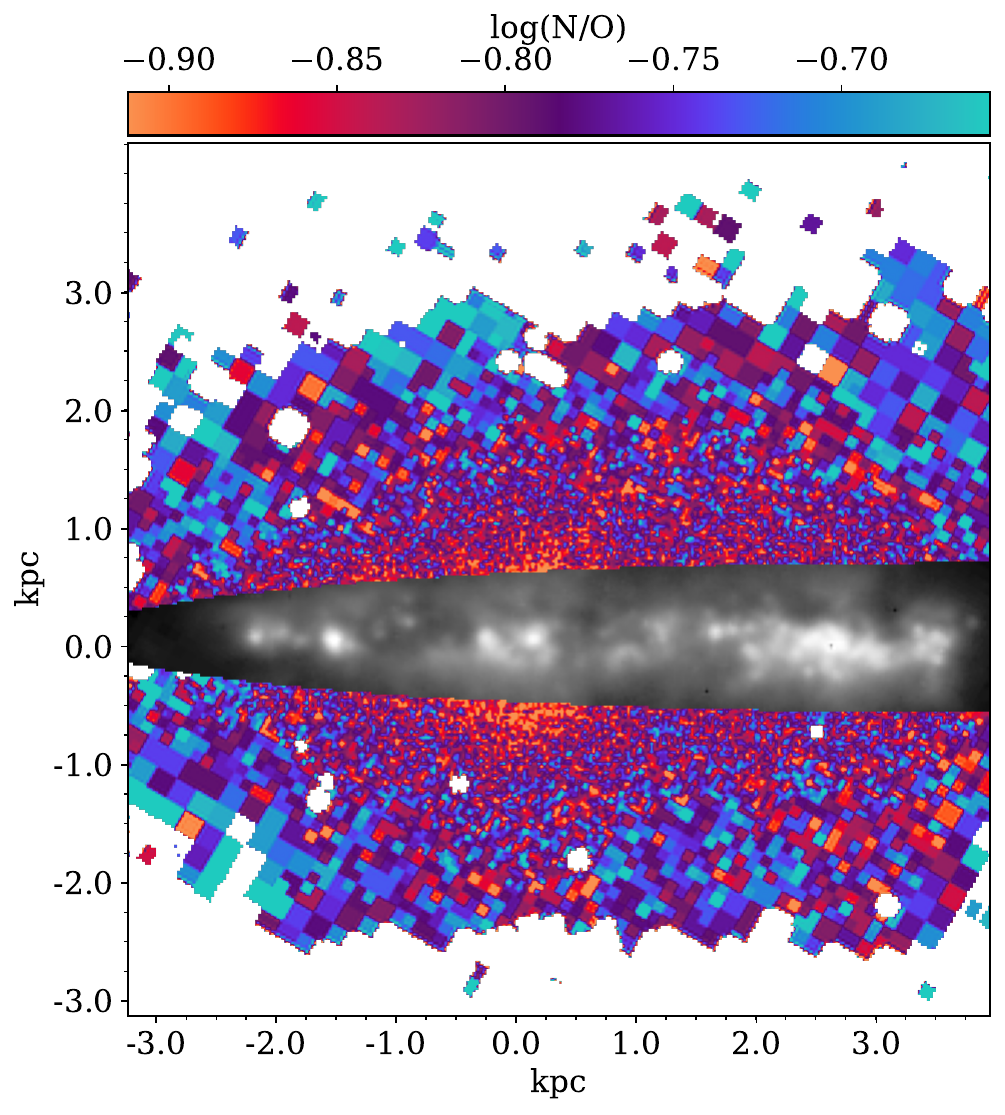}
    \includegraphics[width=0.33\textwidth]{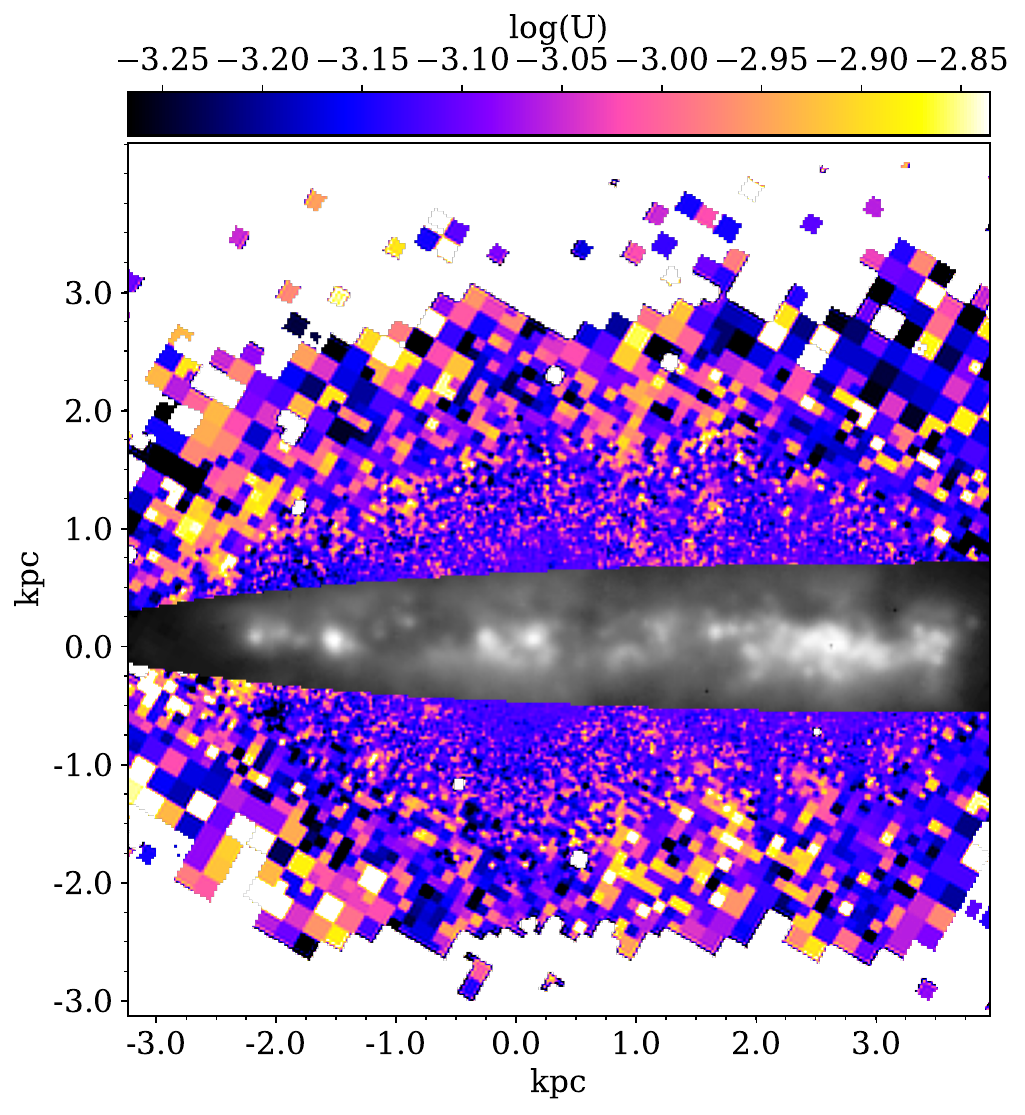}

    \includegraphics[width=0.331\textwidth]{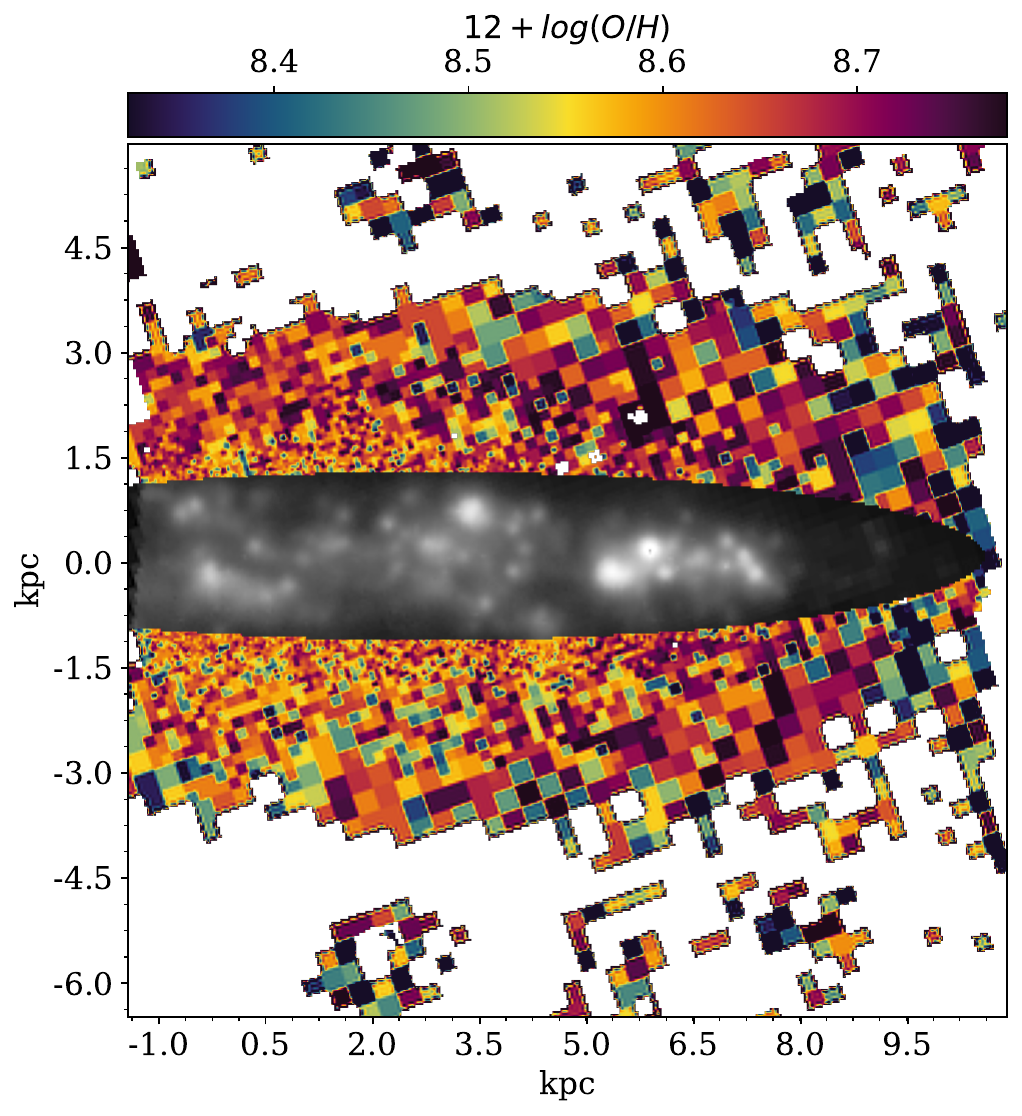}
    \includegraphics[width=0.33\textwidth]{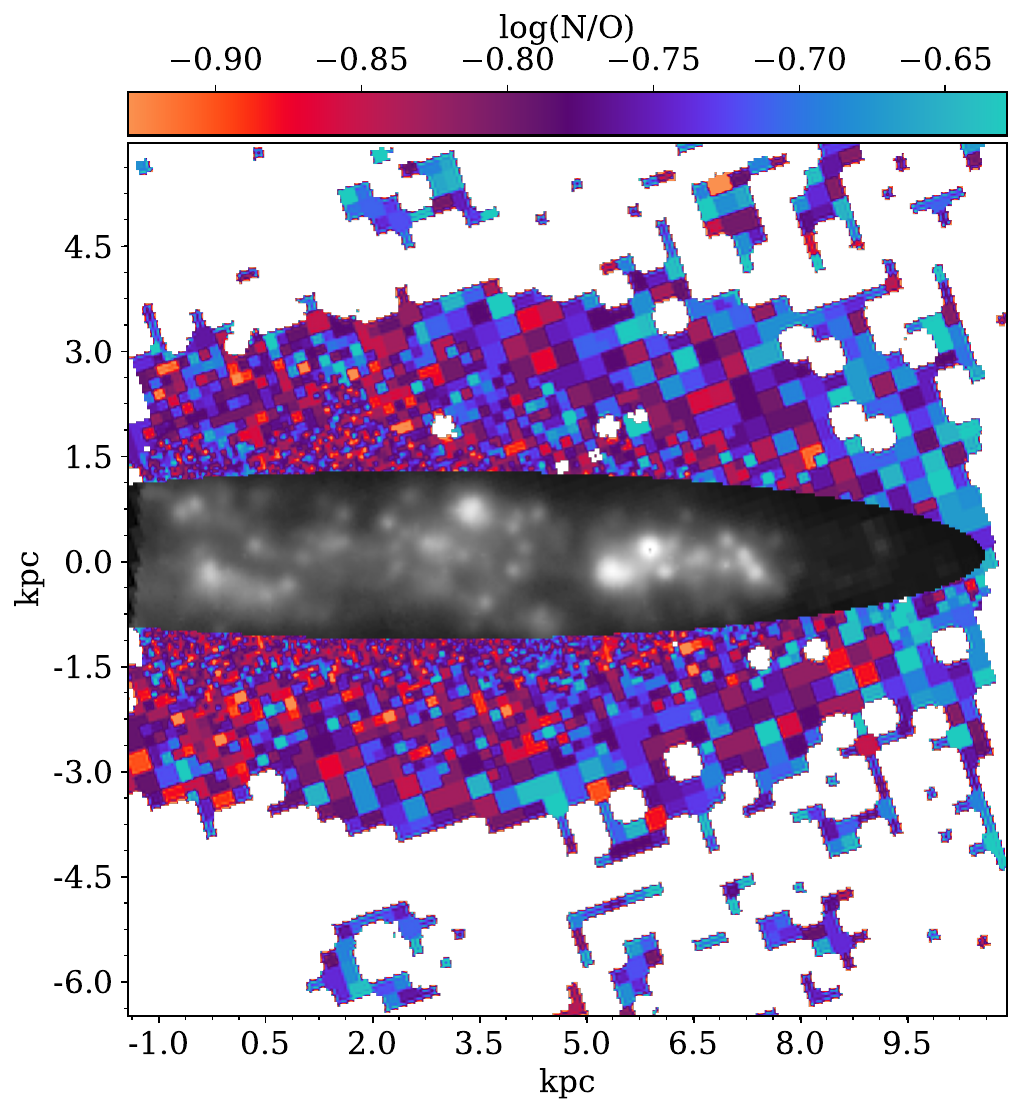}
    \includegraphics[width=0.33\textwidth]{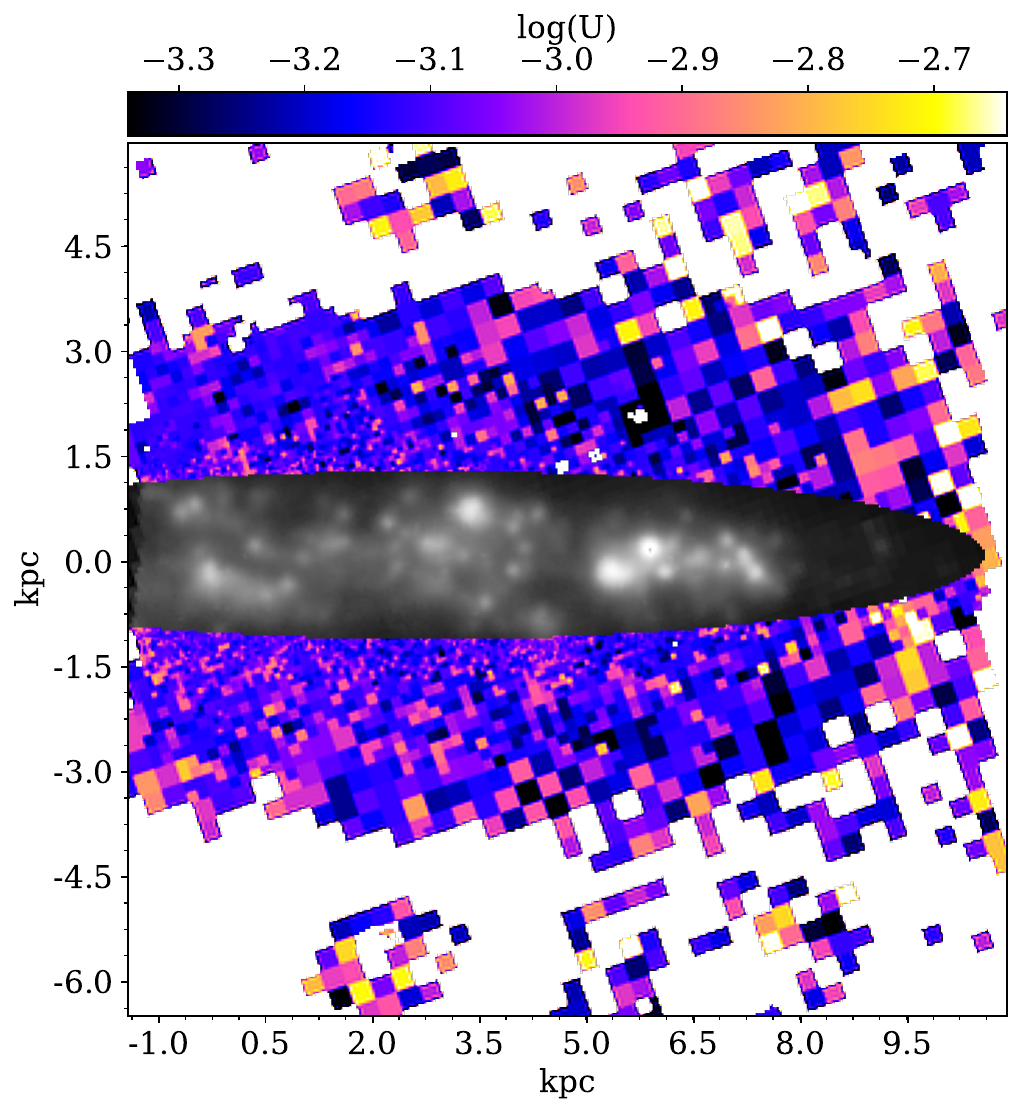}

\caption{Maps of the three free parameters of the \hcm models that best fit the observations. From left to right, the panels show the ISM gas-phase metallicity, $12+\log(\mathrm{O/H})$, the nitrogen-to-oxygen abundance ratio, $\log(\mathrm{N/O})$, and the ionisation parameter, $\log(U)$. From top to bottom, the galaxies are IC 1553, ESO157–49, ESO443–21, and ESO469–15 (in the continued figure). The H$\alpha$ emission from the galactic plane is overlaid on each map, and extraplanar \hii regions in ESO469–15 are masked. Only those bins with a spectroscopic S/N > 2 and a relative error lower than 40\% are shown.}
\label{fig:maps}     
\end{figure*}

\begin{figure*}[ht!]
\addtocounter{figure}{-1}
\centering

    \includegraphics[width=0.331\textwidth]{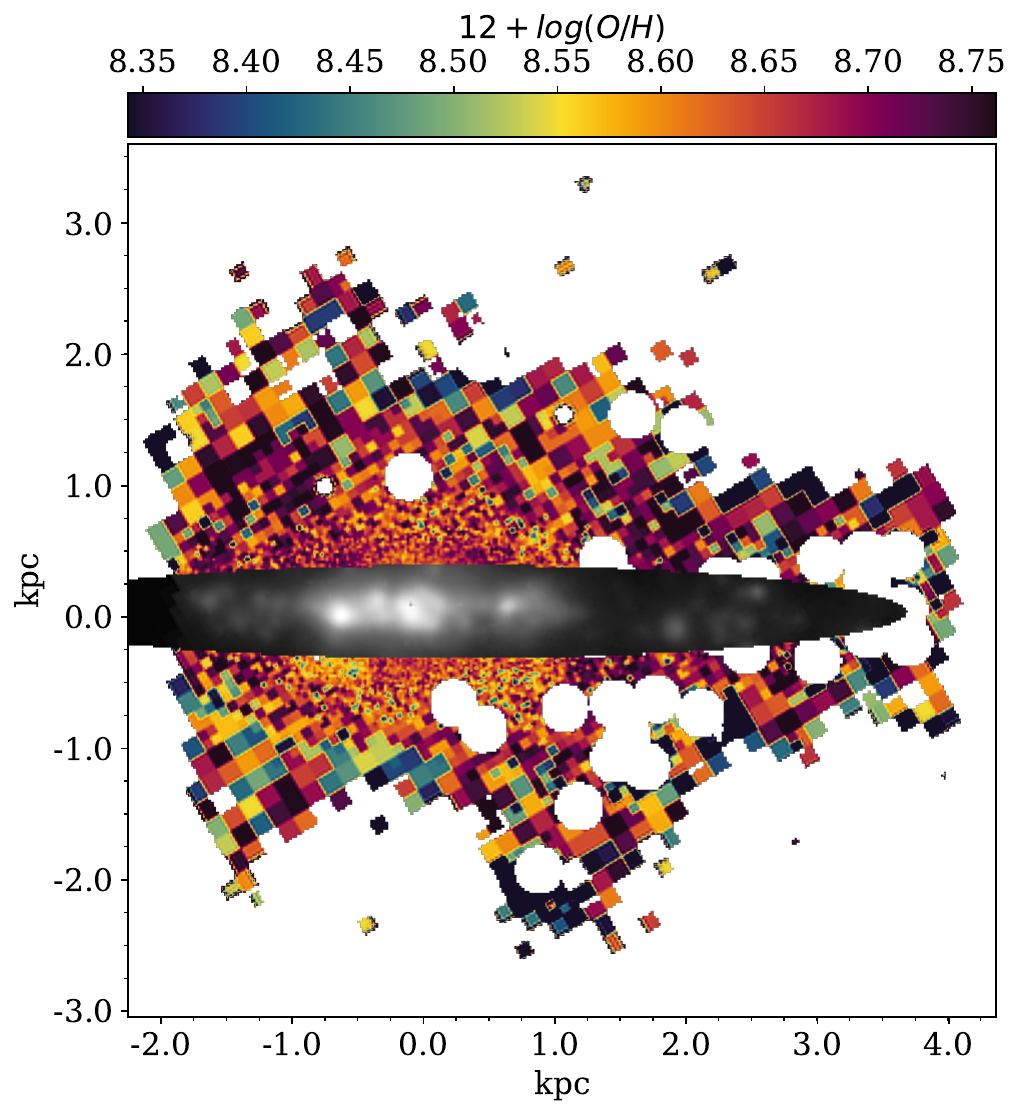}
    \includegraphics[width=0.33\textwidth]{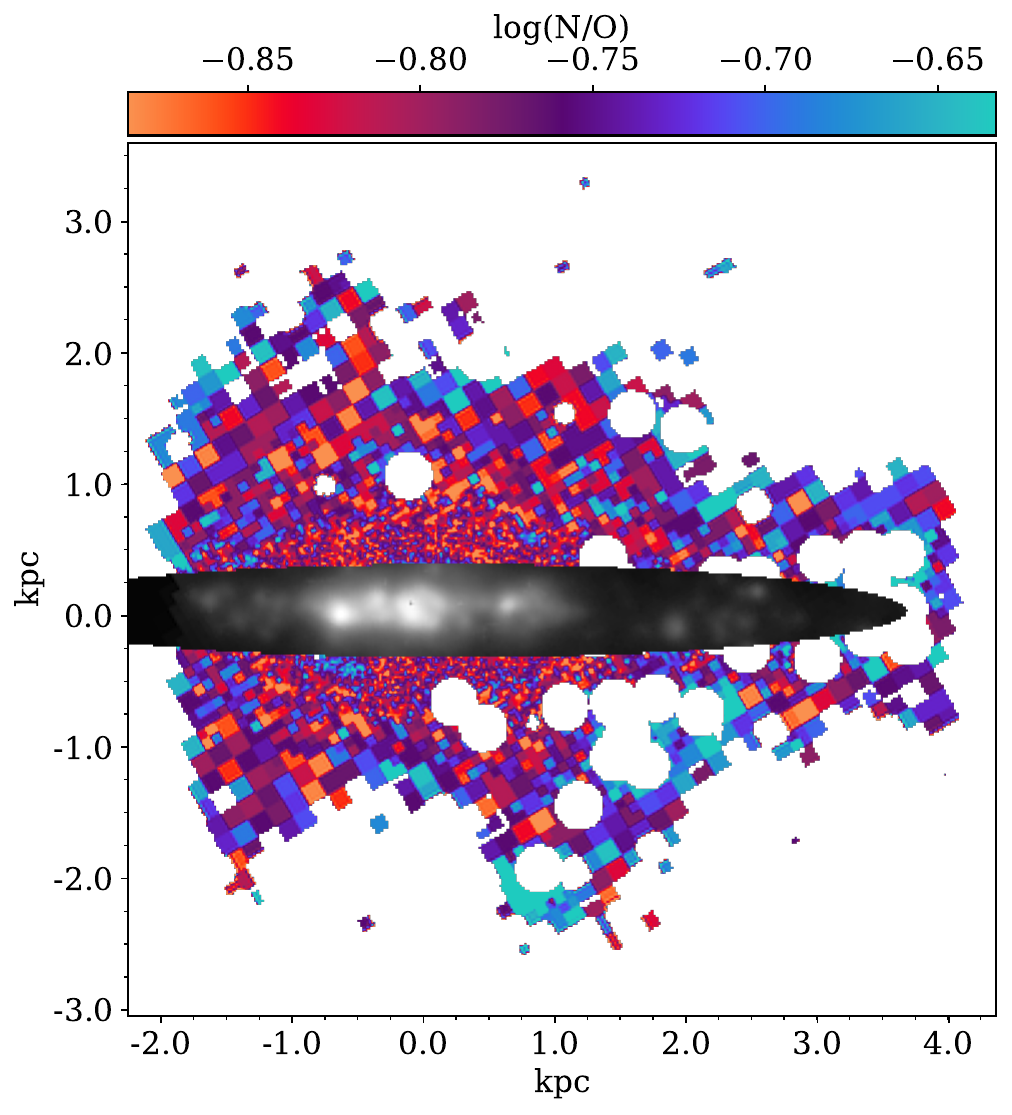}
    \includegraphics[width=0.33\textwidth]{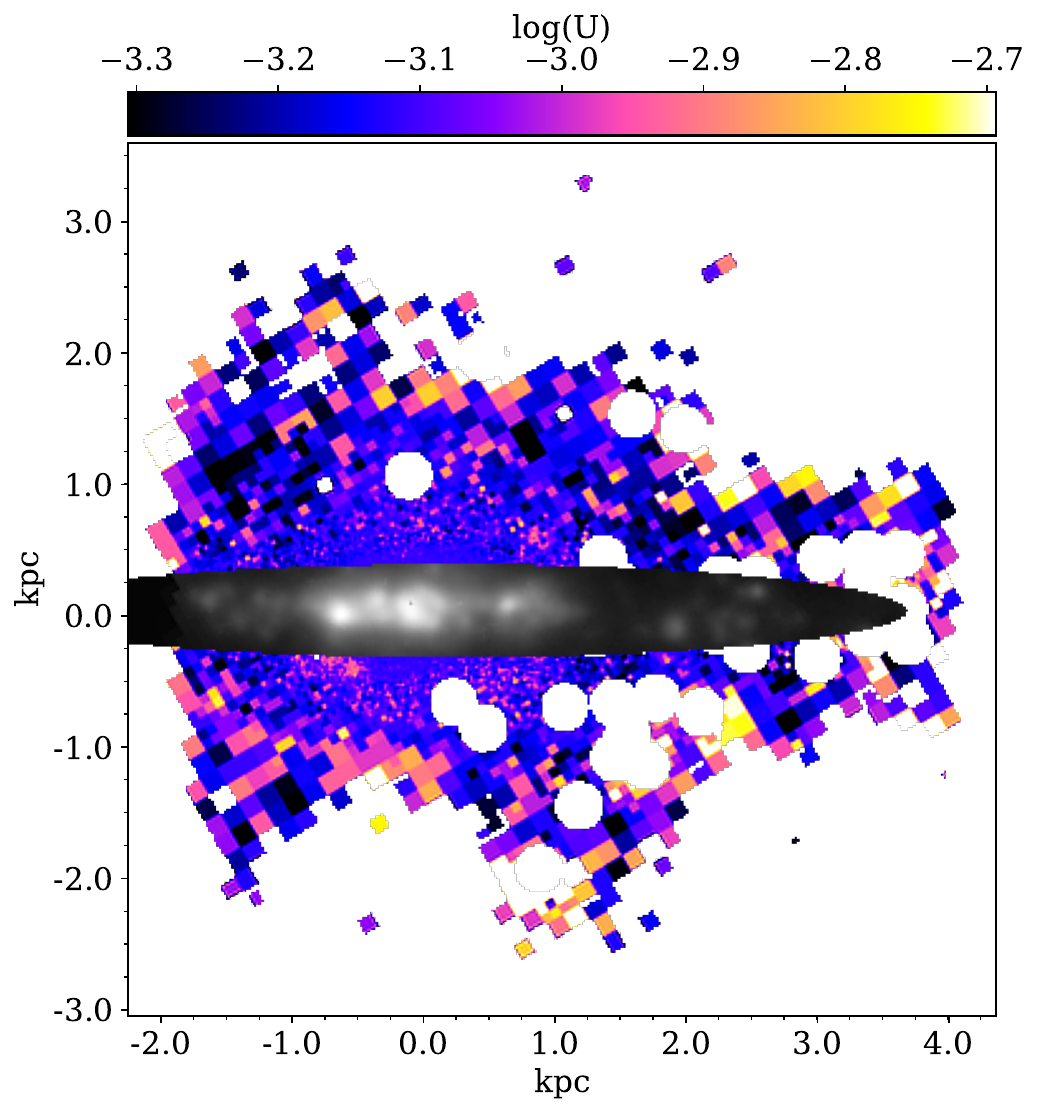}
    
\label{fig:maps_2}
\caption{Continued}
\end{figure*}

\begin{figure*}[ht!]
\centering
    \includegraphics[width=0.331\textwidth]{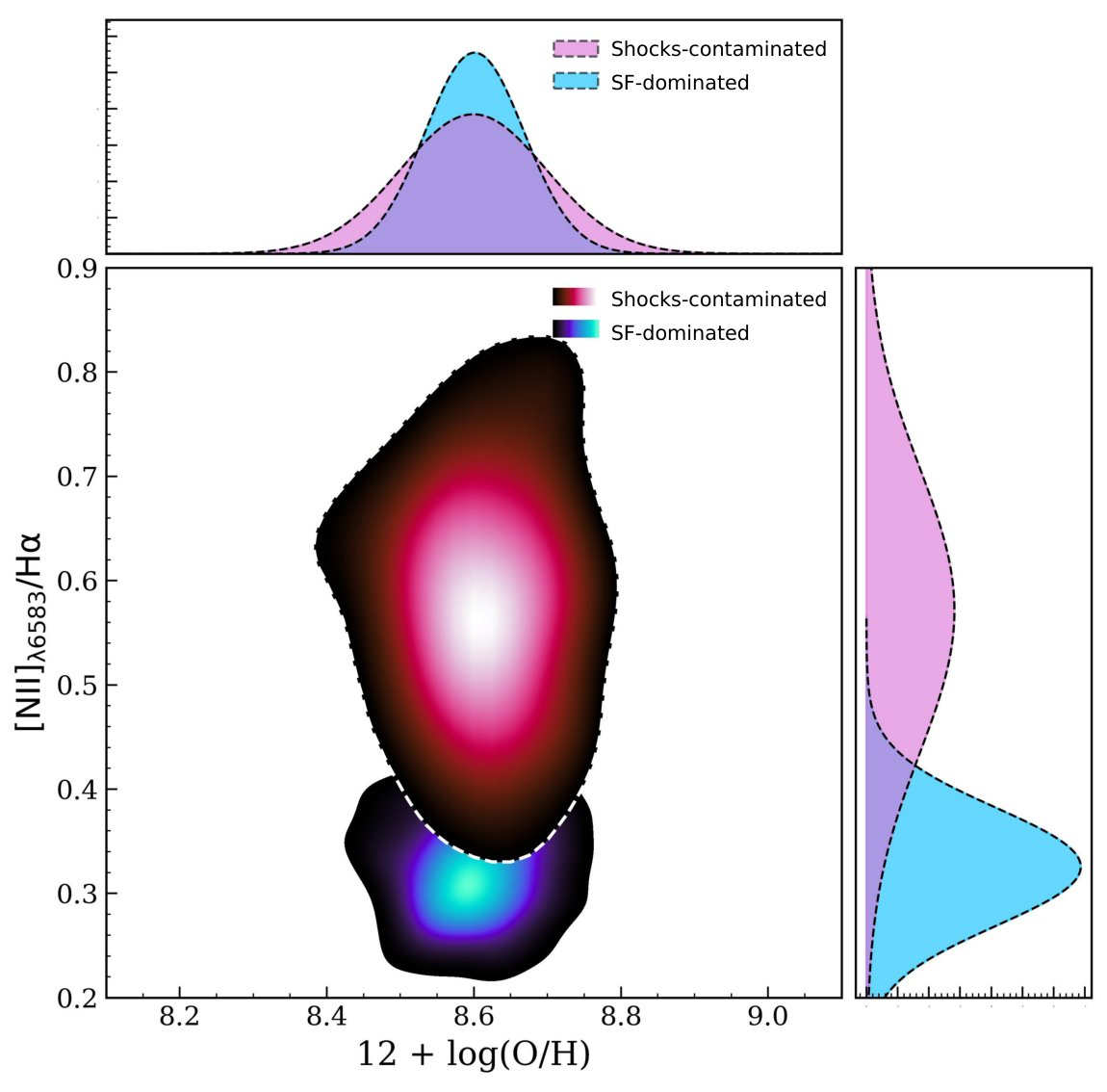}
    \includegraphics[width=0.33\textwidth]{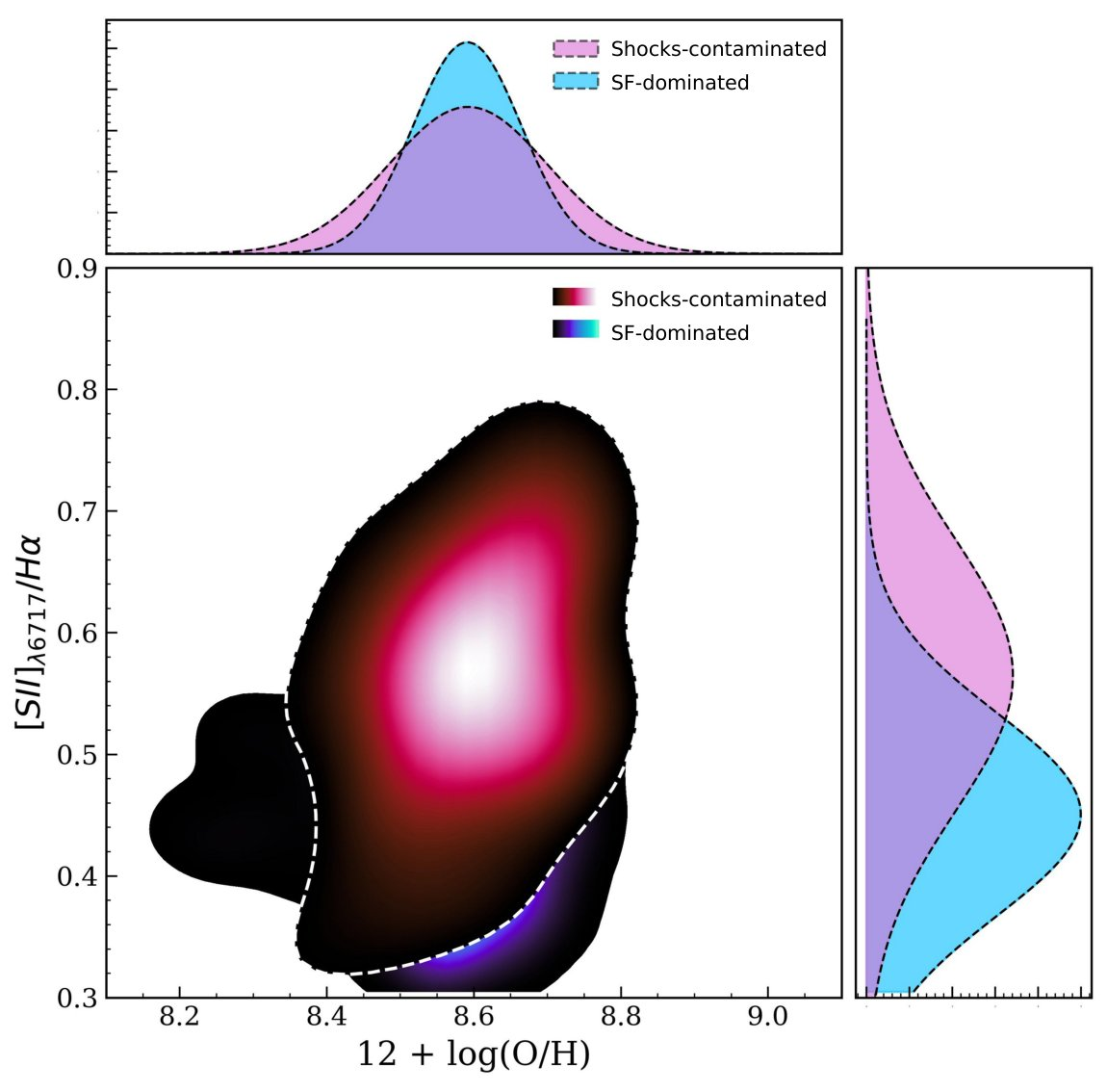}
    \includegraphics[width=0.33\textwidth]{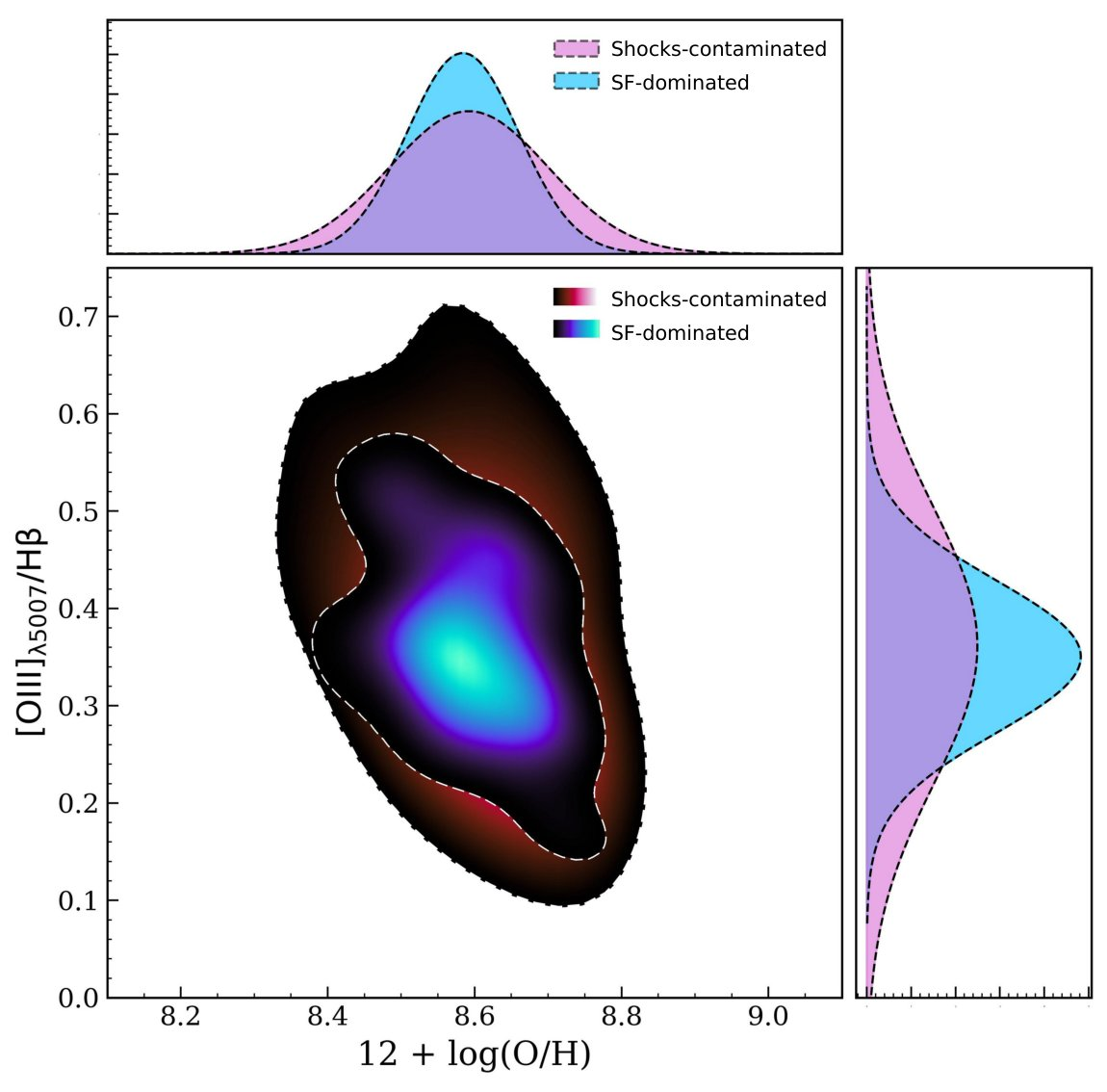}

    \includegraphics[width=0.331\textwidth]{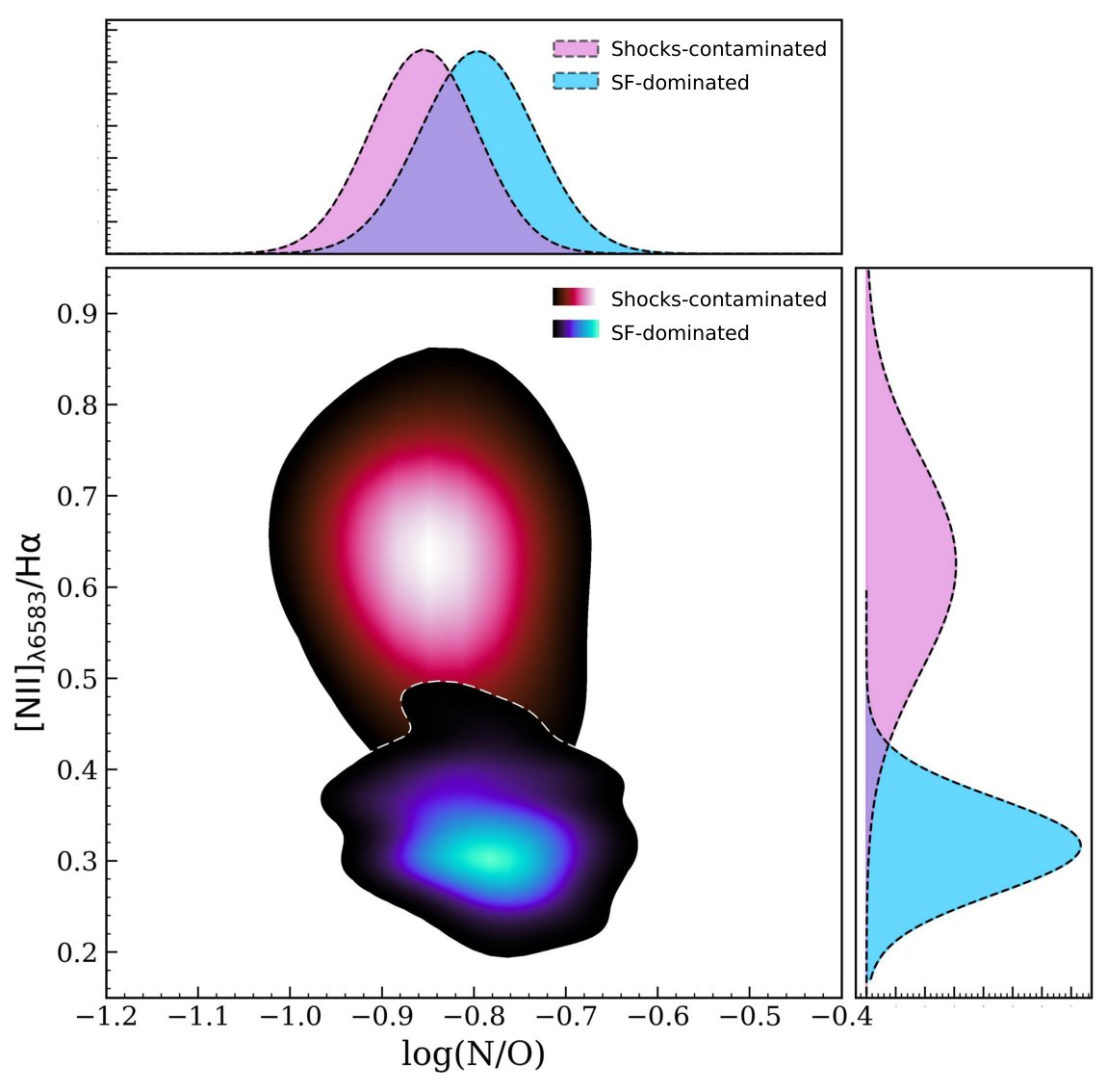}
    \includegraphics[width=0.33\textwidth]{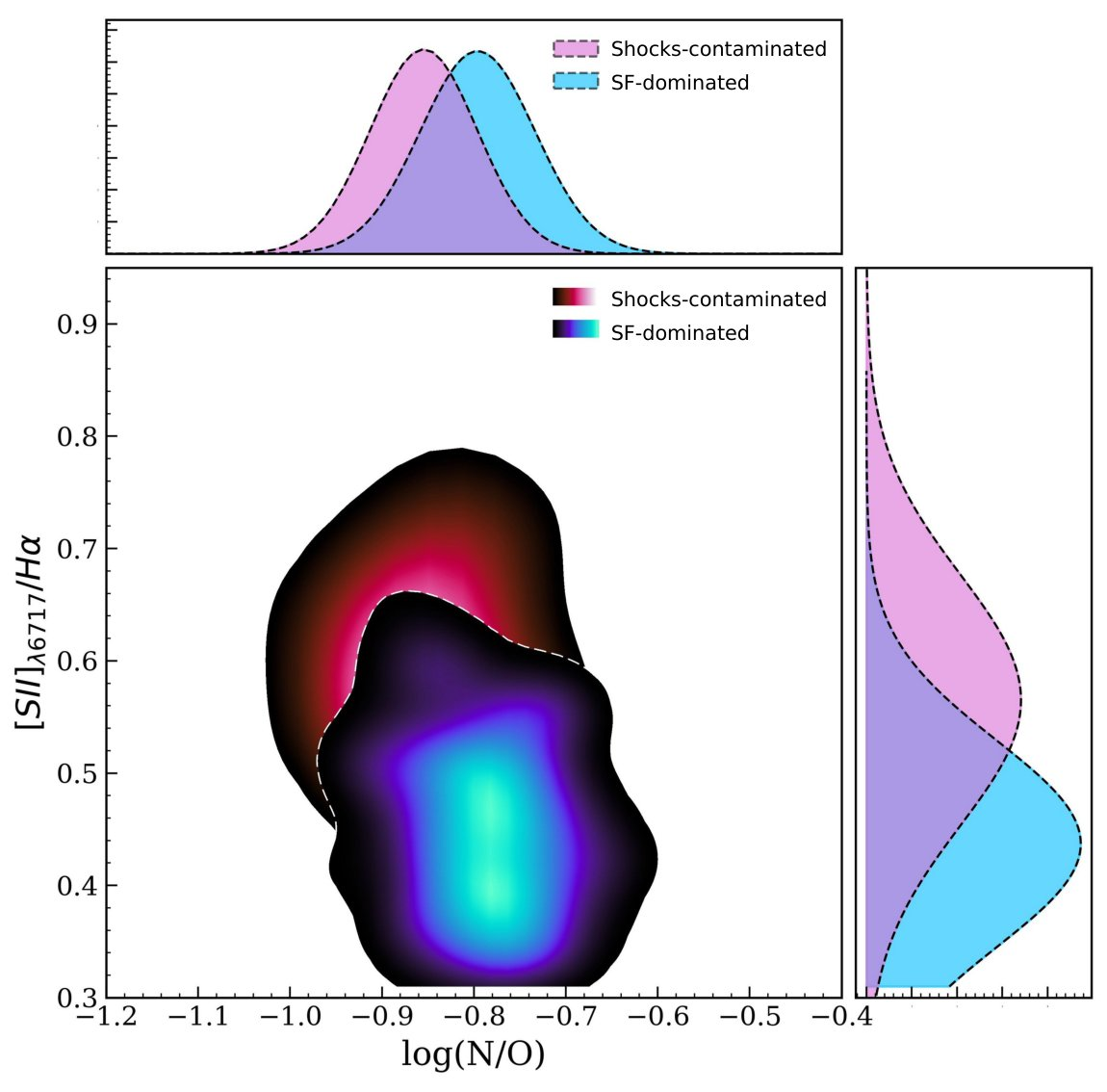}
    \includegraphics[width=0.33\textwidth]{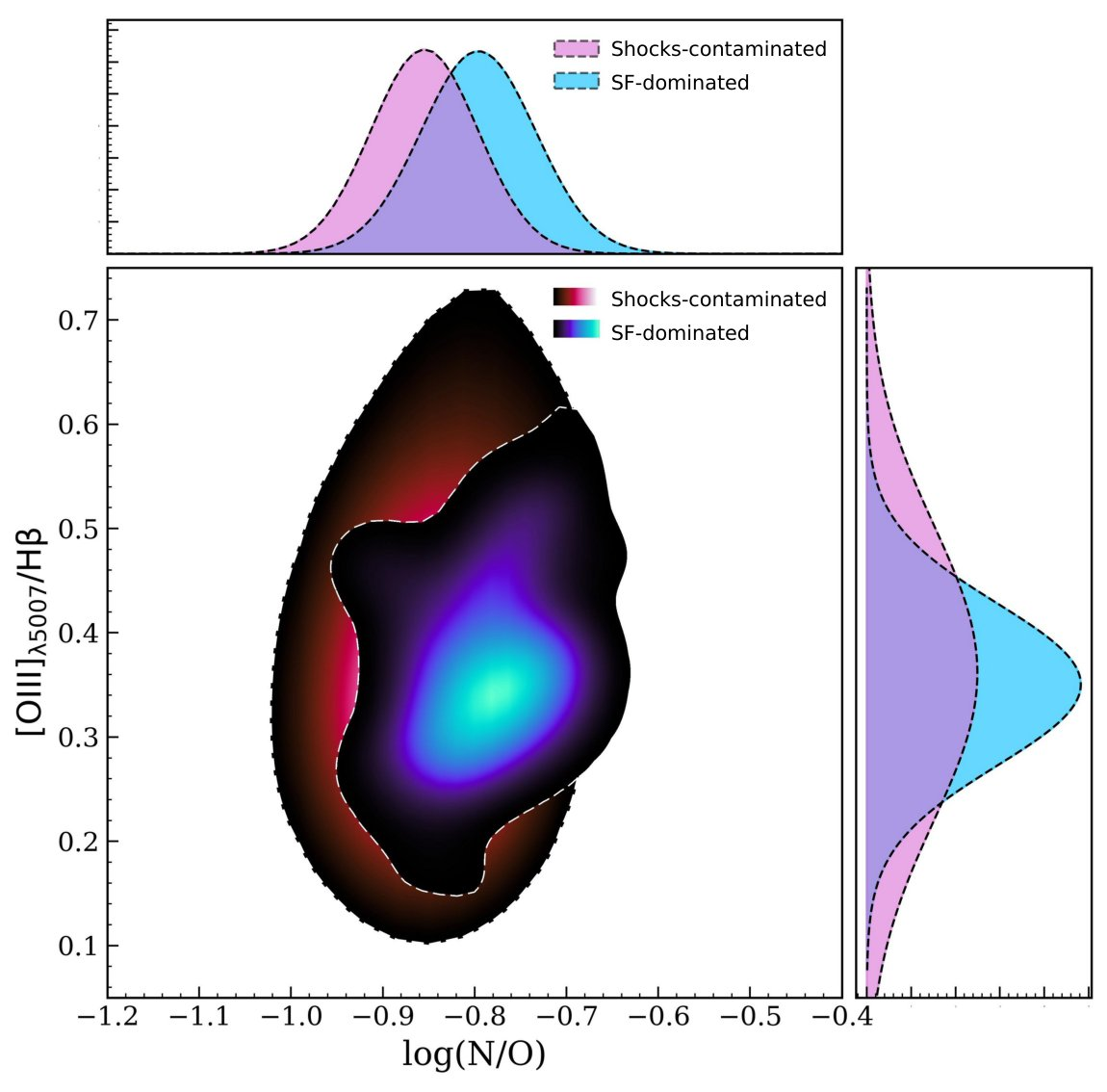}

    \includegraphics[width=0.331\textwidth]{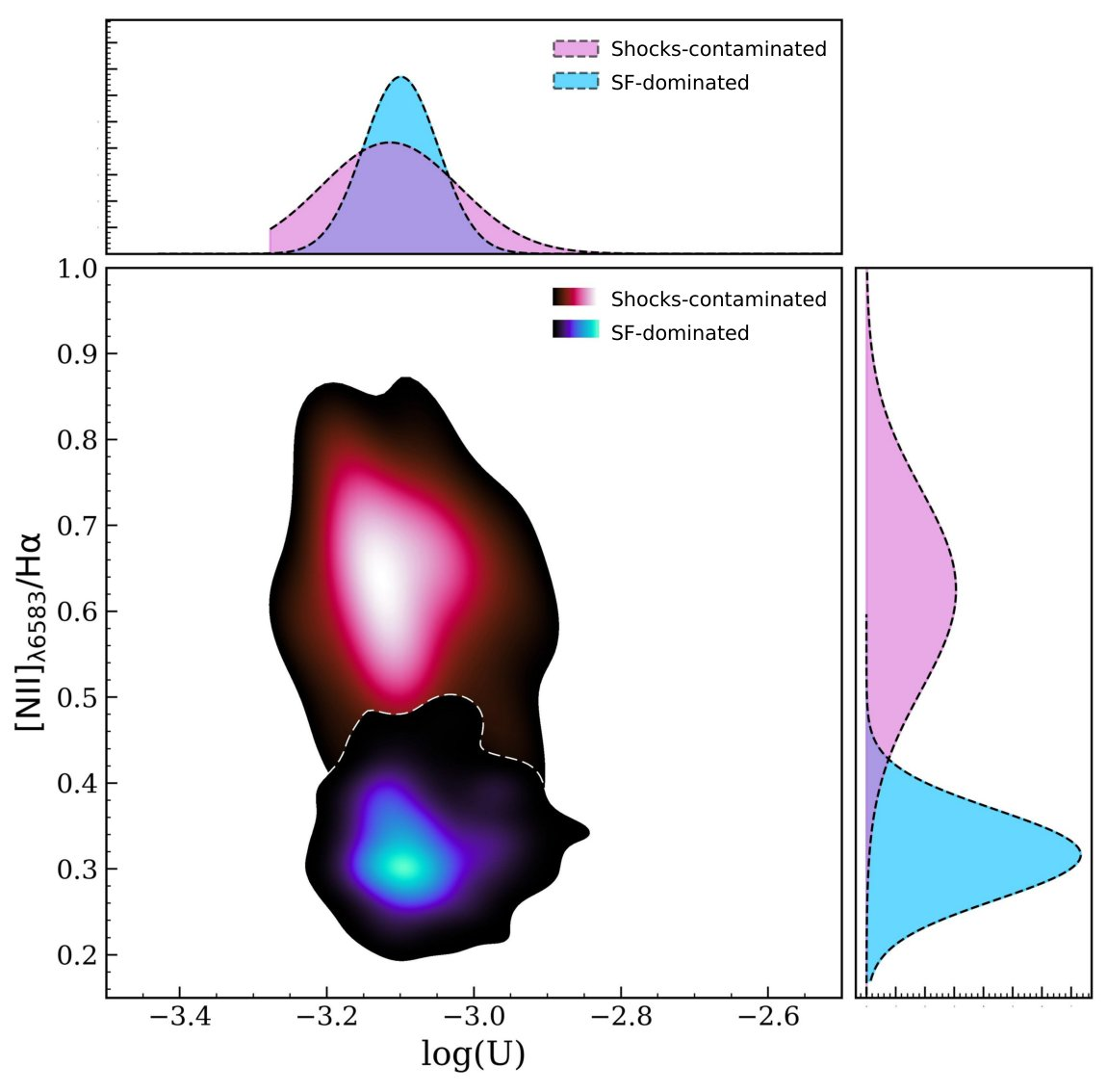}
    \includegraphics[width=0.33\textwidth]{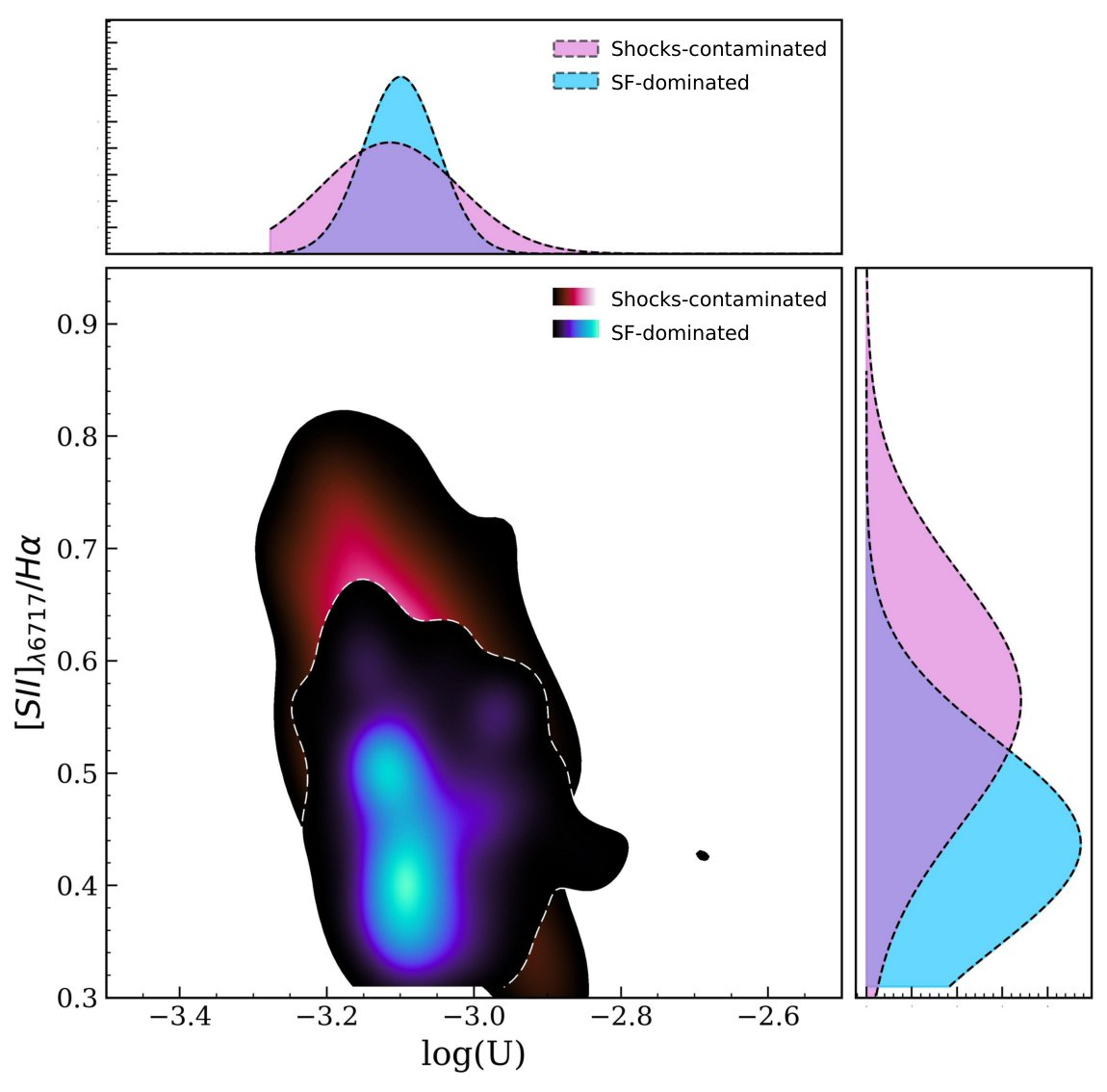}
    \includegraphics[width=0.33\textwidth]{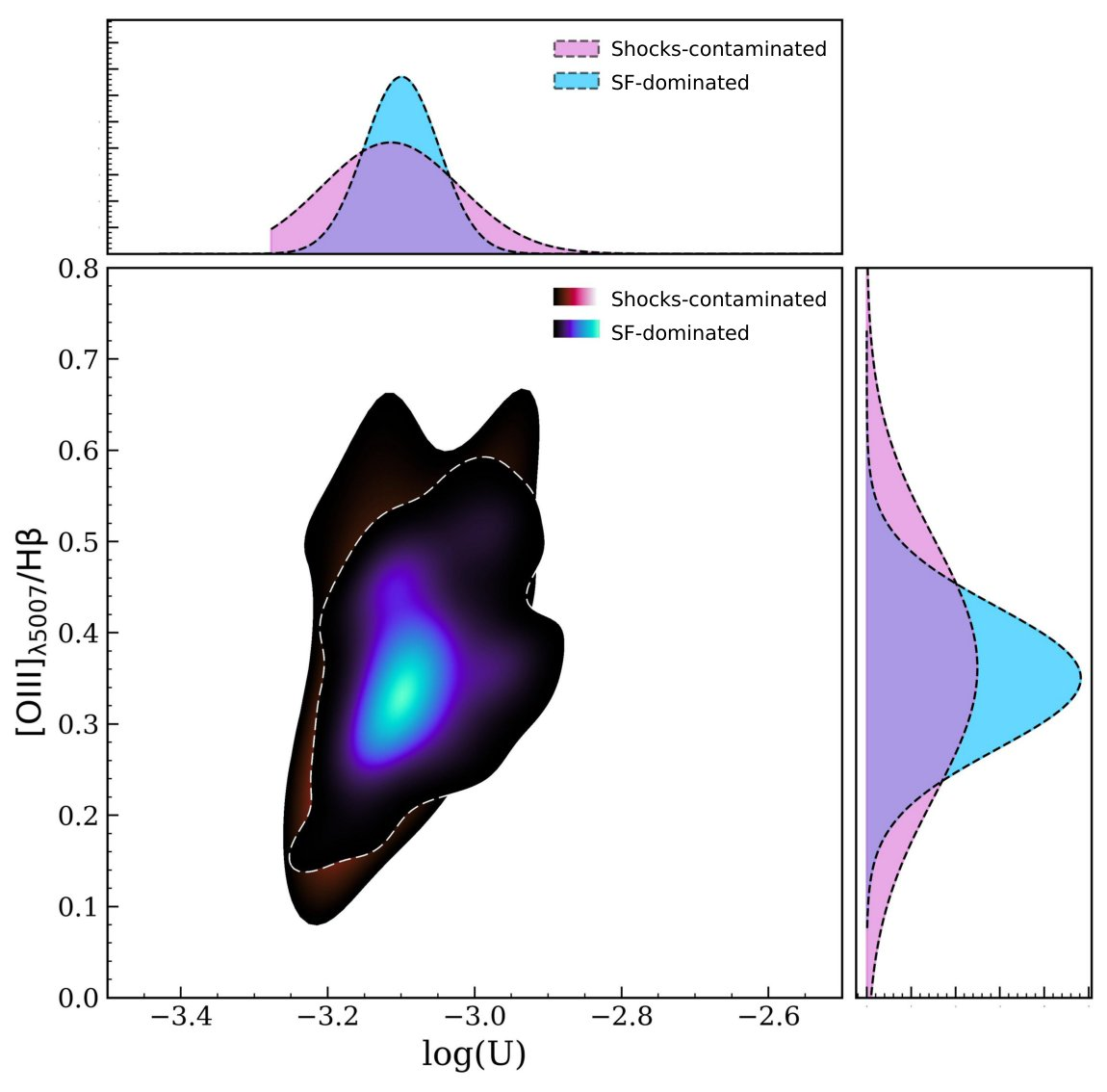}

\caption{2D distributions of the three \hcm free parameters versus line ratios for IC1553. From left to right: \nii/\ha, \sii/\ha, \oiii/\hb. From top to bottom: 12+log(O/H), log(N/O), log(U). Each 2D distribution is smoothed by a Gaussian kernel and divided into two components: the blue region contains eDIG bins dominated by star formation, and the red region contains the bins where the contribution of fast shocks reaches its maximum, as defined in Sect. \ref{sec:3_abundancias}. Additionally, we show in each X and Y margin of every panel the 1D distribution of the corresponding parameter and line ratio for the star-formation-dominated (blue) and shock-contaminated (magenta) regions.}
\label{fig:IC1553_disps}     
\end{figure*}

\begin{figure*}[ht!]
\centering
    \includegraphics[width=0.331\textwidth]{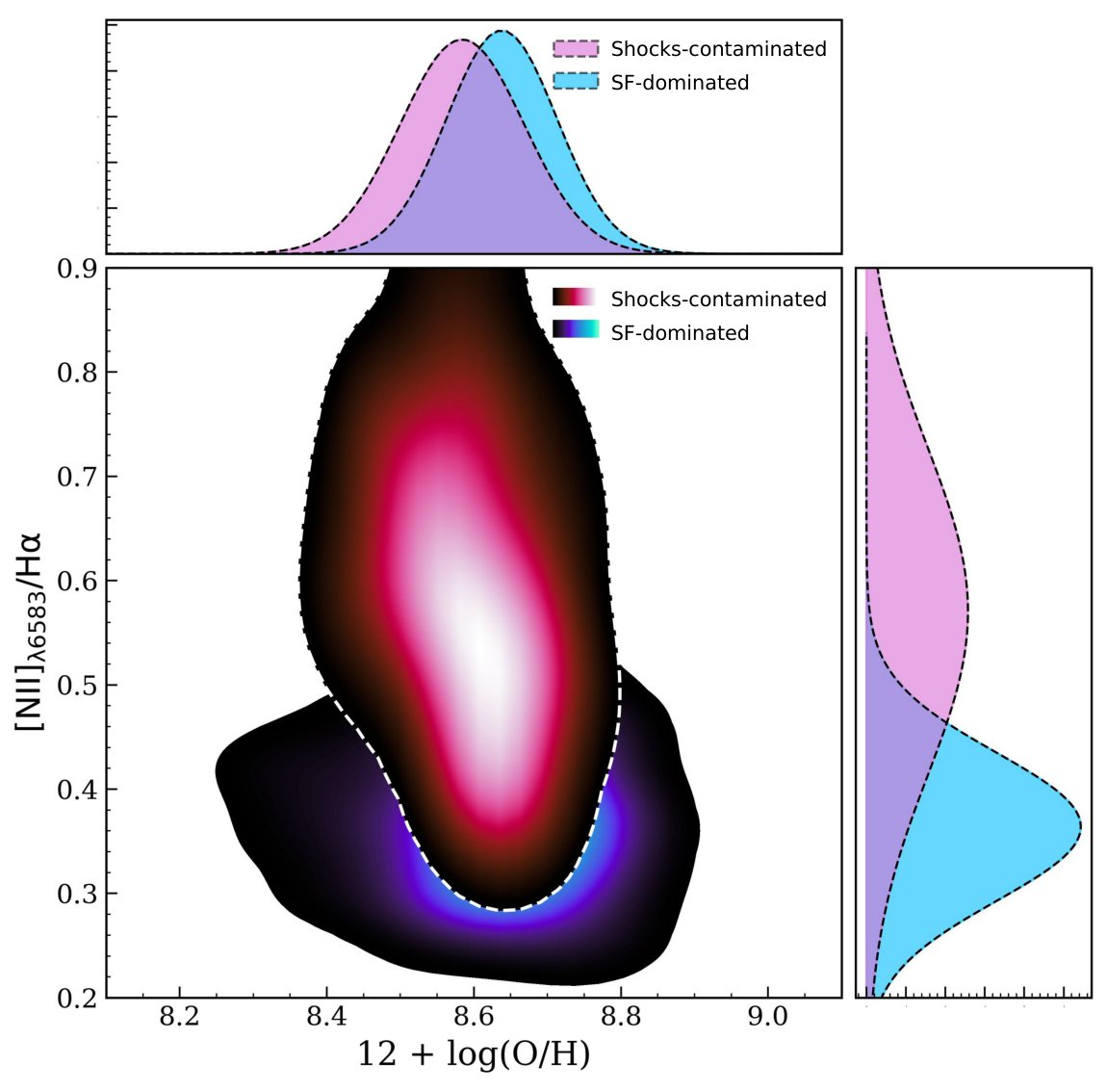}
    \includegraphics[width=0.33\textwidth]{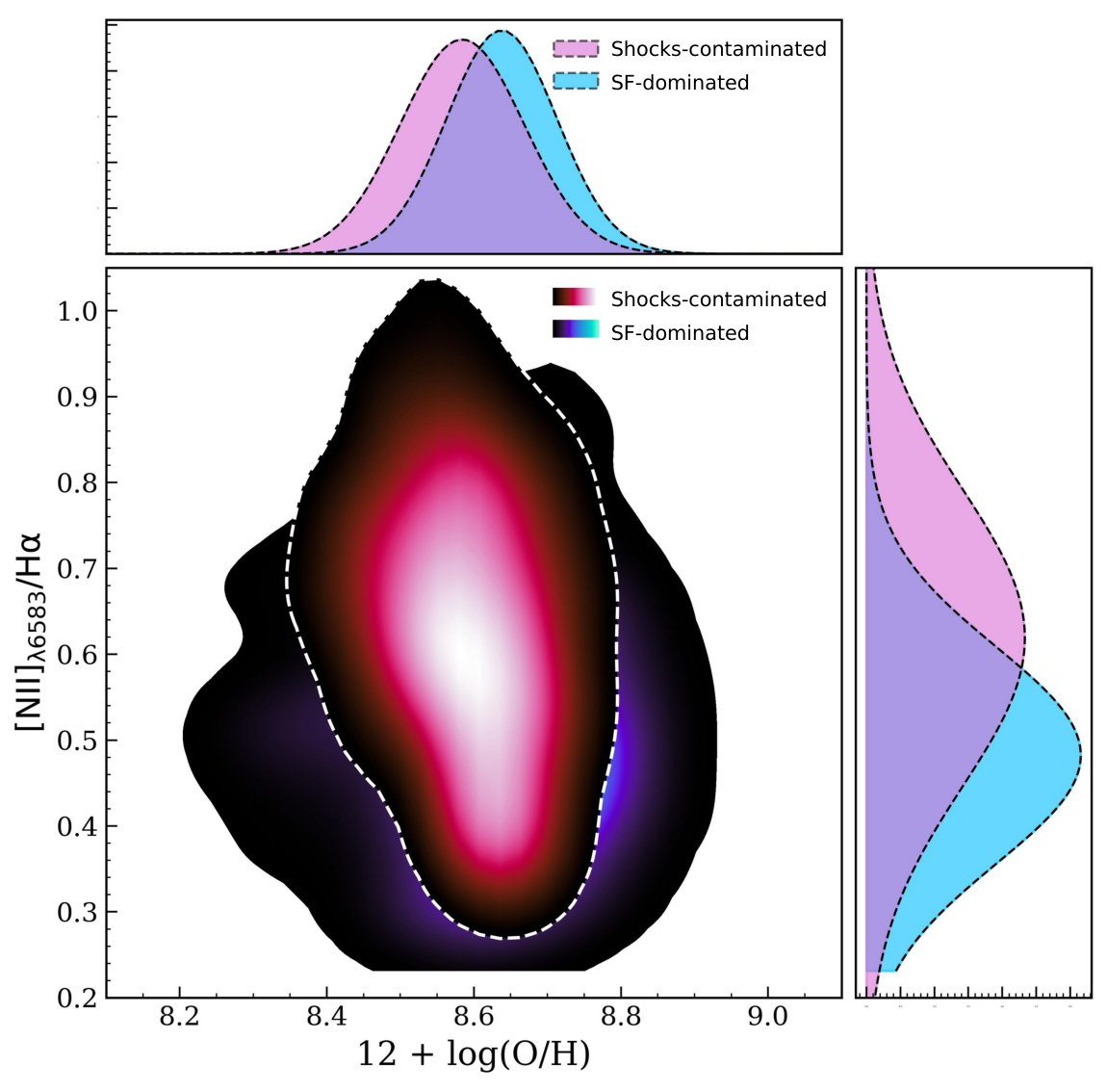}
    \includegraphics[width=0.33\textwidth]{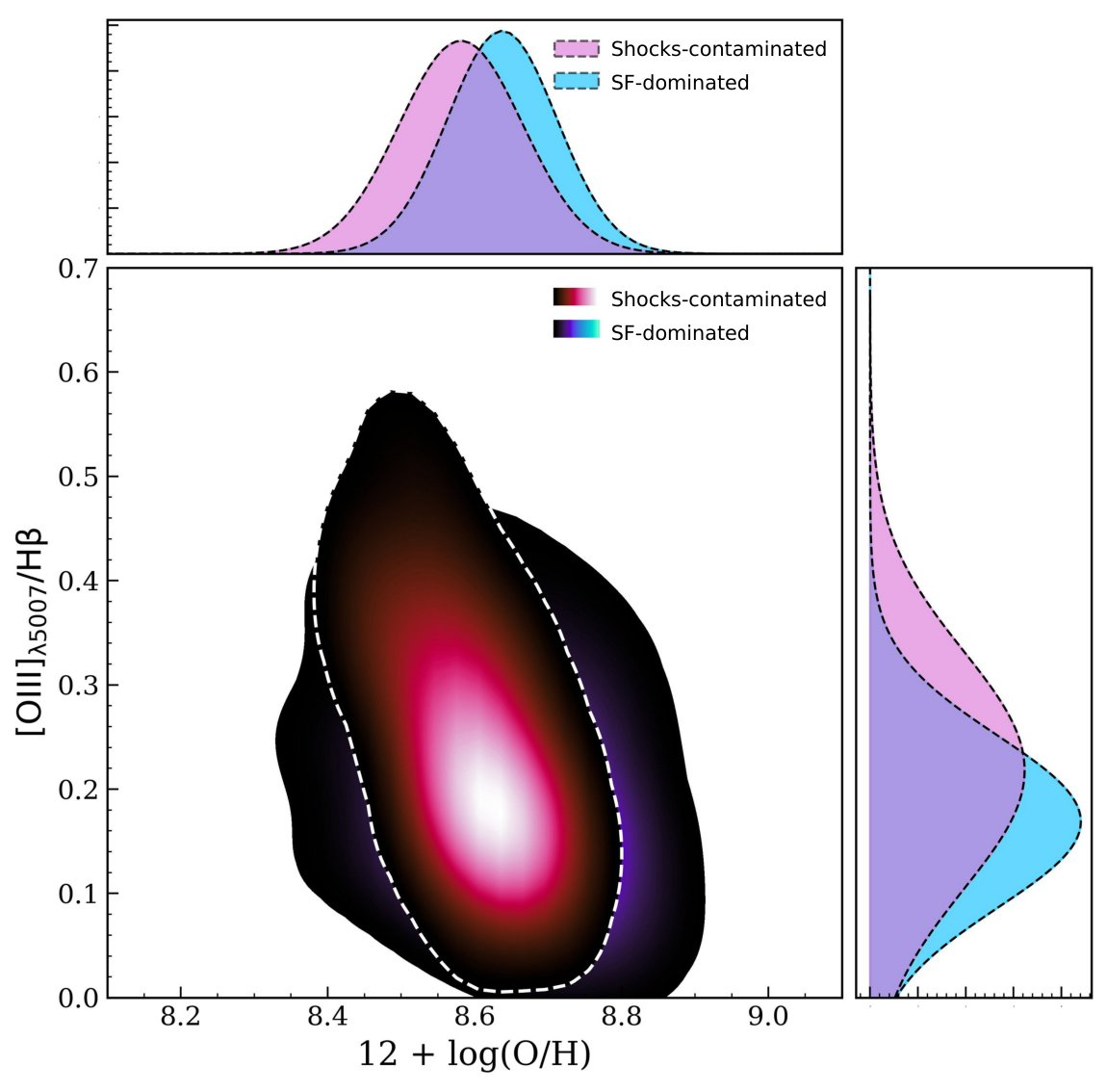}

    \includegraphics[width=0.331\textwidth]{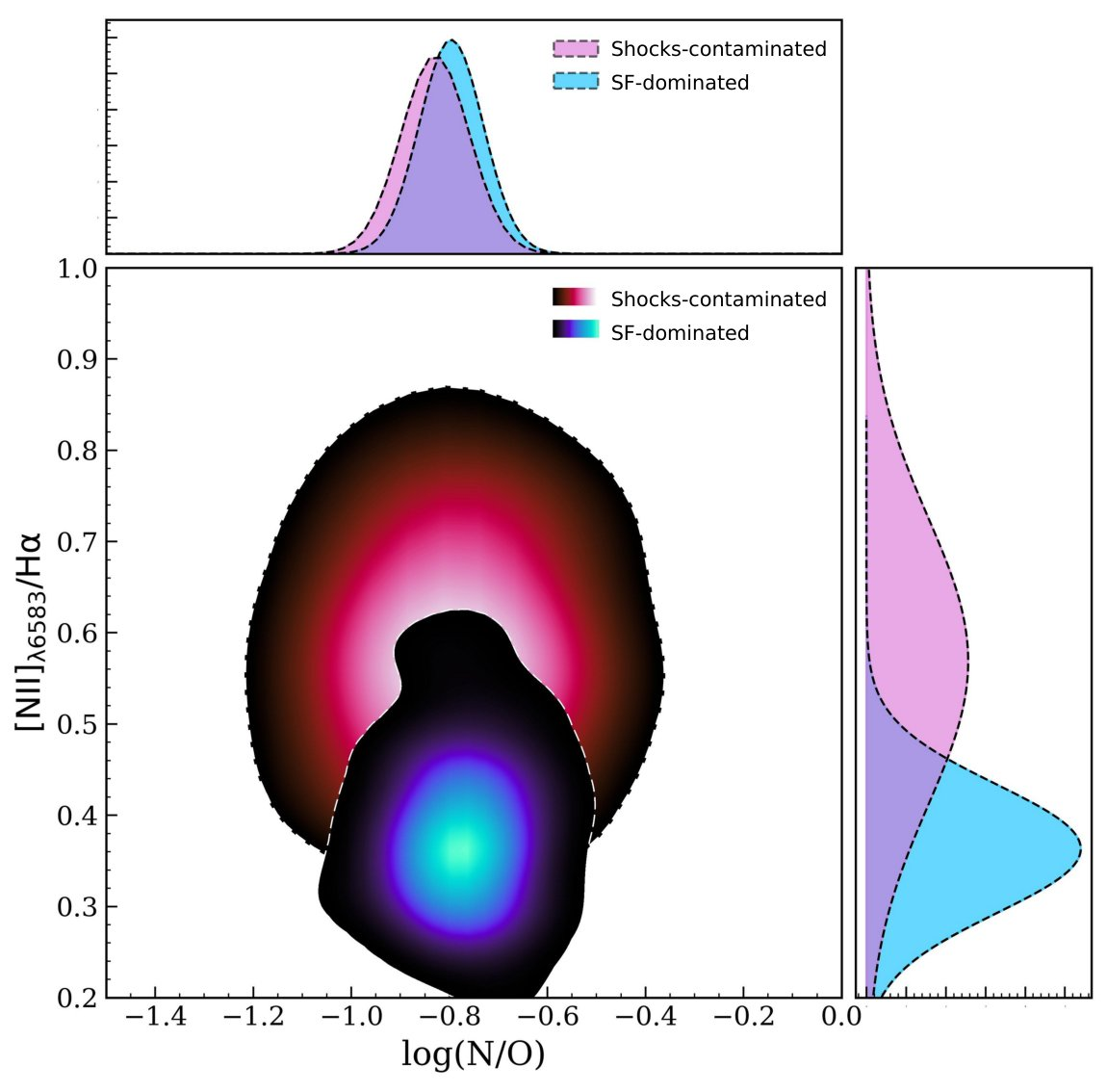}
    \includegraphics[width=0.33\textwidth]{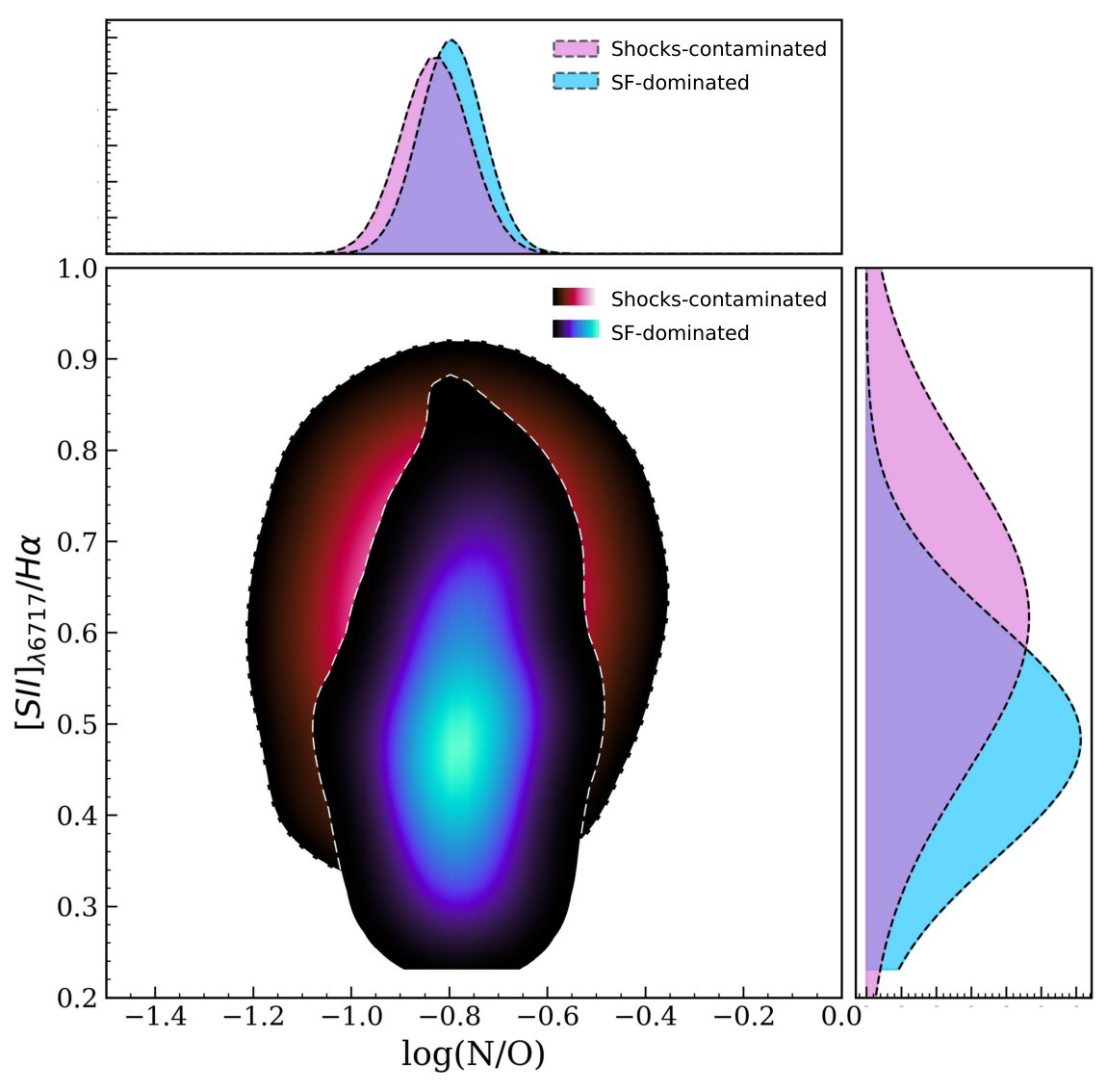}
    \includegraphics[width=0.33\textwidth]{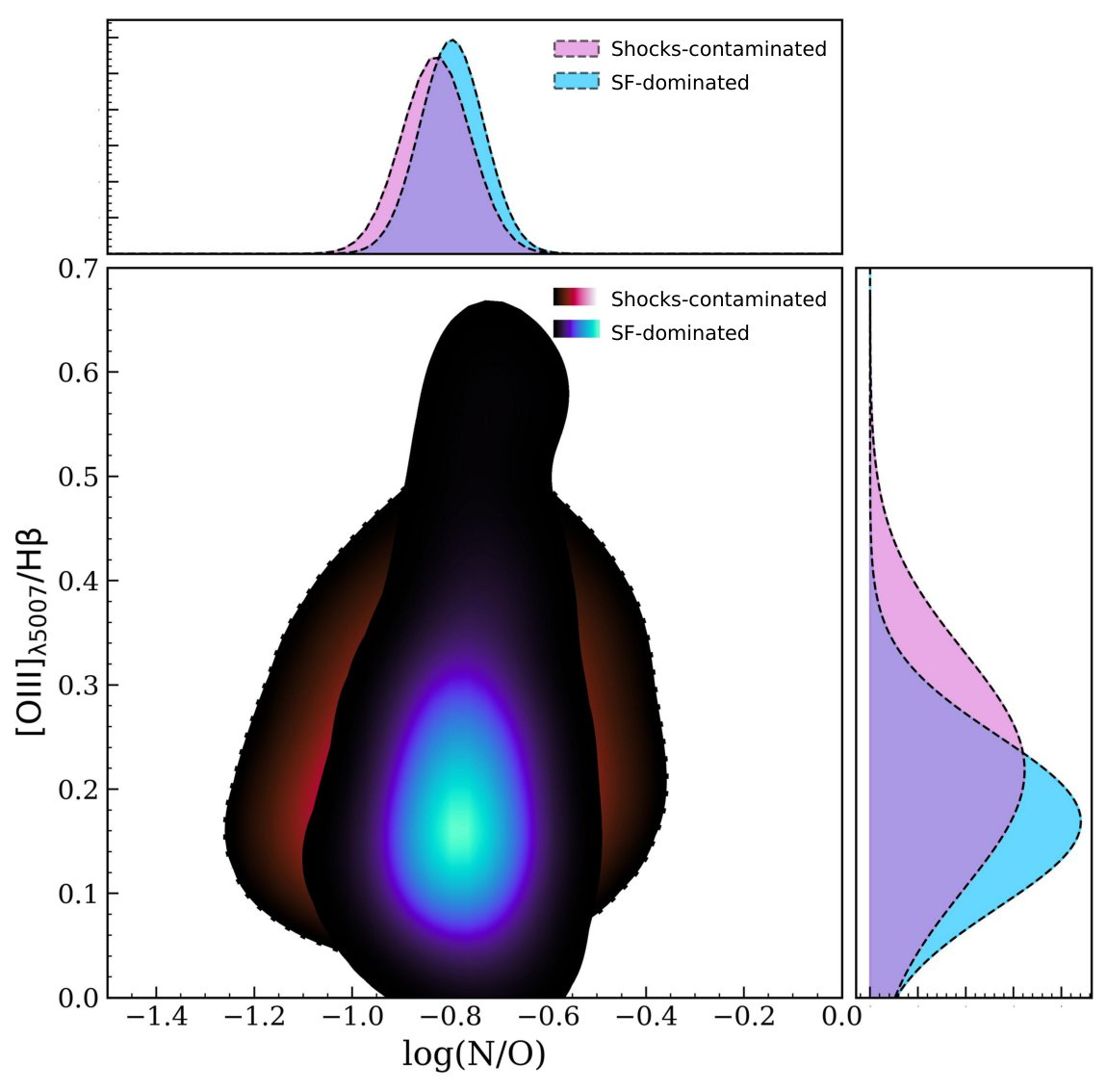}

    \includegraphics[width=0.331\textwidth]{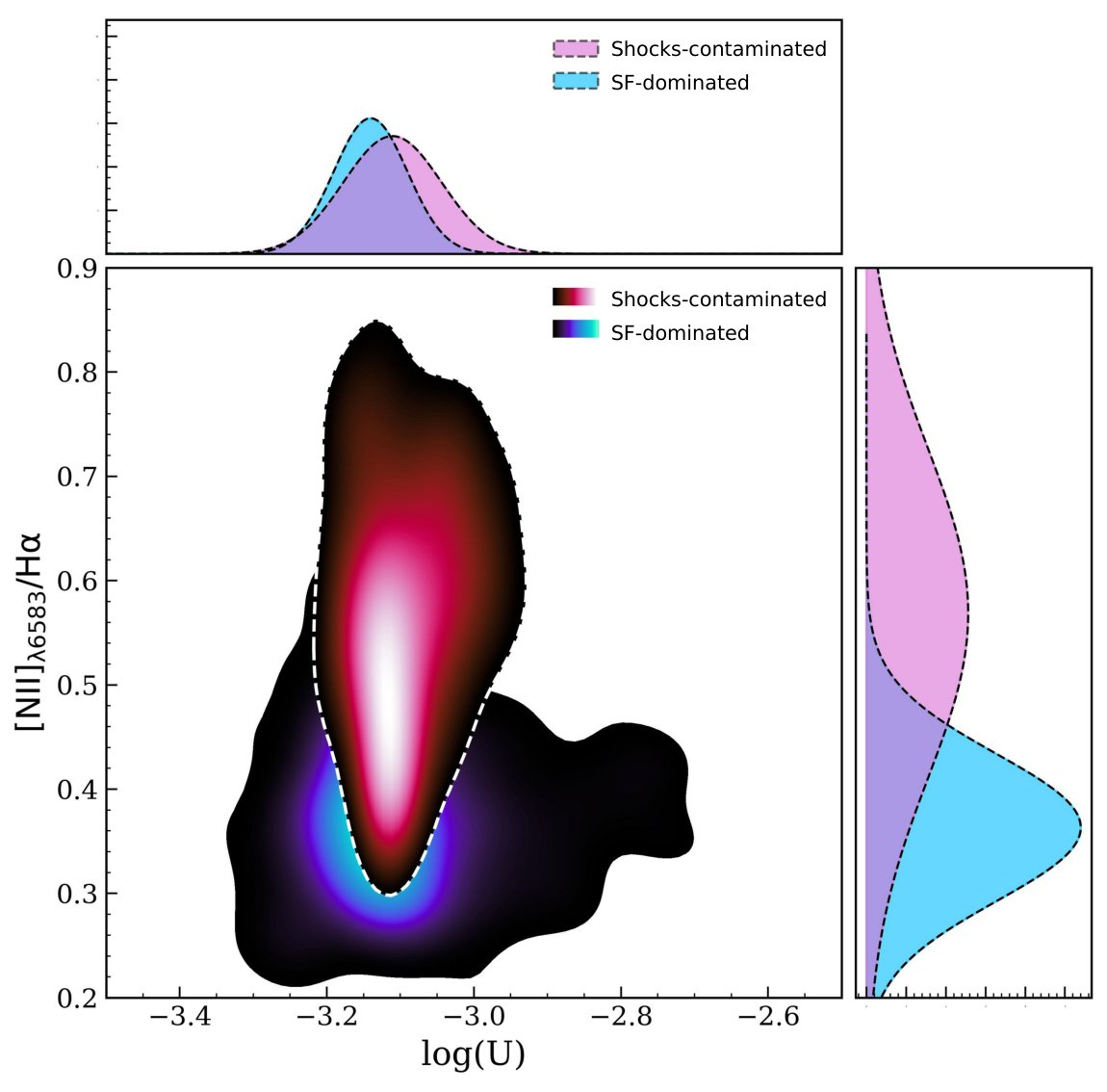}
    \includegraphics[width=0.33\textwidth]{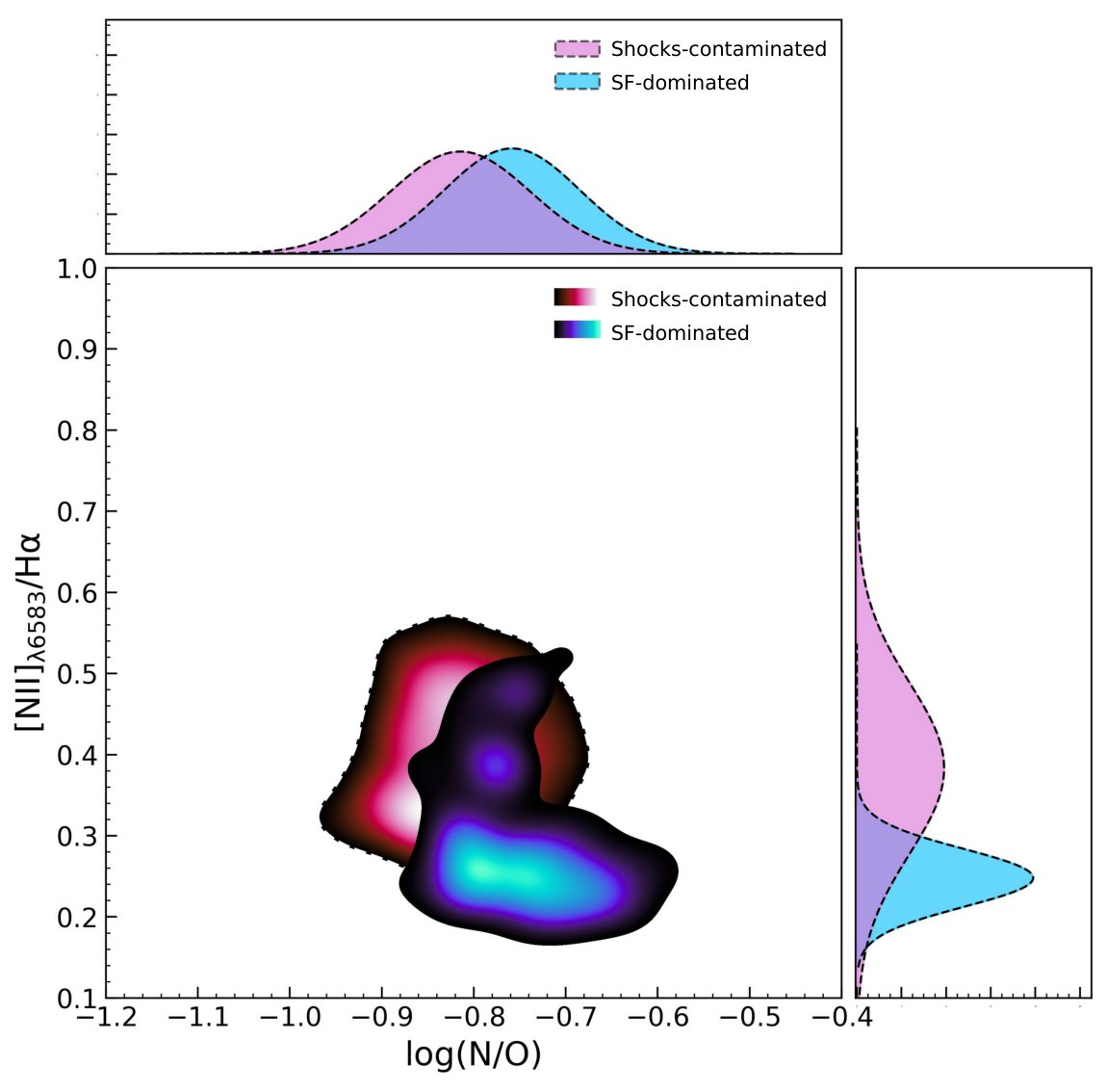}
    \includegraphics[width=0.33\textwidth]{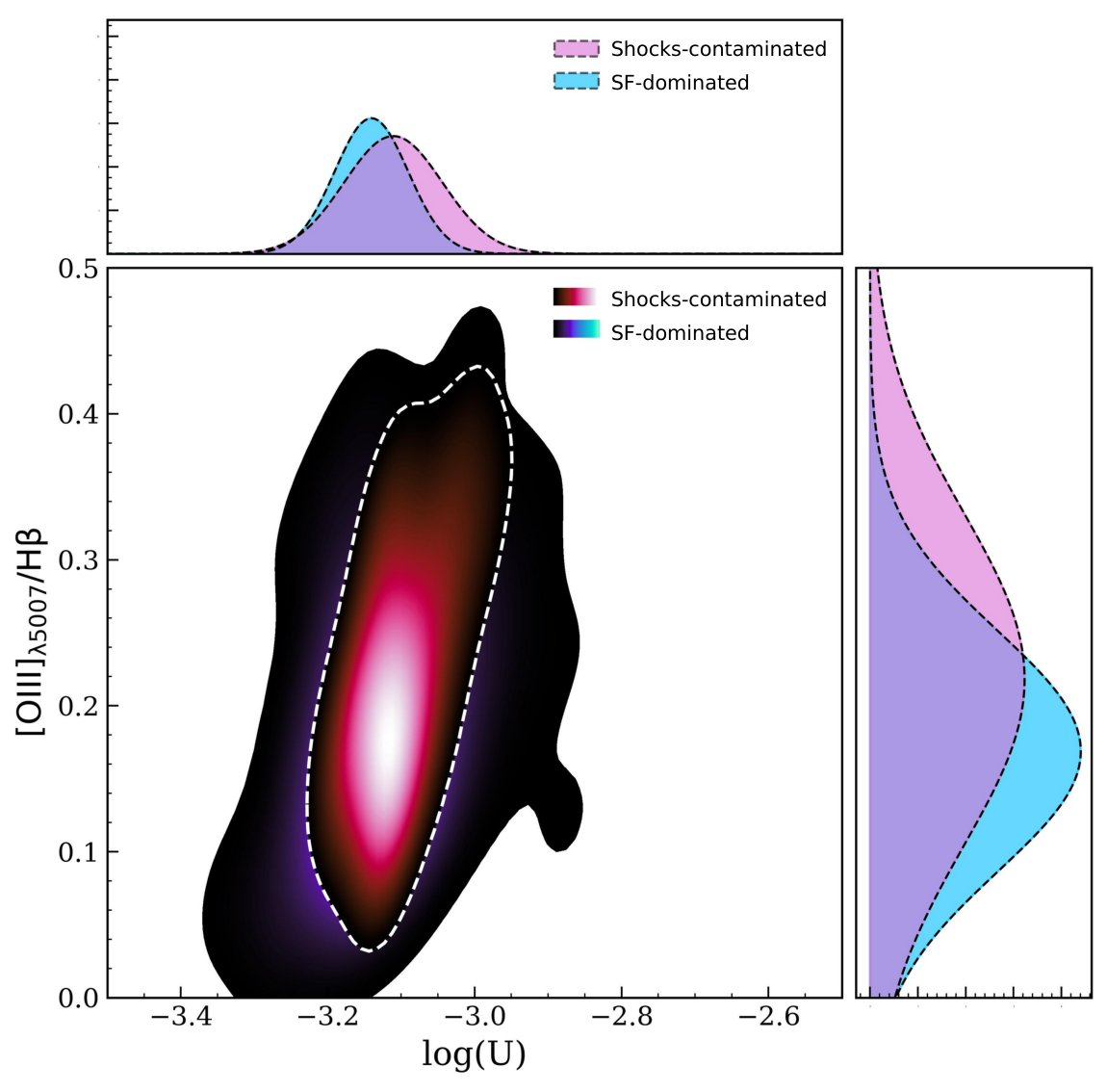}

\caption{Similar to Fig.\ref{fig:IC1553_disps} for ESO157-49.}
\label{fig:ESO157-49_disps}     
\end{figure*}

\begin{figure*}[ht!]
\centering
    \includegraphics[width=0.331\textwidth]{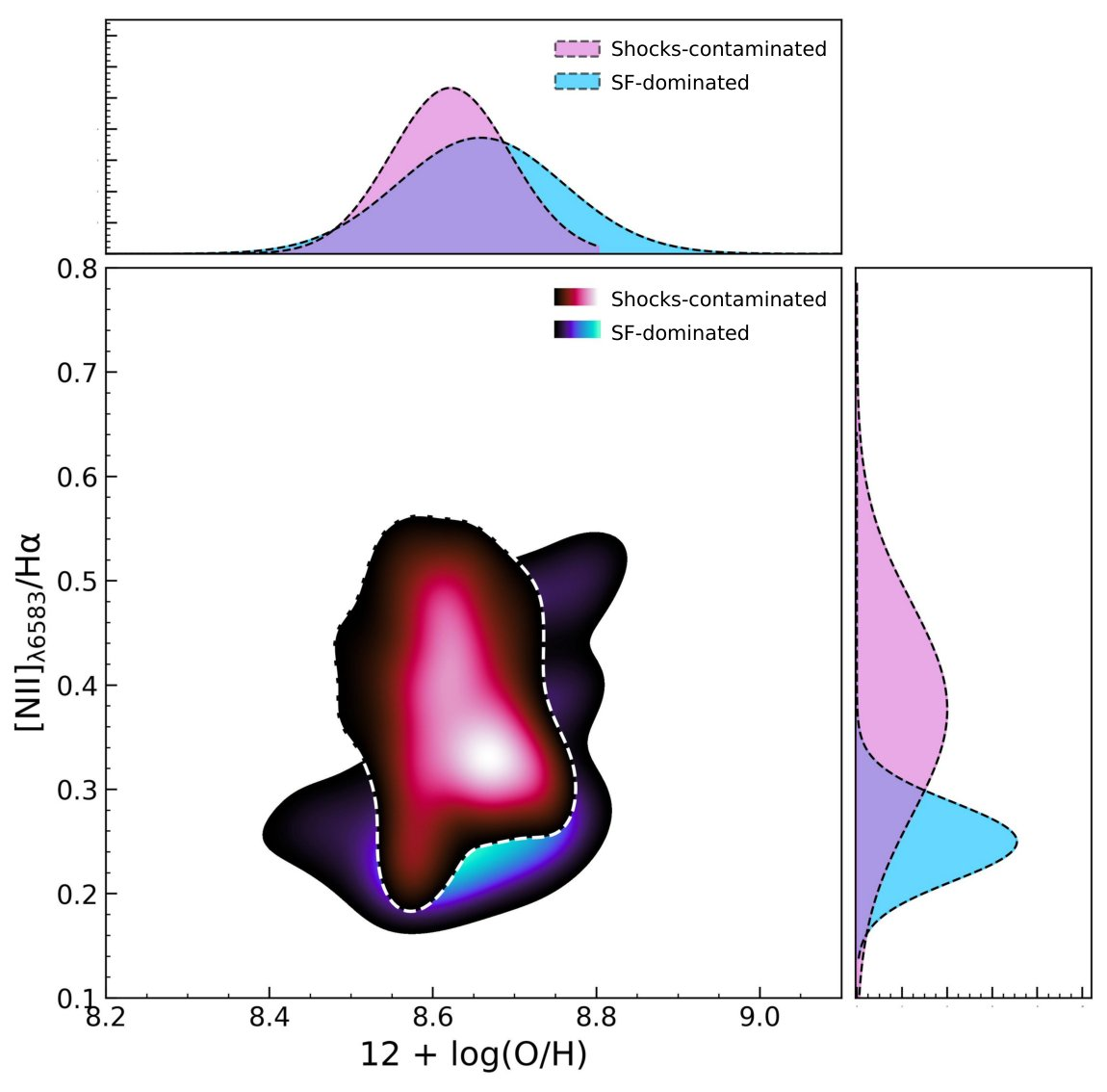}
    \includegraphics[width=0.33\textwidth]{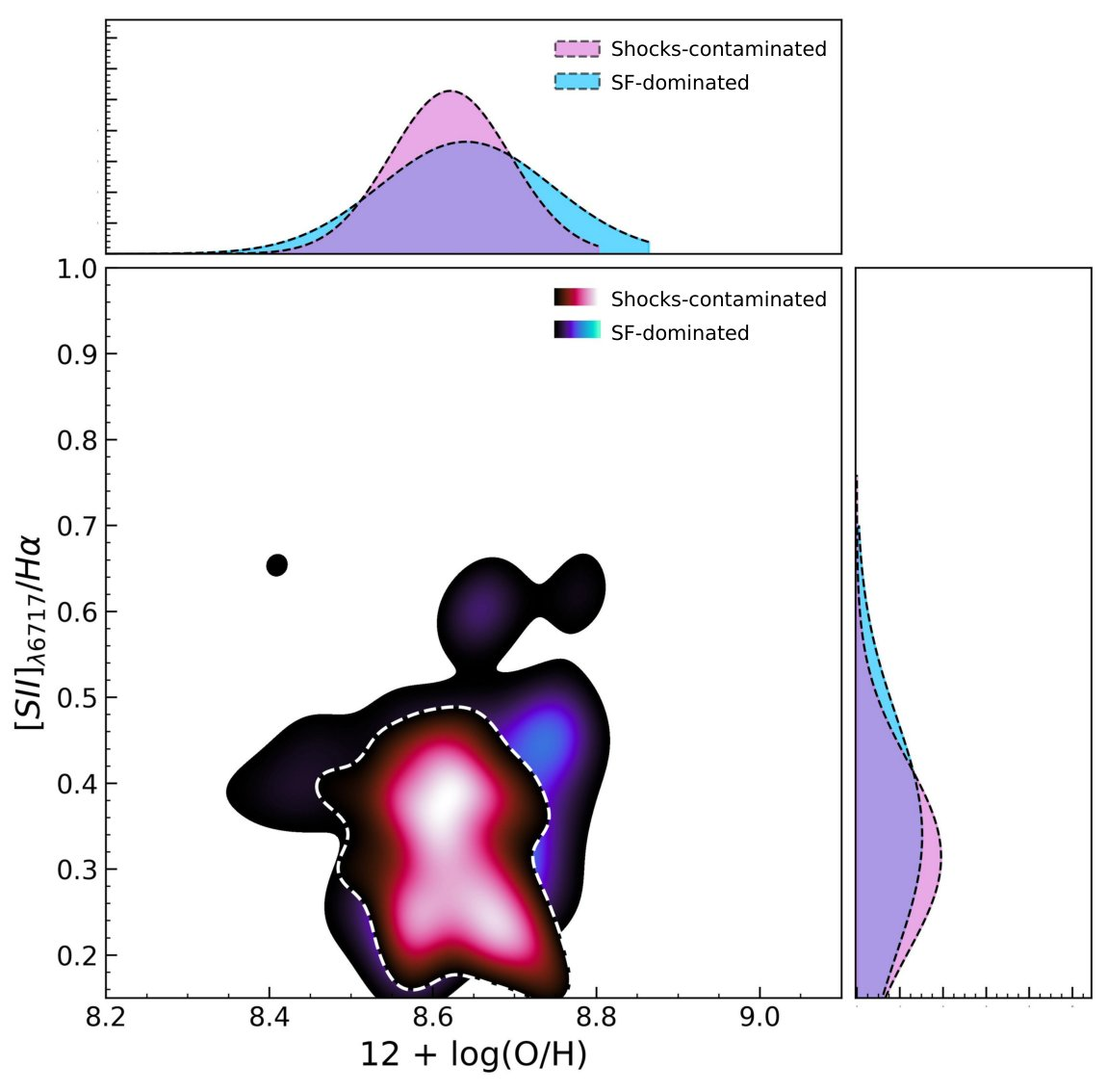}
    \includegraphics[width=0.33\textwidth]{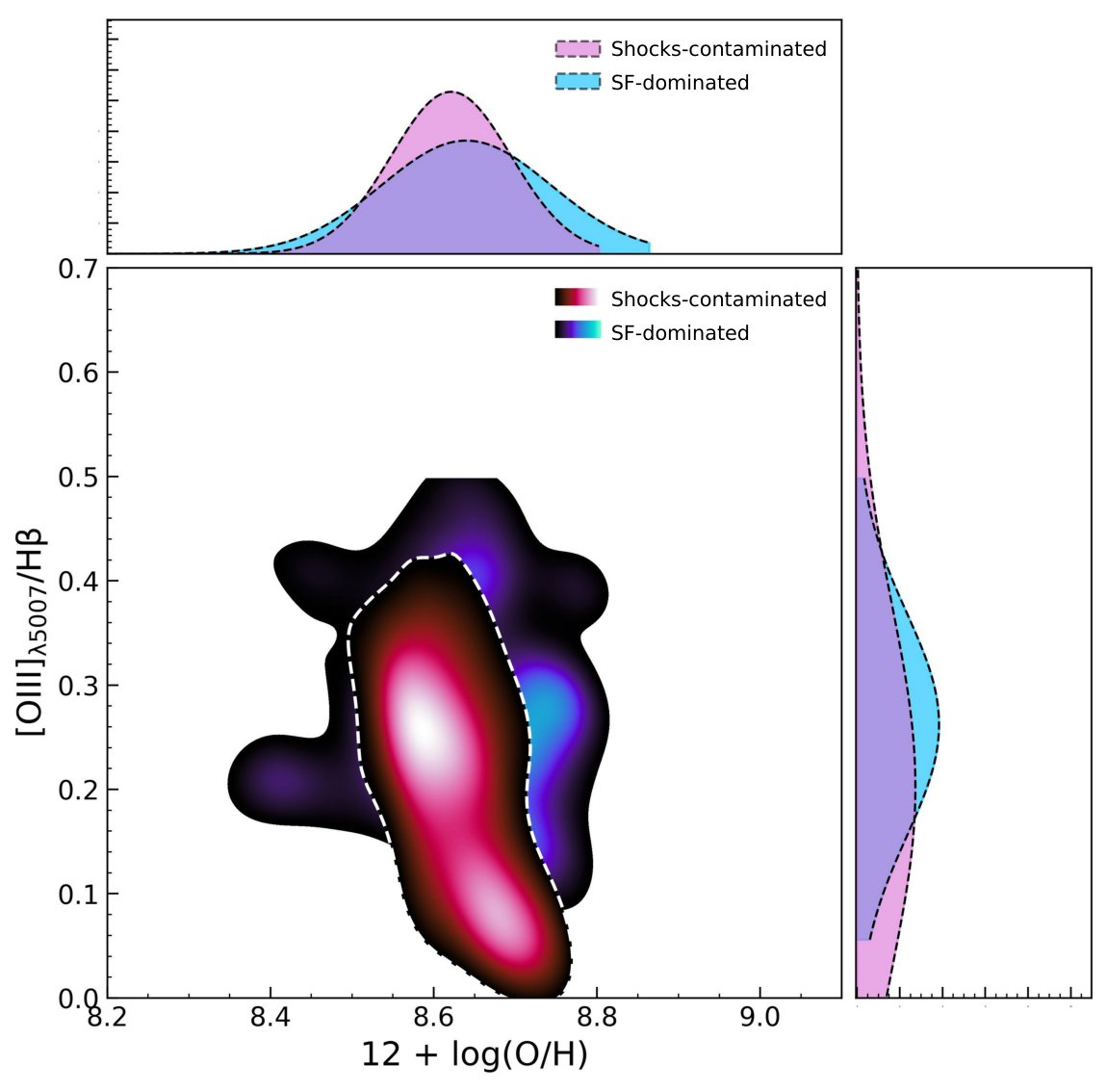}

    \includegraphics[width=0.331\textwidth]{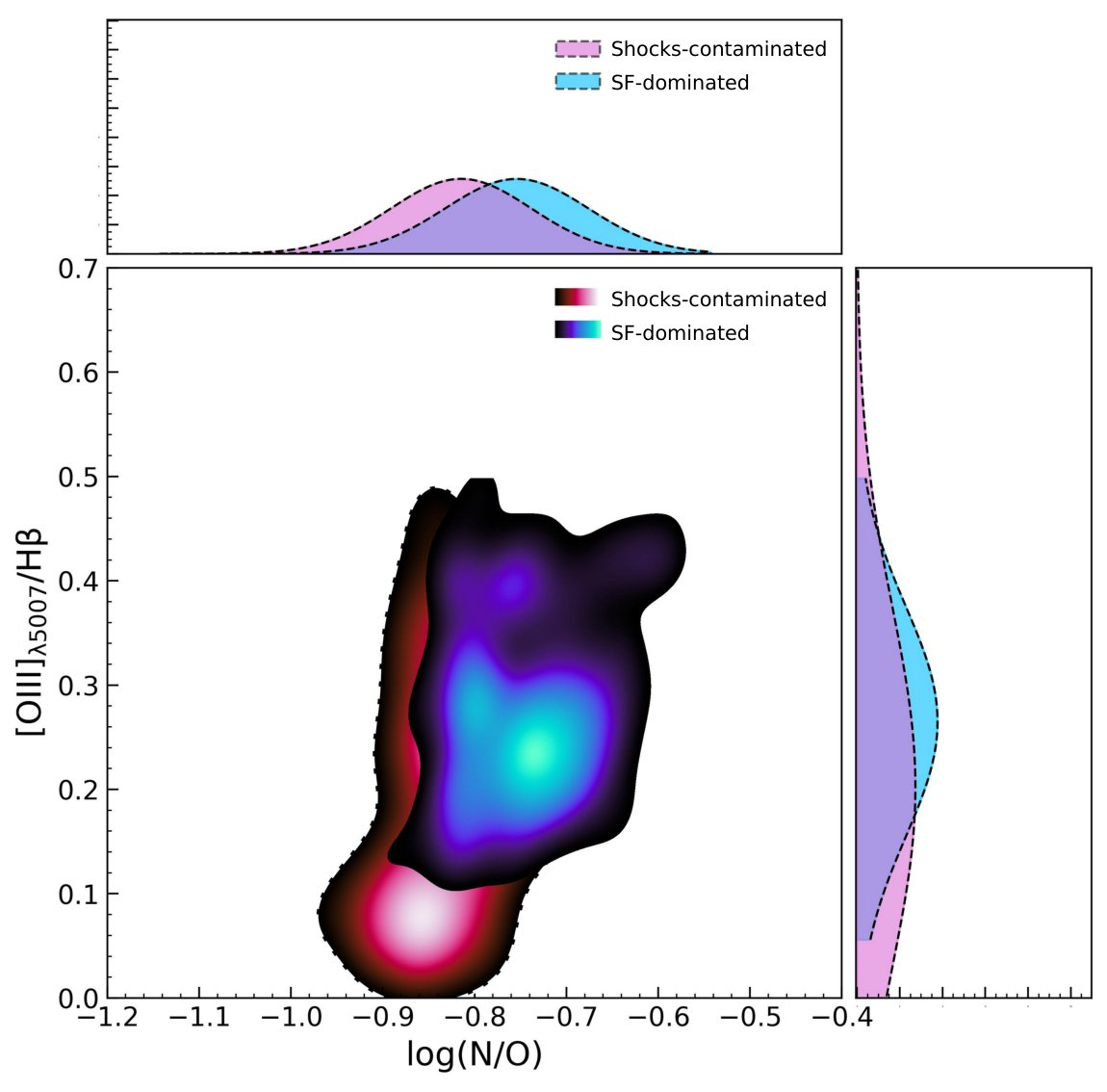}
    \includegraphics[width=0.33\textwidth]{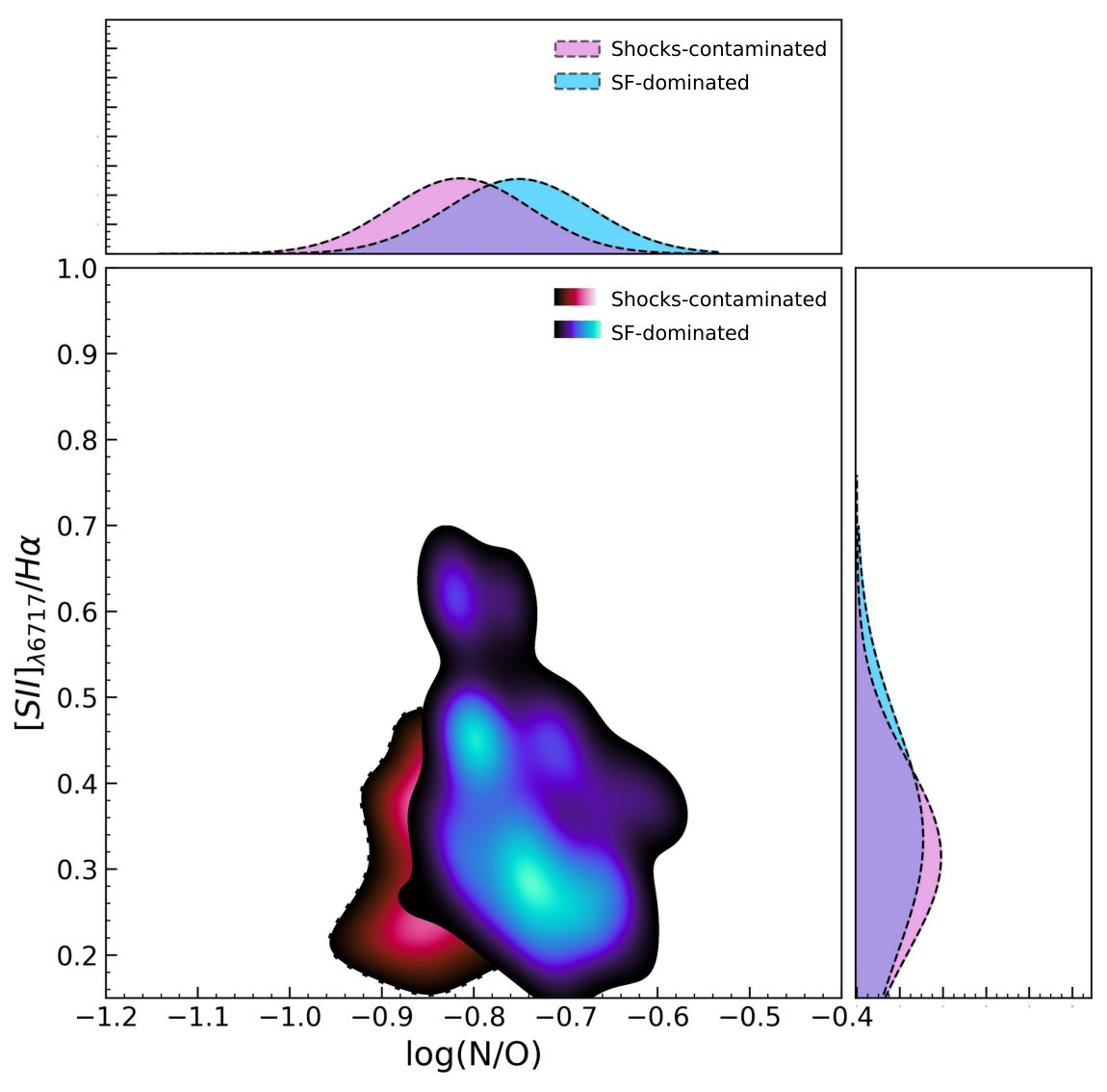}
    \includegraphics[width=0.33\textwidth]{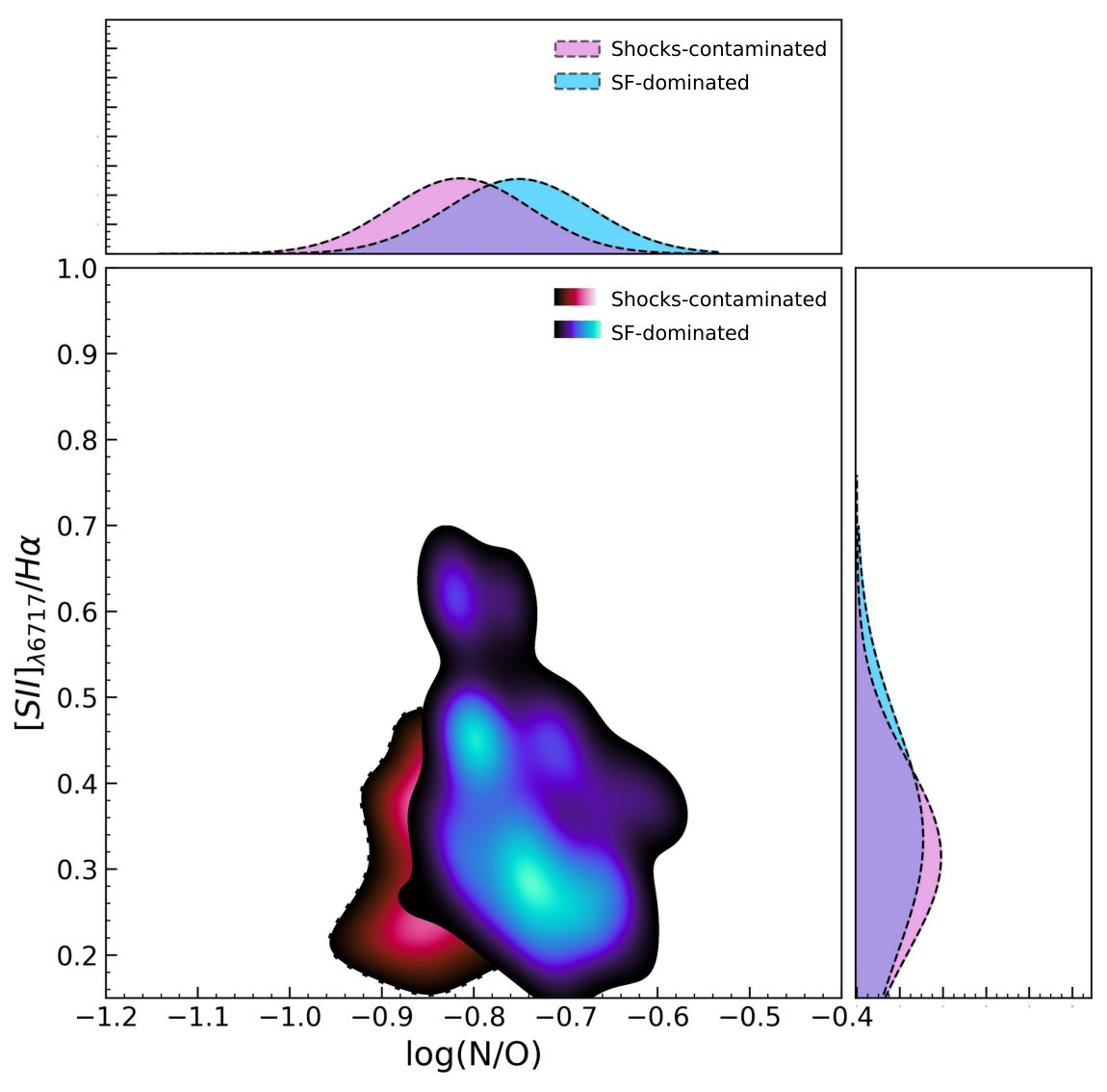}

    \includegraphics[width=0.331\textwidth]{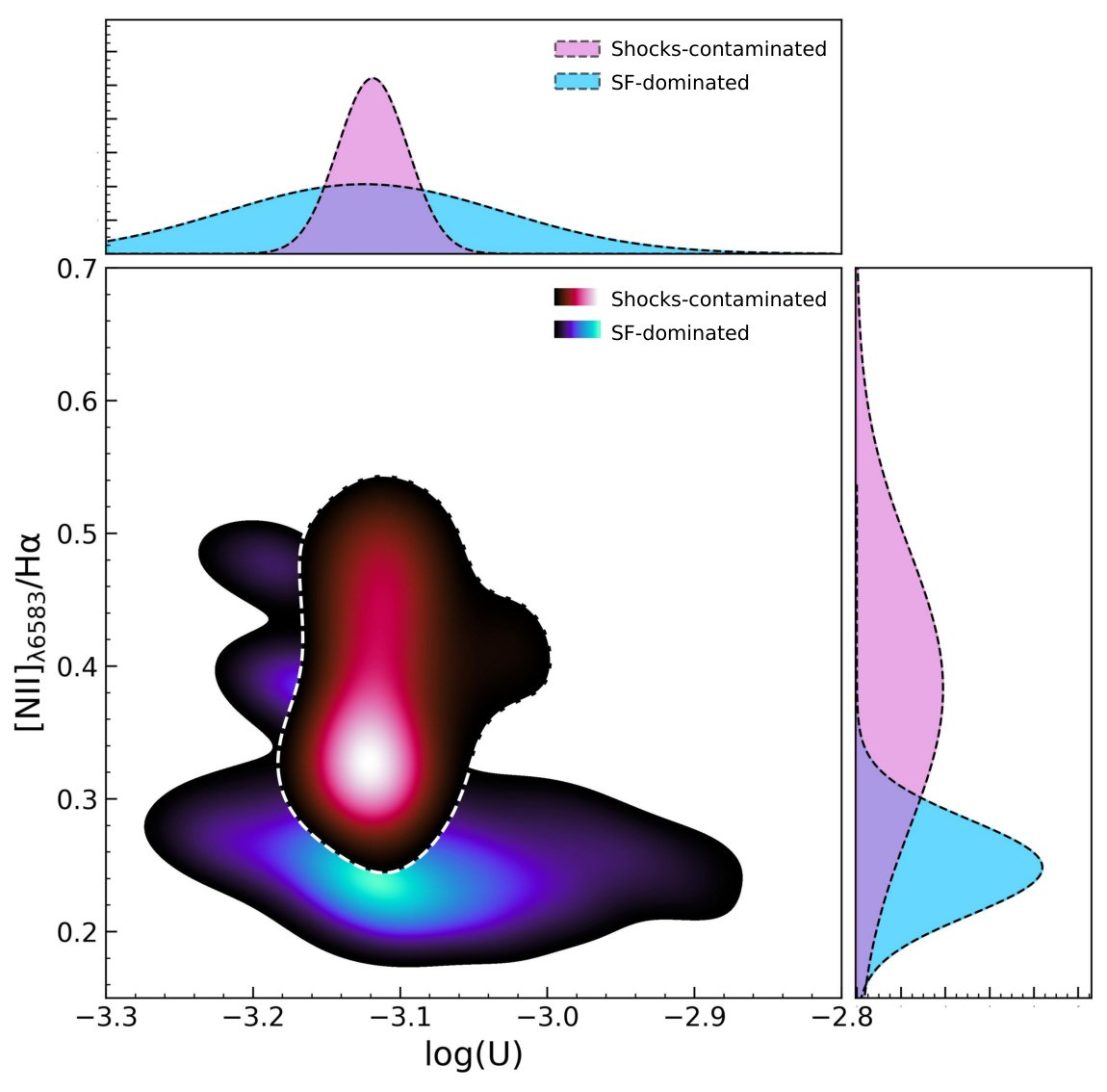}
    \includegraphics[width=0.33\textwidth]{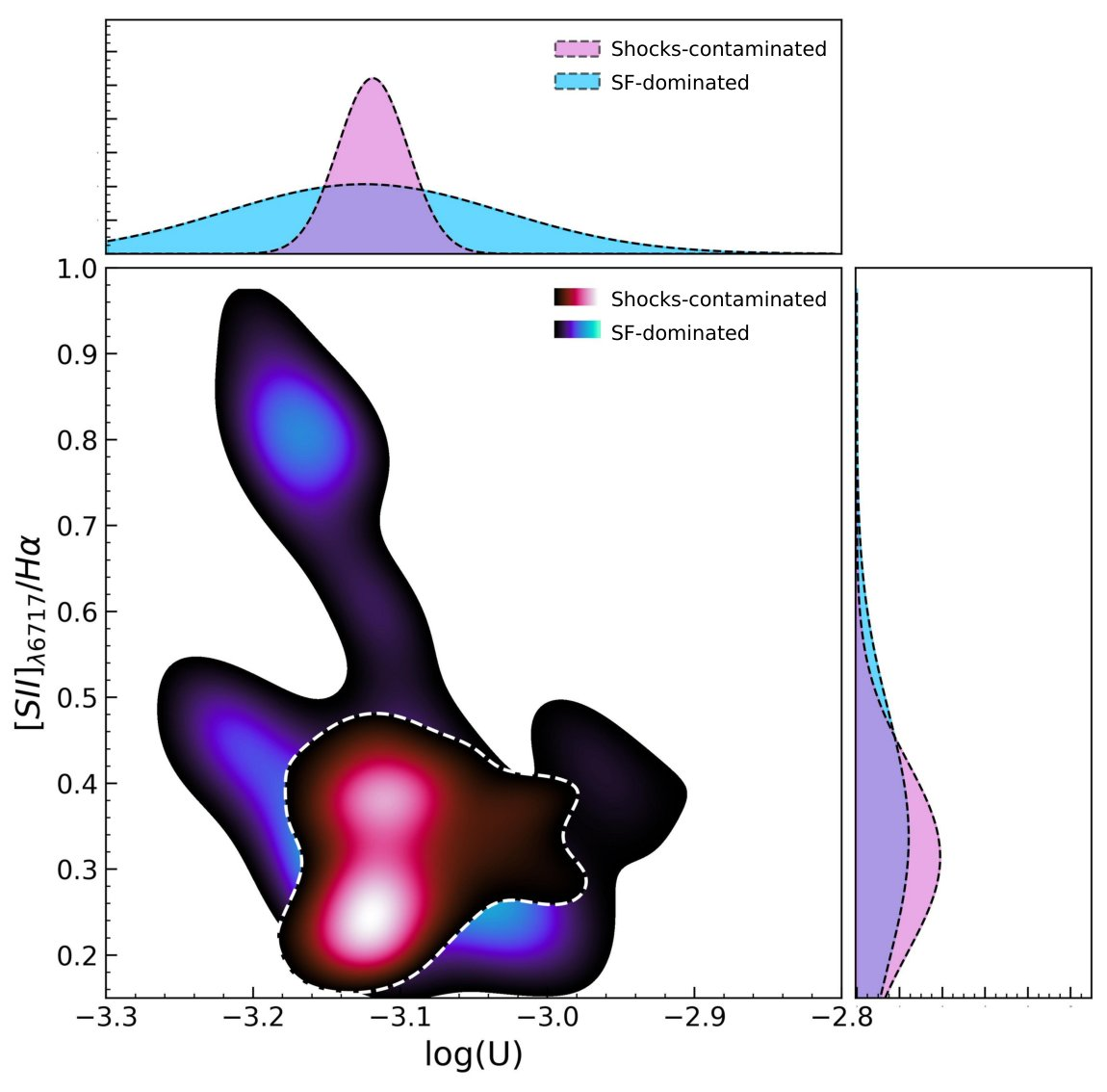}
    \includegraphics[width=0.33\textwidth]{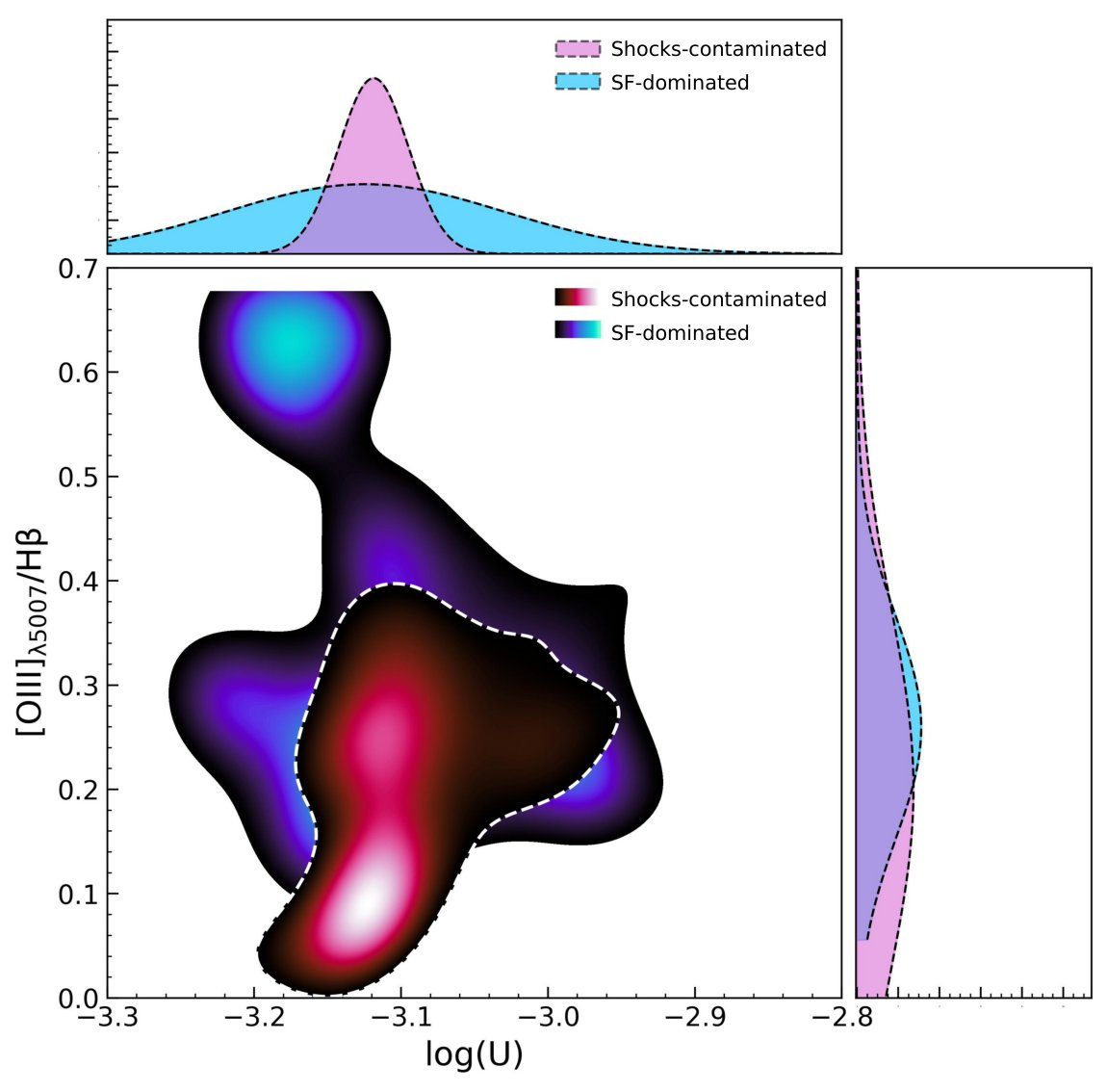}

\caption{Similar to Fig.\ref{fig:IC1553_disps} for ESO443-21.}
\label{fig:ESO443-21_disps}     
\end{figure*}

\begin{figure*}[ht!]
\centering
    \includegraphics[width=0.331\textwidth]{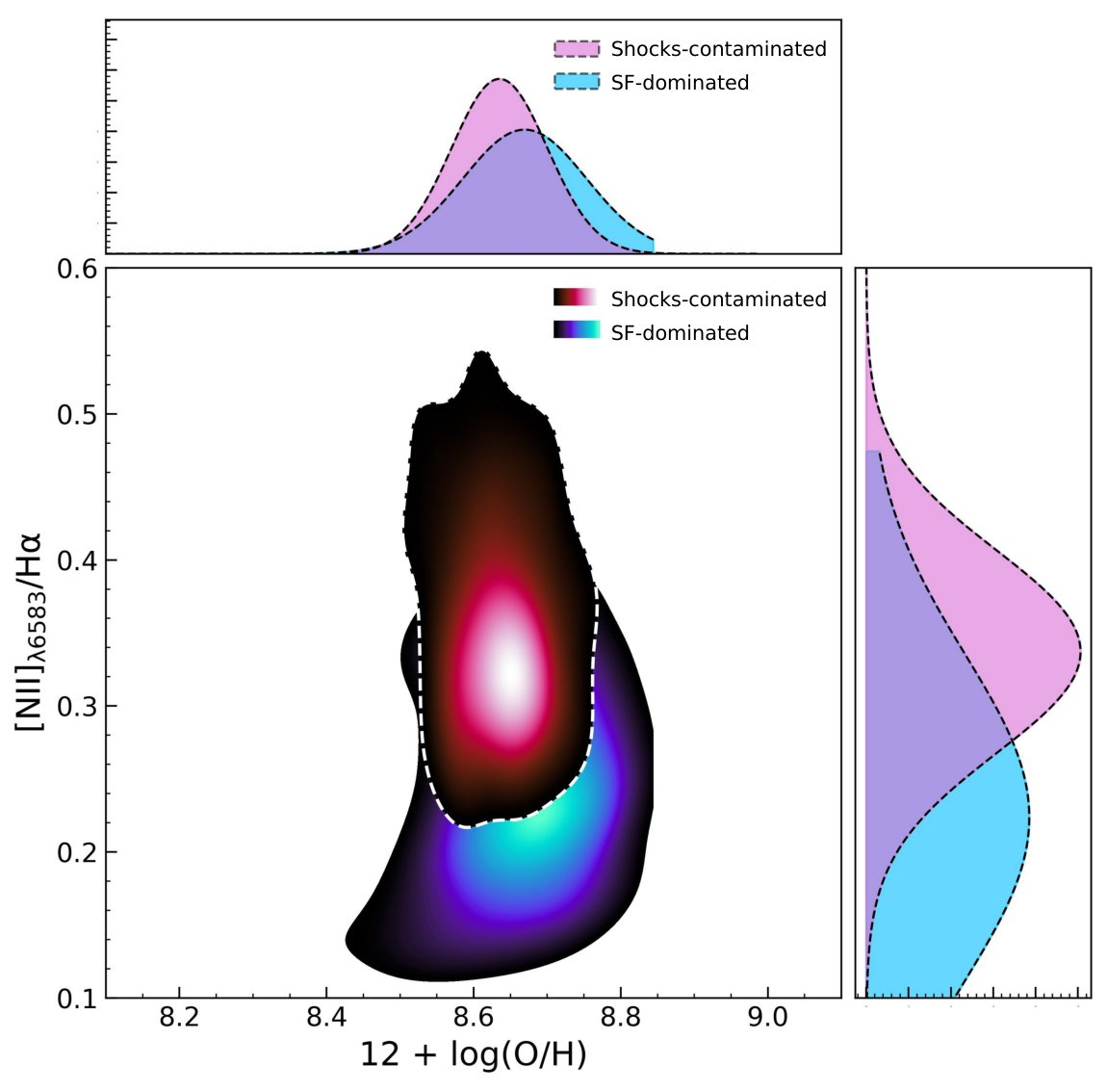}
    \includegraphics[width=0.33\textwidth]{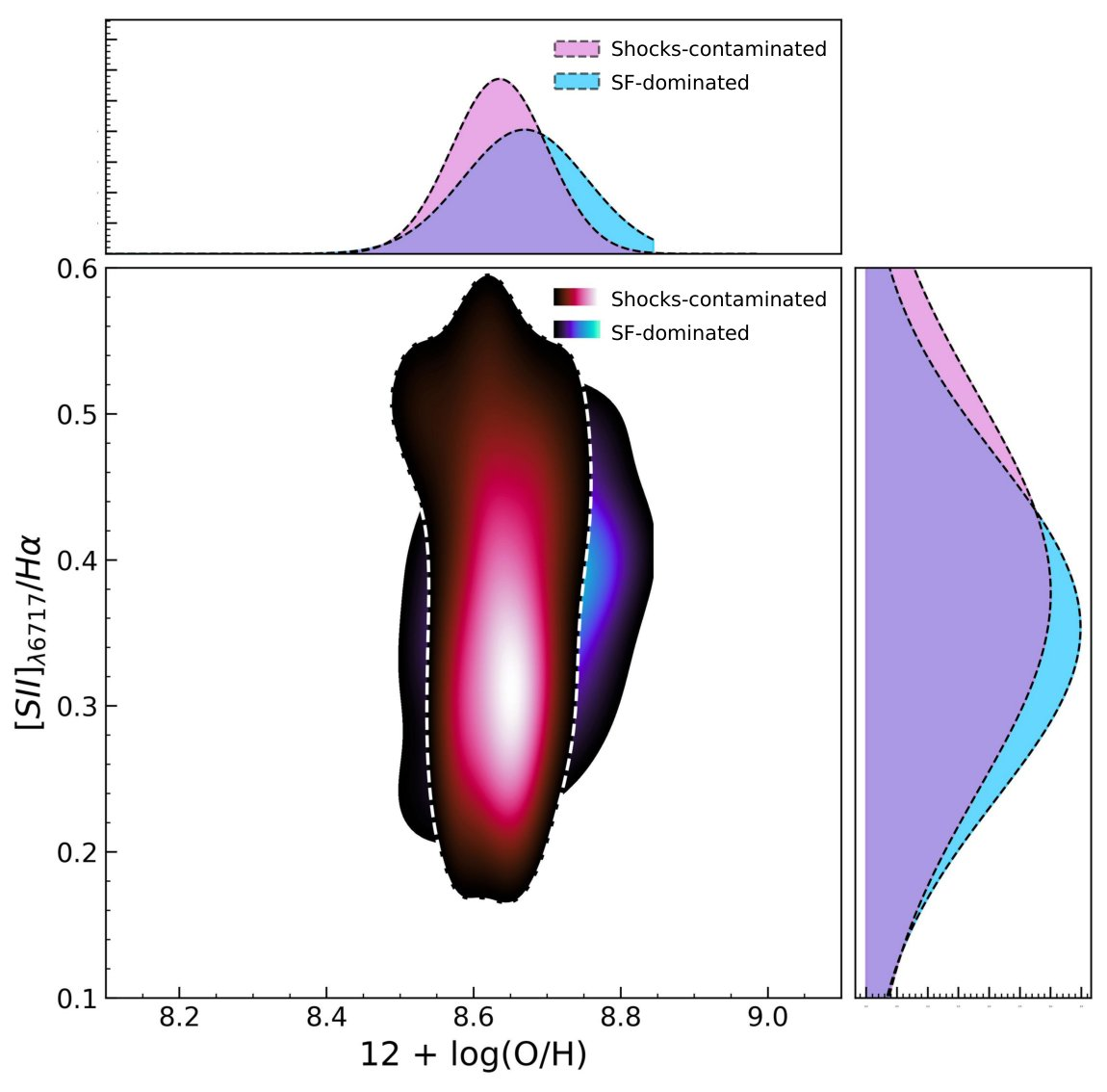}
    \includegraphics[width=0.33\textwidth]{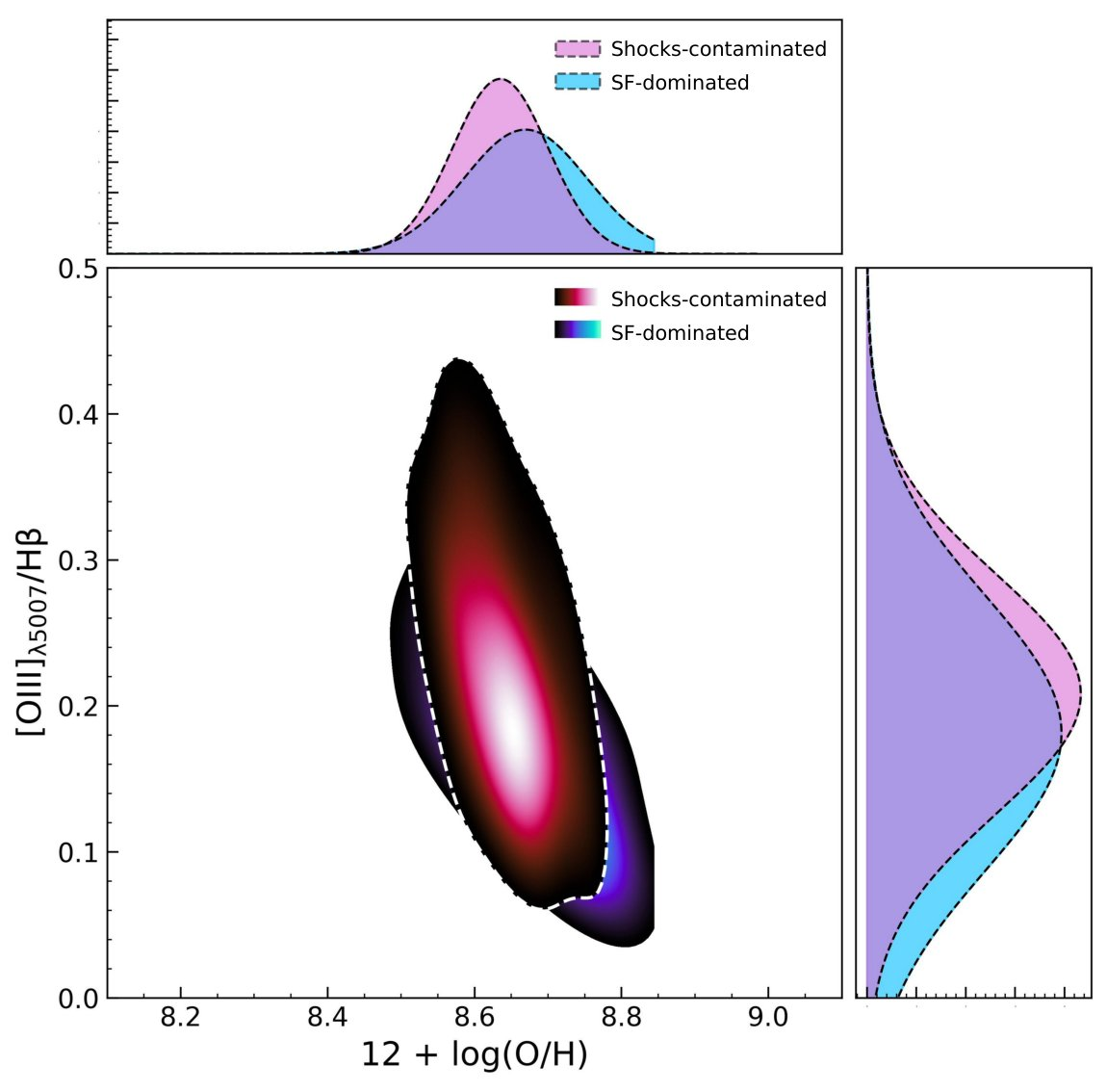}

    \includegraphics[width=0.331\textwidth]{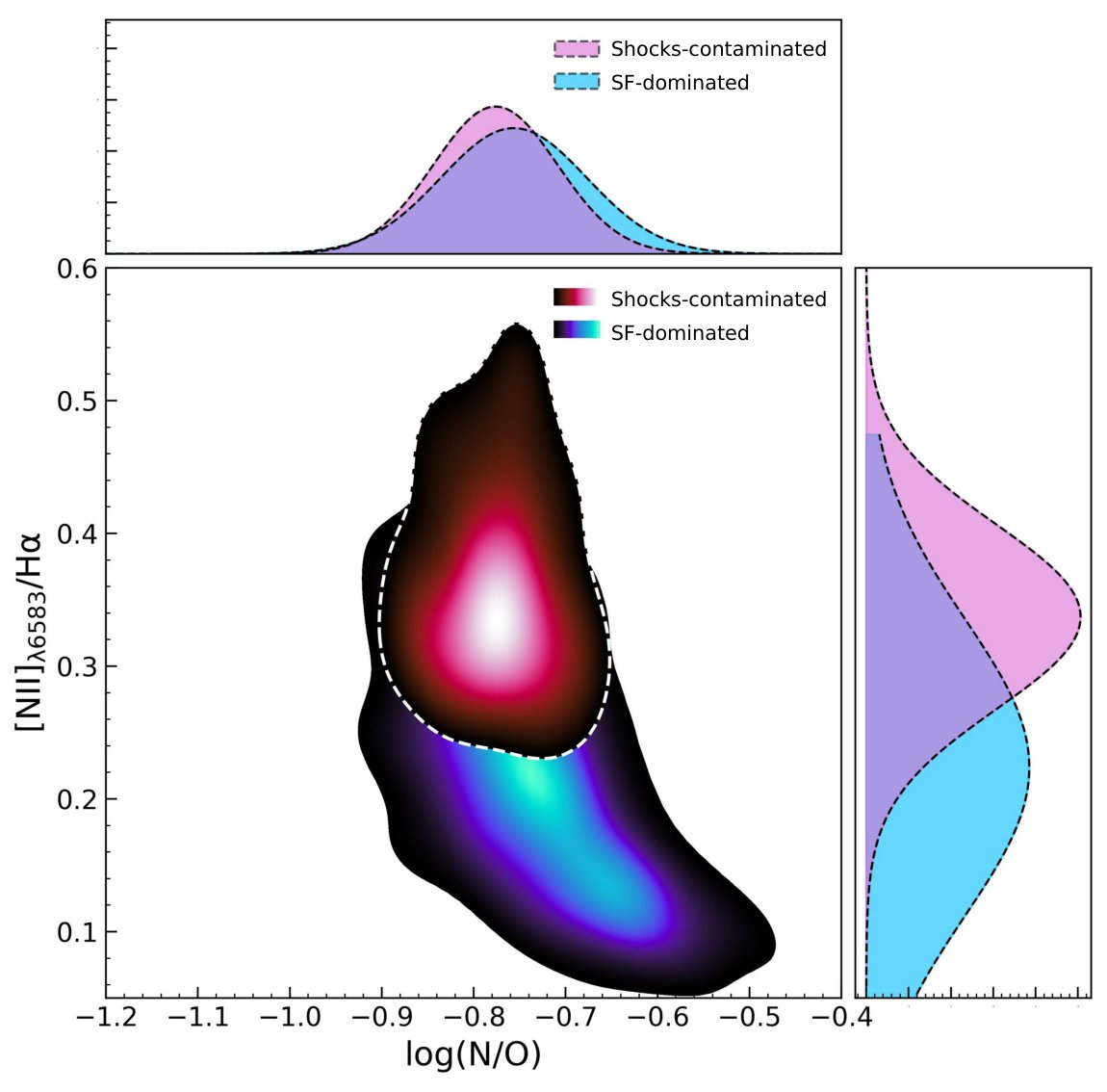}
    \includegraphics[width=0.33\textwidth]{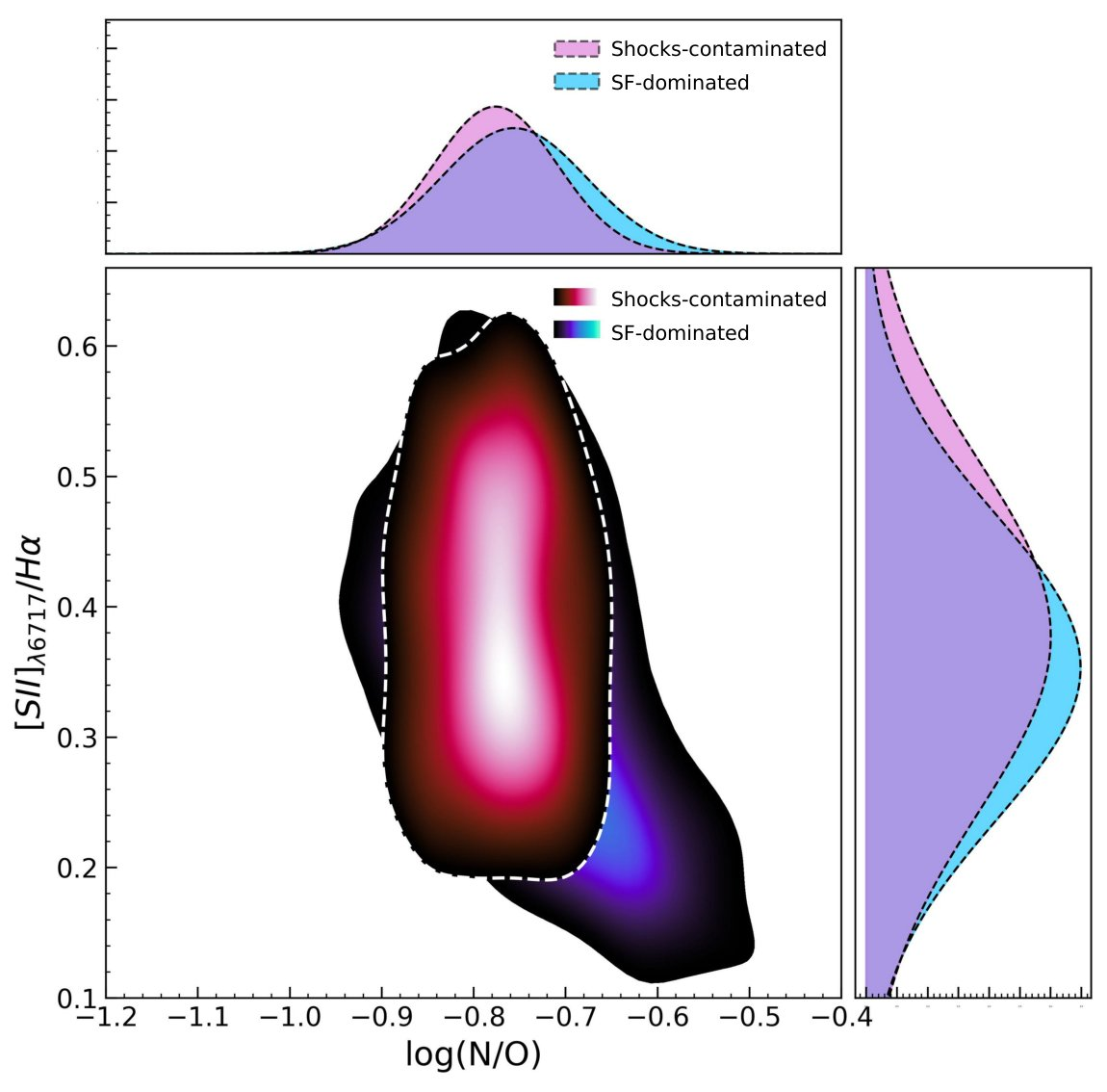}
    \includegraphics[width=0.33\textwidth]{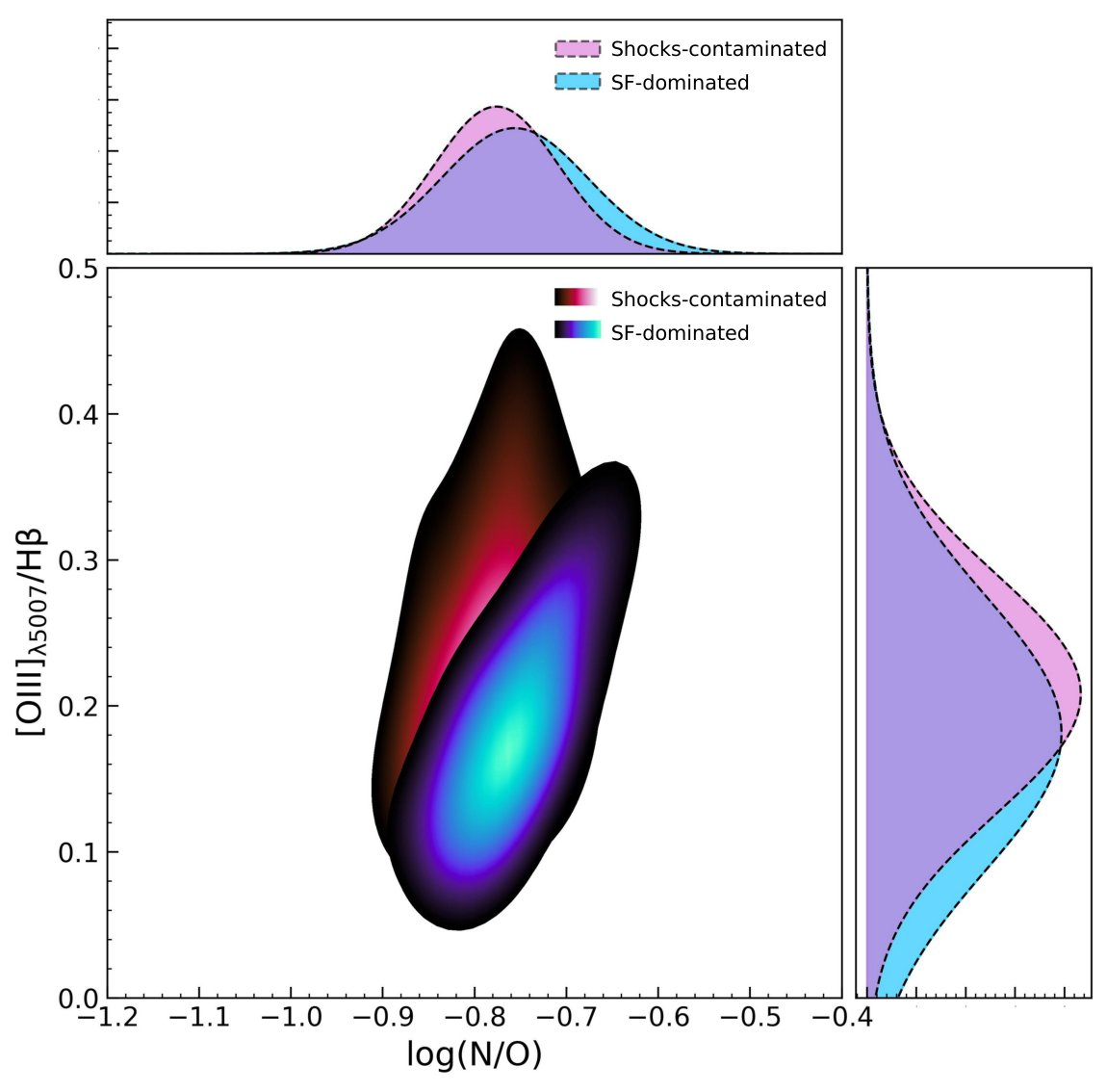}

    \includegraphics[width=0.331\textwidth]{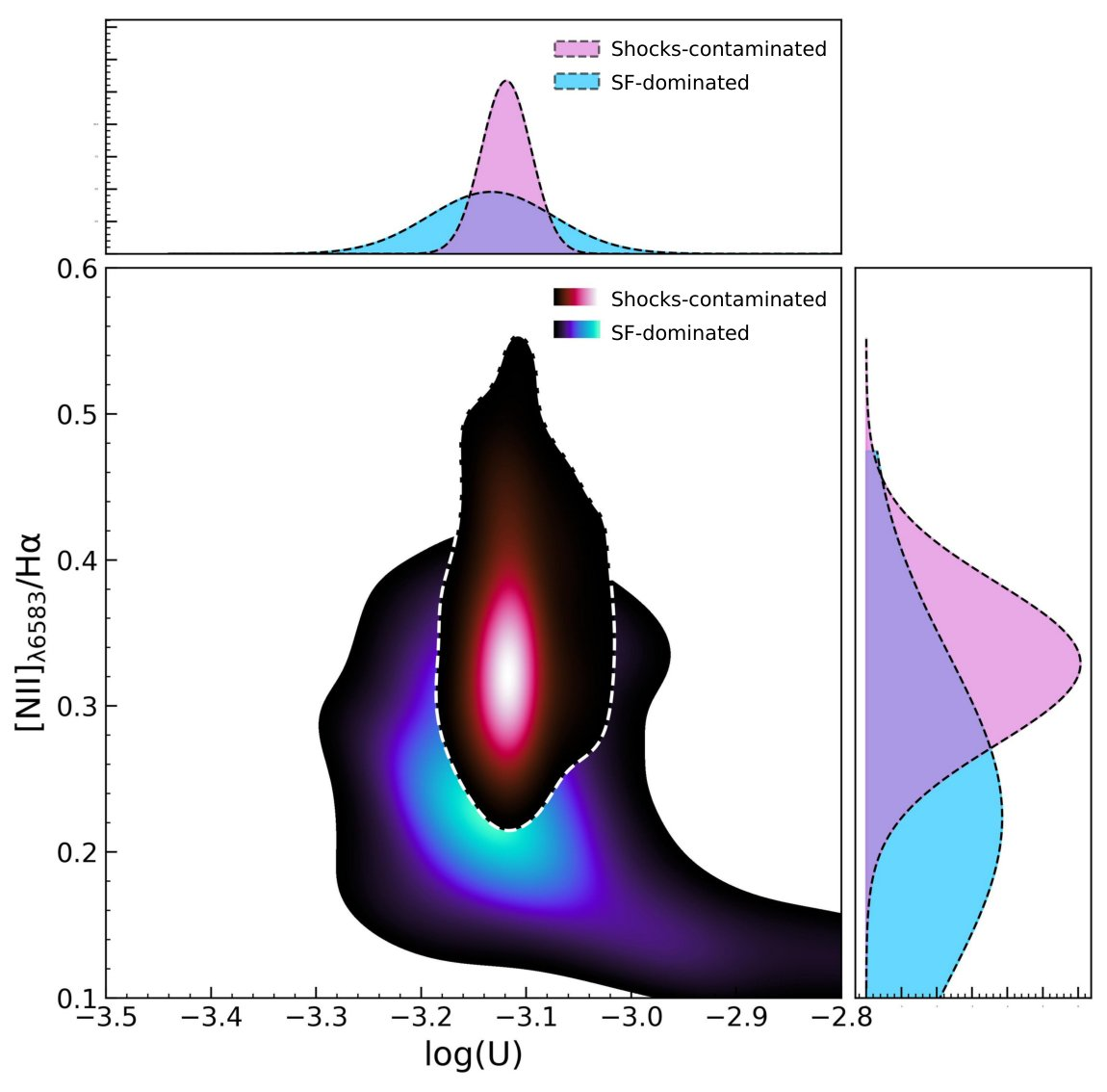}
    \includegraphics[width=0.33\textwidth]{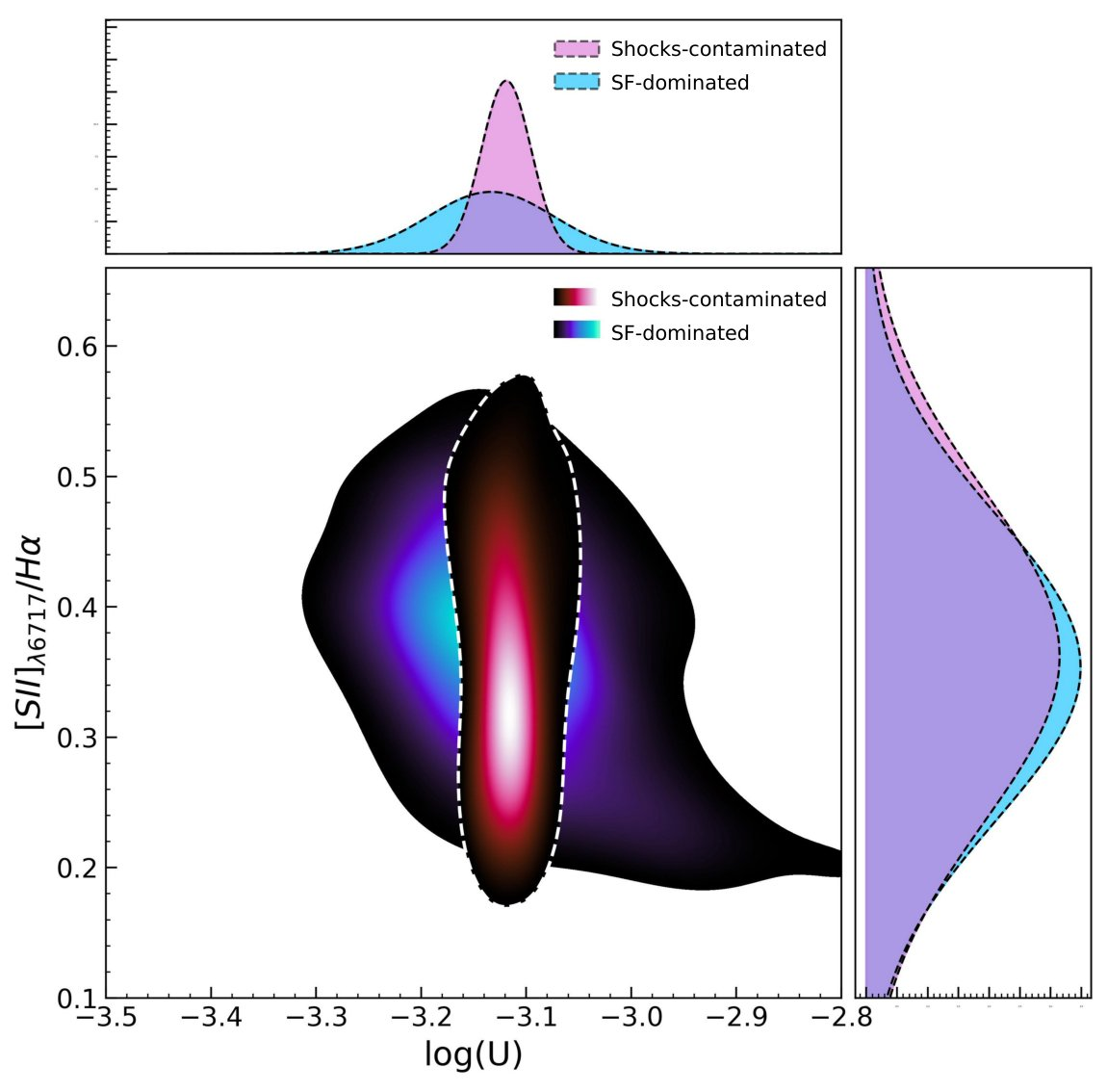}
    \includegraphics[width=0.33\textwidth]{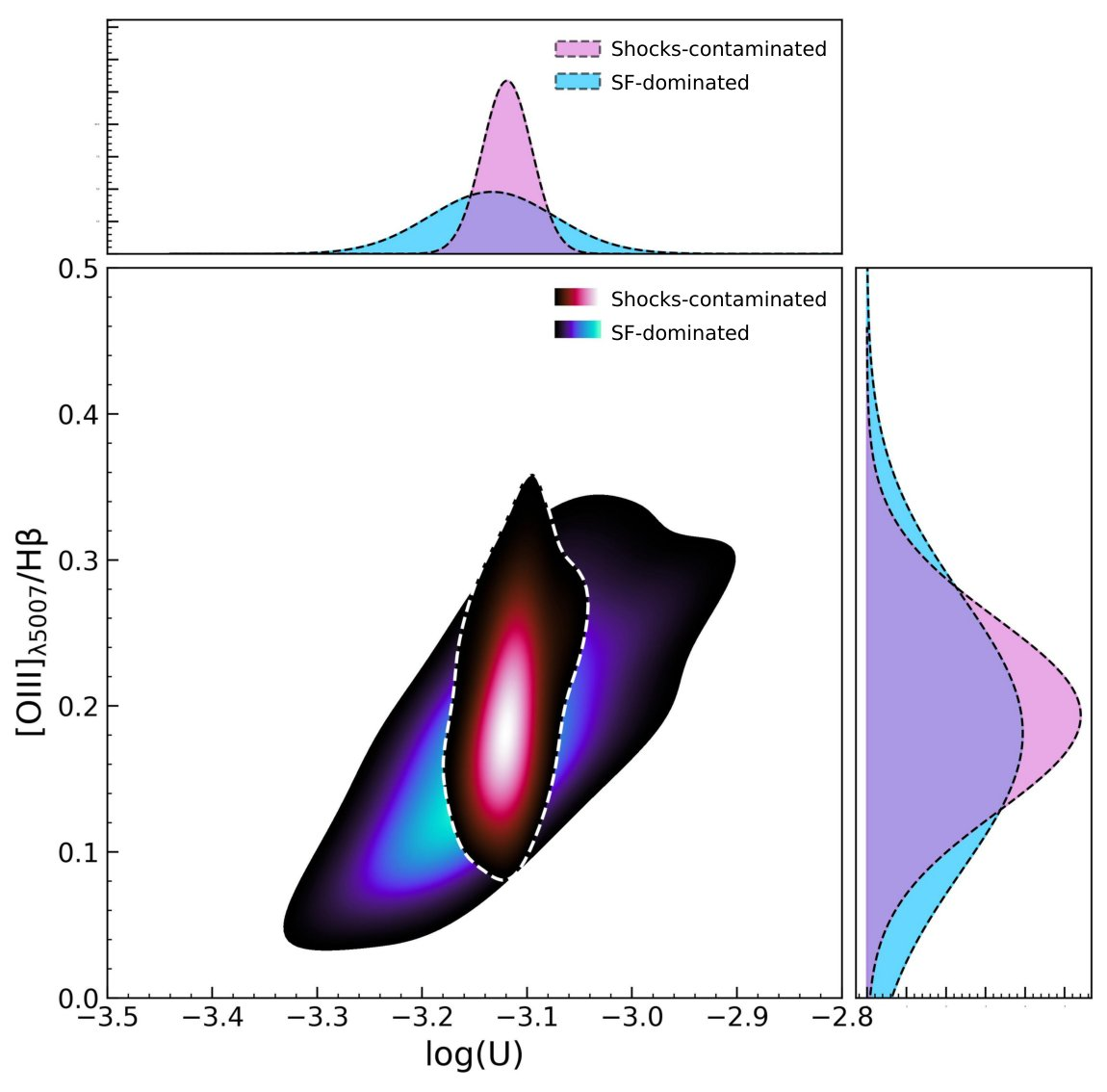}

\caption{Similar to Fig.\ref{fig:IC1553_disps} for ESO469-15.}
\label{fig:ESO469-15_disps}     
\end{figure*}

\section{Sample presentation}\label{sec:sample}
In order to investigate the impact of different DIG regimes on the determination of chemical abundances, we chose the edge-on galaxies for which \citetalias{BETISII} reports the largest differences between the relative contributions of fast shocks and star formation to the ionisation budget of the eDIG. The selected galaxies are IC1553, ESO157–49, ESO443–21, and ESO469–15 (see Table \ref{tab:sample}), and were observed with the MUSE IFS instrument, which provides 1 arcmin$^2$ of field of view (FoV), with a sampling of $0.2\times0.2$ arcsec, a spectral range of $4650-9300$ \r{A}, with a resolution of 1750 at 4650 \r{A} to 3750 at 9300 \r{A}, and a spectral sampling of 1.25 \r{A}. We pre-processed and adaptively binned the MUSE datacubes\footnote{We downloaded the reduced datacubes directly from the \href{https://archive.eso.org/scienceportal/home}{European Southern Observatory archive science portal}.} during the first stage of the BETIS project (\citetalias{BETISI}). The BETIS pipeline\footnote{Available on GitHub: \href{https://github.com/Raulghuelma/BETIS}{BETIS}.} produced rest-frame, de-reddened, and adaptively binned datacubes. We based the binning on the signal-to-noise ratio (S/N) of the \sii emission line, one of the principal tracers of the DIG, combining spatially adjacent spaxels until the bins reached a median S/N of 10. Some bins still had a relatively low S/N after this process, so we retained only bins with a final spectroscopic S/N\,$>$\,2 in the \sii line. Each bin thus corresponded to an integrated spectrum that met this criterion. We then obtained the binned emission-line maps from these datacubes (see \citetalias{BETISI} for further details on the pipeline).

\begin{table*}[ht!]
\centering
\caption{General characteristics of the sample. See \citetalias{BETISI} and \citetalias{BETISII} for more details.}
\resizebox{\textwidth}{!}{%
\begin{tabular}{lrrlccrrccrcc}
\hline
\textbf{Galaxy} & \textbf{RA (J2000)} & \textbf{DEC (J2000)} & \textbf{Type} & \multicolumn{1}{l}{\textbf{z}} & \textbf{D} & \textbf{P.A.} & \textbf{FWHM} & \textbf{Incl.} & \textbf{log SB lim.} & \textbf{Bin size} & \textbf{L$_{eDIG}$(\ha)} & \textbf{log SFR$_{Disc}$} \\
 & (deg) & (deg) &  & \multicolumn{1}{l}{} & (Mpc) & (°) & (pc) & (°) & (erg/s/kpc$^2$) & (pc) & (erg/s) & (M$_\odot$/yr) \\ \hline
IC1553 & 8.1671 & -25.6075 & Irr & 0.00979 & 32.25 & 15.0 & 171.99 & 78.6 & 36.5 & 125.1 & 39.80 & 0.229 \\
ESO443-21 & 194.9412 & -29.6002 & Scd & 0.00941 & 40.31 & 160.8 & 195.45 & 79.0 & 36.6 & 156.4 & 39.71 & 0.184 \\
ESO469-15 & 347.2317 & -30.8579 & Sb & 0.00545 & 19.81 & 149.2 & 86.44 & 90.0 & 35.9 & 76.9 & 38.90 & 0.029 \\
ESO157-49 & 69.9036 & -53.0126 & Sc & 0.00559 & 24.82 & 30.4 & 96.26 & 79.3 & 35.3 & 96.3 & 39.46 & 0.060 \\ \hline
\end{tabular}%
}
\tablefoot{The columns represent, from left to right: the designation of the galaxy, the RA and DEC in the J2000 epoch, the morphological Hubble type, the redshift, the distance in Mpc, the position angle (in deg), the full-width at half maximum of the point spread function (PSF FWHM) in pc, inclination with respect to the line of sight (in deg), the 2$\sigma$ \ha surface brightness (\SBha) limit and the average physical size of the bins in pc. The RA, DEC, the morphological type, the redshift, and distances are obtained from NED. The position angles and inclinations are obtained directly from \citet{2019A&A...623A..89C} and Hyperleda. The last columns represent the integrated \ha luminosity within the eDIG and the integrated SFR of the disc (see \citetalias{BETISII}).}
\label{tab:sample}
\end{table*}

\section{Abundance determinations}\label{sec:2_mapas}

\subsection{Inputs and {\sc HII-CHI-Mistry} setup}

In this section, we estimate the chemical abundances in these galaxies using {\sc HII-CHI-Mistry} v5.5 (\hcm hereafter; \citealt{PerezMonteroHCm14}). This tool estimates abundances from measured ratios of both strong and auroral emission lines by comparing the measurements and their errors with the emission-line fluxes predicted by an extensive, user-selectable library of photoionisation models. Abundances obtained with this approach have been shown to be consistent with those derived using the direct method (see \citealt{PerezMonteroHCm14} for details). Moreover, because none of the classical auroral lines are detected in our dataset, this code is particularly well suited to our analysis. The modelling framework includes three free parameters: the gas-phase metallicity of the ISM, $12+\log(\mathrm{O/H})$, the relative nitrogen-to-oxygen abundance ratio, $\log(\mathrm{N/O})$, and the ionisation parameter, $\log(U)$. These parameters are estimated through comparison between the observed line ratios and the model predictions.

\noindent
As reported in \citetalias{BETISII} and previously introduced, ionisation driven by OB stars in the disc is the primary source of the observed eDIG features. Although it is not the sole mechanism responsible for powering the eDIG, the influence of the star formation in the disc is present in all scenarios within this sample, contributing at least 50\% of the total ionisation of the halo. For this reason, one of the constraints selected for \hcm is an ionising field of young star clusters with an instantaneous burst at an age of 1 Myr, given by the POPSTAR stellar population models \citep{POPSTAR2009} and assuming a \citet{2003PASP..115..763C} initial mass function (IMF), representing a photoionisation due to massive stars, and independent of the age of the populations \citep{PerezMonteroHCm14,PerezMontero21}. In order to assess the impact of fast shocks on model-based metallicity determinations, in Sect. \ref{sec:3_abundancias} we explore the potential small biases that can be introduced when shocks partially contribute to the ionisation conditions of the gas.

The values of the three free parameters in all model grids extend over a range that covers typical values observed in star-forming galaxies, i.e. 12+log(O/H)$\in$[6.9, 9.1] with bins of 0.1 dex, log(N/O)$\in$[-2.0, 0] with bins of 0.125 dex and log(U)$\in$[-4.0, -1.5] with bins of 0.25 dex. Since no auroral lines are detected and the electron temperature cannot be directly determined, the code adopts an empirical relation between O/H and log(U), whereby metal-poor systems exhibit higher values of U and vice versa (see Fig. 1 of \citealt{PerezMonteroHCm14}). 

\noindent
\hcm takes as input the de-reddened fluxes of the selected emission lines, which in our case we use as input for \hcm the de-reddened fluxes and errors of \nii, \sii, and \oiii relative to \hb. The measured fluxes and errors correspond to the individual bins in the binned emission-line maps presented in \citetalias{BETISII}, with a spectroscopic S/N greater than 2 and a relative error of the measured lines lower than 40\%. The fluxes are corrected for interstellar extinction using the law from \citet{1989ApJ...345..245C} with $R_V = 3.1$  and assuming a Balmer decrement of H$\alpha$/H$\beta = 2.87$. Besides, as we measure both \nii and \sii lines, the code can provide a solution for N/O, which implies that the code does not need to subsequently assume any empirical relation between O/H and N/O, allowing us to obtain an independent measure of O/H based on \nii lines. 

\subsection{Results}

Fig.\ref{fig:maps} presents the reconstructed maps of the $12+\log(\mathrm{O/H})$, $\log(\mathrm{N/O})$, and $\log(U)$ values calculated by \hcm given the previous conditions for the four galaxies in the sample. In each map, the galactic plane is masked using the elliptical apertures defined in \citetalias{BETISII}. The masked region is shown in greyscale, representing the H$\alpha$ emission from the galactic plane.
\noindent

The first top three panels, corresponding to IC1553, show a noticeable increase in O/H in regions located farther from the \hii regions in the plane, reaching solar values (12+log(O/H) $\sim$ 8.69; \citealt{2021A&A...653A.141A}). This value decreases monotonically up to 8.45 as we approach the main concentration of \hii regions, at major-axis distances (MAD) between 2 and 4 kpc. No significant structure is observed in the eDIG in the O/H map, as similar to what is seen in the \oiii/H$\beta$ map presented in \citetalias{BETISII}, the only noticeable variation in O/H occurs in the proximity of the \hii region overdensity within the galactic plane. The N/O map exhibits super-solar values ($\log(\mathrm{N/O}) > -0.86$; \citealt{2021A&A...653A.141A}), reaching up to $\sim -0.70$ at higher distances. An exception is found in the central region ($-1\,\mathrm{kpc} < \mathrm{MAD} < 1\,\mathrm{kpc}$), where the inferred abundance decreases to approximately solar values, forming a conical structure similar to that observed in the low-excitation line-ratio maps presented in \citetalias{BETISII} (i.e. the \ha, \nii/\ha, \sii/\ha and \oi/\ha maps). The ionisation parameter remains close to $\log(U) \approx -3.0$, with a dispersion not exceeding 0.3 dex. Moreover, considering that the bin width of the model grids is of the order of 0.25 dex, we conclude that the eDIG in this galaxy exhibits a spatially homogeneous, low-ionisation state characterised by $\log(U) \approx -3.0$. 

\noindent
A similar scenario is observed in ESO157-49, which exhibits a structureless 12+log(O/H) map comparable to the high-excitation line \oiii/\hb map presented in \citetalias{BETISII}, with the highest values reaching solar abundances in different points of the halo. A clearer trend is observed in the N/O map, where the ratio increases with distance from the midplane, ranging from solar values near the midplane (up to log(N/O) $\approx$-0.9) to values above -0.70 at $\textbar z\textbar\sim$2.5 kpc. Similarly, lower values of log(U) are found near the midplane, increasing from -3.25 to -2.85 at $\textbar z\textbar\sim$2.5 kpc, suggesting an enhancement of the ionisation conditions at greater distances from the midplane.

\noindent
ESO443-21 represents a case characterised by a large knot extending to distances up to 4.5 kpc, visible only in the low-excitation maps of \citetalias{BETISII}, such as \ha, \sii, and \nii. However, this knot is not visible in any of the maps presented in this work, suggesting that this structure does not affect the inferred abundance measurements. Instead, we find trends similar to those described above: $12+\log(\mathrm{O/H})$ increases with distance from the midplane, reaching values of up to 8.75; $\log(\mathrm{N/O})$ rises to approximately $-0.65$; and $\log(U)$ remains nearly constant at $\sim -3.0$ throughout the halo. 

\noindent
ESO469-15 represents a particularly interesting case, as it exhibits extraplanar \hii regions (masked using the {\sc python} routine {\sc DAOStarFinder} from the {\sc PhotUtils} package\footnote{\href{https://photutils.readthedocs.io/en/stable/api/photutils.detection.DAOStarFinder.html}{{\sc PhotUtils}} documentation.}). While the O/H, N/O, and log(U) values behave similarly to previous cases, regions in close proximity to the extraplanar \hii regions appear to show enhanced values for all three parameters. Although this enhancement could be attributed to the presence of these \hii regions, it appears to be more pronounced in bins located near the \hii regions at greater distances from the midplane. This suggests that the observed variations are not solely due to changes in the ionisation conditions, but may instead reflect a transition in the prevailing eDIG regime. We note that \hcm was calibrated on grids of pure photoionisation by young stellar clusters, and its application to eDIG bins with non-negligible fast-shock contribution lies outside its calibration regime. Furthermore, each bin in edge-on geometry integrates emission along a line of sight through a possibly multi-phase medium at varying projected galactocentric radii. The values from Fig. \ref{fig:maps} should therefore be interpreted as the parameters of a single-zone photoionisation model that best reproduces the observed line ratios, not as local chemical abundances of the gas. We return to this point in the following sections.

\section{The role of eDIG regimes in chemical abundances}\label{sec:3_abundancias}

As the \citetalias{BETISII} results suggest, the eDIG exhibits at least two different regimes depending on the contribution of reprocessed gas due to star-formation feedback from the disc, with the difference between these two regimes being more pronounced in regions farther from the star-formation complexes of the disc, we explore in detail in this section whether the results from Sect. \ref{sec:2_mapas} can be attributed to this transition between eDIG regimes. We divided the analysis into two different subsets: one corresponding to the eDIG bins dominated by star formation and the other to the eDIG bins dominated by fast shocks. The bins are selected by considering the contribution of fast shocks to the ionisation budget using the hybrid models presented in \citetalias{BETISII}, which combine the photoionisation models of \citet{2010AJ....139..712L} and the fast shock models of \citet{2008ApJS..178...20A}. This framework enables us to constrain the relative contribution of each component to the total ionisation of the region.

\subsection{IC1553}

In Fig. \ref{fig:IC1553_disps} we show the 2D distributions of both star-formation (SF) and shocks-contaminated bins for this galaxy arranged in a $3 \times 3$ grid of panels. Each panel represent one of the three \hcm free parameters (12+log(O/H), log(N/O) and log(U)) as a function of one of the three line ratio (\nii/\ha, \sii/\ha and \oiii/\hb). In the case of IC1553, we find in \citetalias{BETISII} that the halo regions above the asymmetric distribution of \hii regions at 2 kpc $\leq$ MAD $\leq$4 kpc exhibit a lower degree of ionisation compared to regions where the feedback emission from the \hii regions is weaker (-2 kpc $\leq$ MAD $\leq$ 2 kpc). In the latter case, fast shocks contribute 40\% ($f_{\rm shocks}$=0.4) to the total observed ionising flux, whereas when approaching the \hii region concentration at 2 kpc $\leq$ MAD $\leq$ 4 kpc, this contribution drops to 20\%. We define bins where $f_{\rm shocks}$ reaches its minimum value as SF-dominated eDIG bins (in this case, at 2 kpc $\leq$ MAD $\leq$ 4 kpc with $f_{\rm shocks}$=0.2), and bins where it reaches its maximum value as shock-contaminated eDIG bins (in this case, $f_{\rm shocks}$=0.4 at -2 kpc $\leq$ MAD $\leq$ 2 kpc). Throughout the entire work, the terms SF-dominated and shock-contaminated do not refer to ionisation mechanisms that are independent of SF. In all bins, photoionisation by young massive stars is the main ionisation source, and the fast shocks are also powered by SF, for example through stellar winds, supernovae, and disc--halo feedback (see \citetalias{BETISII}). The shock contribution is never the dominant one, since it reaches at most $\sim$50\% of the ionising flux in the shock-contaminated bins.
For the $12+\log(\mathrm{O/H})$ distributions, the locus remains centred around a value of $\sim$8.6 for both SF-dominated and shock-contaminated bins. In contrast, although the peaks of the two distributions are located at similar values, the shock-contaminated bins exhibit a broader spread, with a higher population at both sub-solar (down to $\sim$8.4) and super-solar (up to $\sim$8.9) metallicities compared to the SF-dominated bins. In contrast, clearer differences are observed in the \hcm derived N/O abundance. Most $\log(\mathrm{N/O})$ values are lower in shock-contaminated regions than in SF-dominated regions, shifting the locus from $\log(\mathrm{N/O}) \approx -0.8$ in the latter to $\approx -0.9$. In addition, although the ionisation parameter spans a relatively narrow range ($-3.25 \lesssim \log(U) \lesssim -2.90$), shock-contaminated regions tend to exhibit lower values. This is consistent with the low-ionisation state inferred for these regions in \citetalias{BETISII}, as well as with the enhanced emission of low-ionisation species such as \sii and \nii relative to H$\alpha$. This behaviour is also evident in Fig. \ref{fig:IC1553_disps}, where the \sii/H$\alpha$ and \nii/H$\alpha$ ratios are systematically higher in the shock-contaminated eDIG compared to the SF-dominated eDIG, by up to $\sim$0.2 dex and $\sim$0.4 dex, respectively. The locus is at approximately the same value for both distributions ($\sim$0.35). However, similarly to the $12+\log(\mathrm{O/H})$ distributions, the shock-contaminated regions exhibit a broader range of values, spanning from $\sim$0 to $\sim$0.7, whereas the SF-dominated eDIG regions are more tightly clustered around the central value, within a narrower range of $\sim$0.25 dex.

\subsection{ESO157-49}

In the case of ESO157–49, the contribution of fast shocks to the ionisation budget is higher than in IC1553, reaching $\sim$30\% in SF-dominated eDIG bins (-1.5 kpc $\leq$ MAD $\leq$ 3 kpc) and up to $\sim$50\% in shock-contaminated bins (-1 kpc $\leq$ MAD $\leq$ 1 kpc) \citepalias{BETISII}. Fig.\ref{fig:ESO157-49_disps} presents the corresponding 2D distributions. In this galaxy, a clearer separation is observed in the $12+\log(\mathrm{O/H})$ distributions, with the shock-contaminated component shifted towards lower values, peaking at $\sim$8.55, while the SF-dominated component shows a locus centred near solar metallicity ($\sim$8.69). The log(N/O) distribution is also shifted; however, in this case, the magnitude of the shift is much smaller than in IC1553, with the shock-contaminated distribution reaching, on average, values only 0.1 dex lower than the SF-dominated distribution and being centred around -0.85. Another notable difference with respect to the previous case is observed in log(U): here, the shock-contaminated regions exhibit on average a higher ionisation parameter than the SF-dominated regions (log(U) = -3.0 and -3.2, respectively), as well as higher \oiii/\hb ratios on average (0.25 and 0.15 respectively). This suggests a change in the ionisation conditions, implying a less diluted and more intense radiation field compared to the scenario found in IC1553, where shocks play a less significant role. The \nii/\ha and \sii/\ha ratios also consistently exhibit higher values in the shocks-contaminated regions; however, their differences are less pronounced than in the previous case. In particular, the \nii/\ha locus shifts from 0.35 in the SF-dominated regions to 0.6 in the shocks-contaminated regions, corresponding to a difference of 0.25 dex, in contrast to the 0.4 dex difference observed in IC1553. A similar trend is found for \sii/\ha, where the locus moves from 0.5 in the SF-dominated regions to 0.65 in the shocks-contaminated regions, yielding a difference of 0.15 dex, compared to 0.2 dex in IC1553. Moreover, these ratios generally reach higher values than in IC1553 under both ionisation scenarios.

\subsection{ESO443-21}

ESO443-21 is a particularly peculiar case due to the presence of a gas knot between -1 kpc $\leq$ MAD $\leq$ 2.5 kpc and extending up to 4.5 kpc. The hybrid models of \citetalias{BETISII} cannot reproduce the full ionisation of the halo unless a stellar metallicity twice the solar value is assumed, in contrast to the previous cases, for which a solar metallicity is sufficient. This is well reflected in the \hcm results, which are independent of the hybrid models assumed to explain the eDIG ionisation budget, and we consistently obtain higher gas-phase metallicity values than in the previous cases. The SF-dominated regions where the ionised gas knot is located present 12+log(O/H) values spanning 8.6–8.8, while the shock-contaminated regions reach values $\sim$0.1 dex lower. Similarly, log(N/O) also presents higher values in comparison with the previous cases, spanning from -1.0 up to -0.6, with a locus around -0.75 in the SF-dominated bins and a shock-contaminated distribution $\sim$0.1 dex lower. Regarding the ionisation parameter, while the shock-contaminated distribution remains centred around log(U) = -3.15 with a dispersion of no more than 0.1 dex, the SF-dominated distribution consistently presents lower values in comparison with the shock-contaminated distribution and those observed in the other galaxies, with a locus around log(U) = -3.2 and a broad dispersion spanning -3.3 to -2.9. Notably, an additional locus can be identified in the log(U) panels for the SF-dominated distribution, centred at log(U)$\approx -3.2$, and associated with an enhancement in \nii/\ha (up to 0.5), as well as notably higher values of \sii/\ha and \oiii/\hb for this locus (centred at 0.8 and 0.75, respectively). This is compatible with a low ionisation regime that enhances the \sii and \nii emission in the DIG and eDIG, also seen in \citetalias{BETISII} and reaching the same values, where the filamentary knot, commonly observed in galaxies experiencing strong SF feedback and disc–halo interactions (e.g. \citealt{1996ApJ...462..712R,2003A&A...406..493R,2003A&A...406..505R,2011MNRAS.415.1534M,2019ApJ...882...84L,2019A&A...623A..89C}), has a small influence on the ionisation of the halo, inducing this low ionisation regime. Although this structure is not apparent in the metallicity maps (Fig. \ref{fig:maps}), its indirect impact is detected in the ionisation parameter, as the change in the ionisation conditions does not lead to significant variations in the derived gas-phase abundances from \hcm.

\begin{figure*}[ht!]
\centering
    \includegraphics[width=0.5\textwidth]{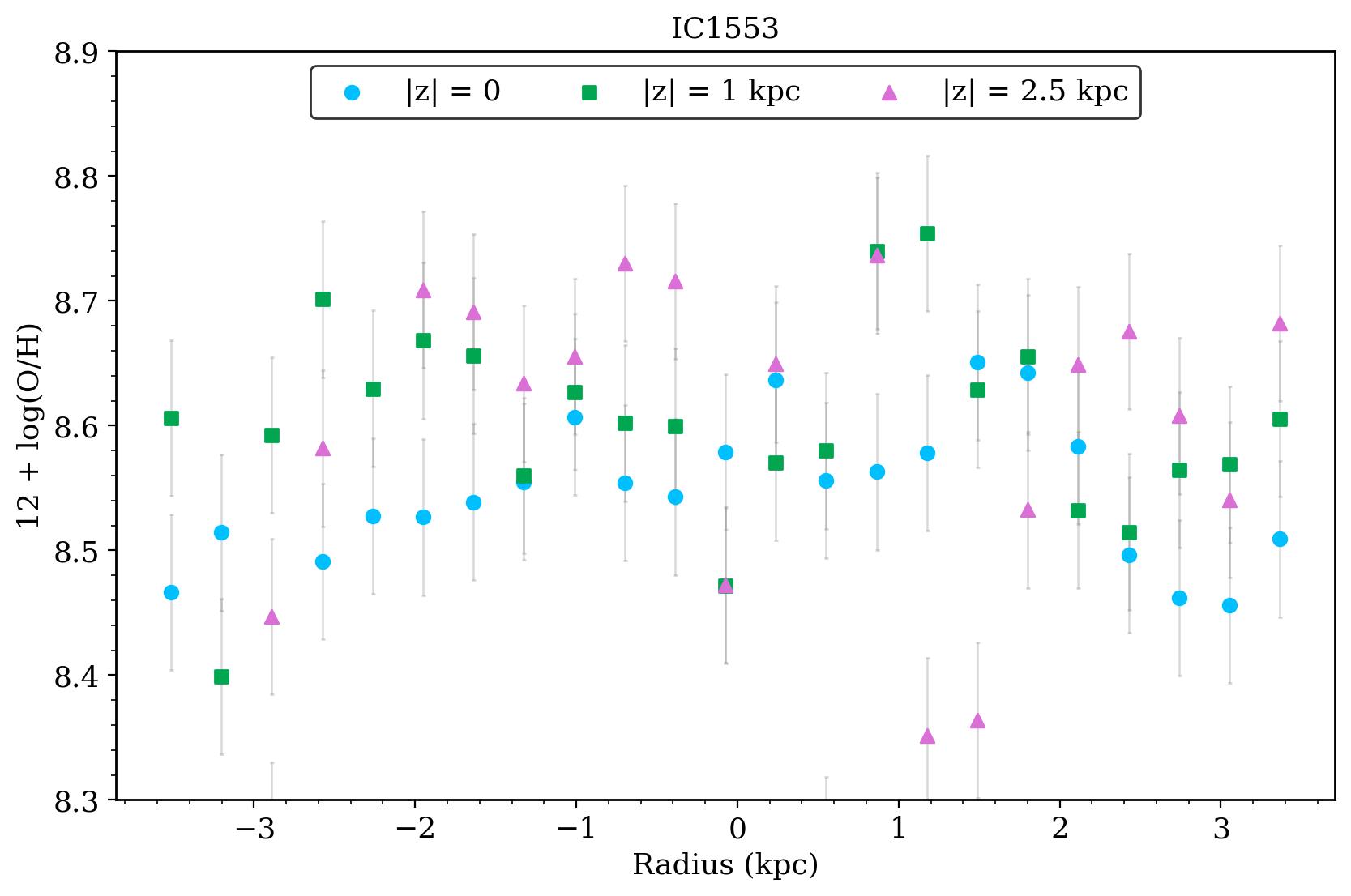}
    \includegraphics[width=0.49\textwidth]{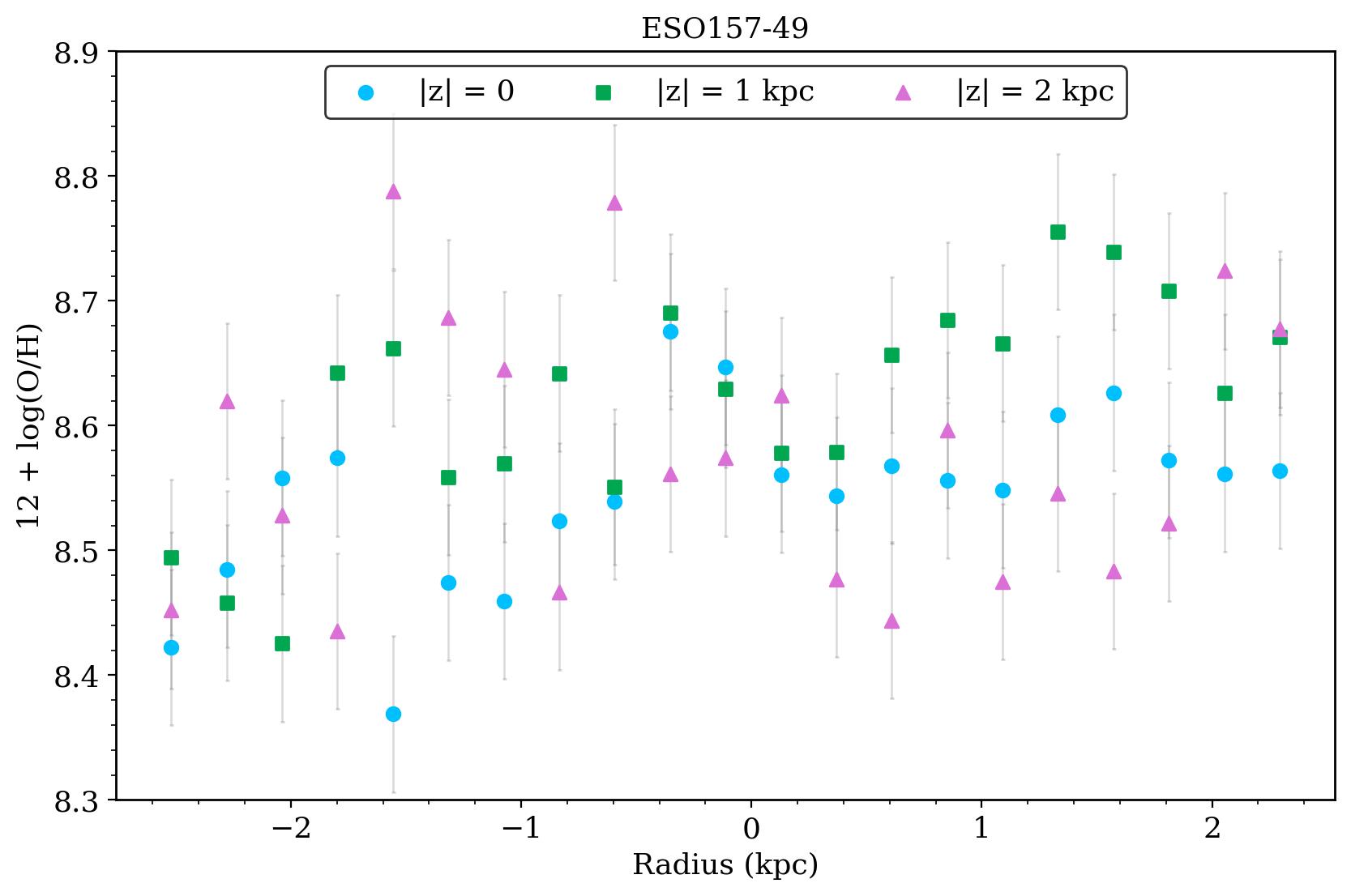}
    \includegraphics[width=0.5\textwidth]{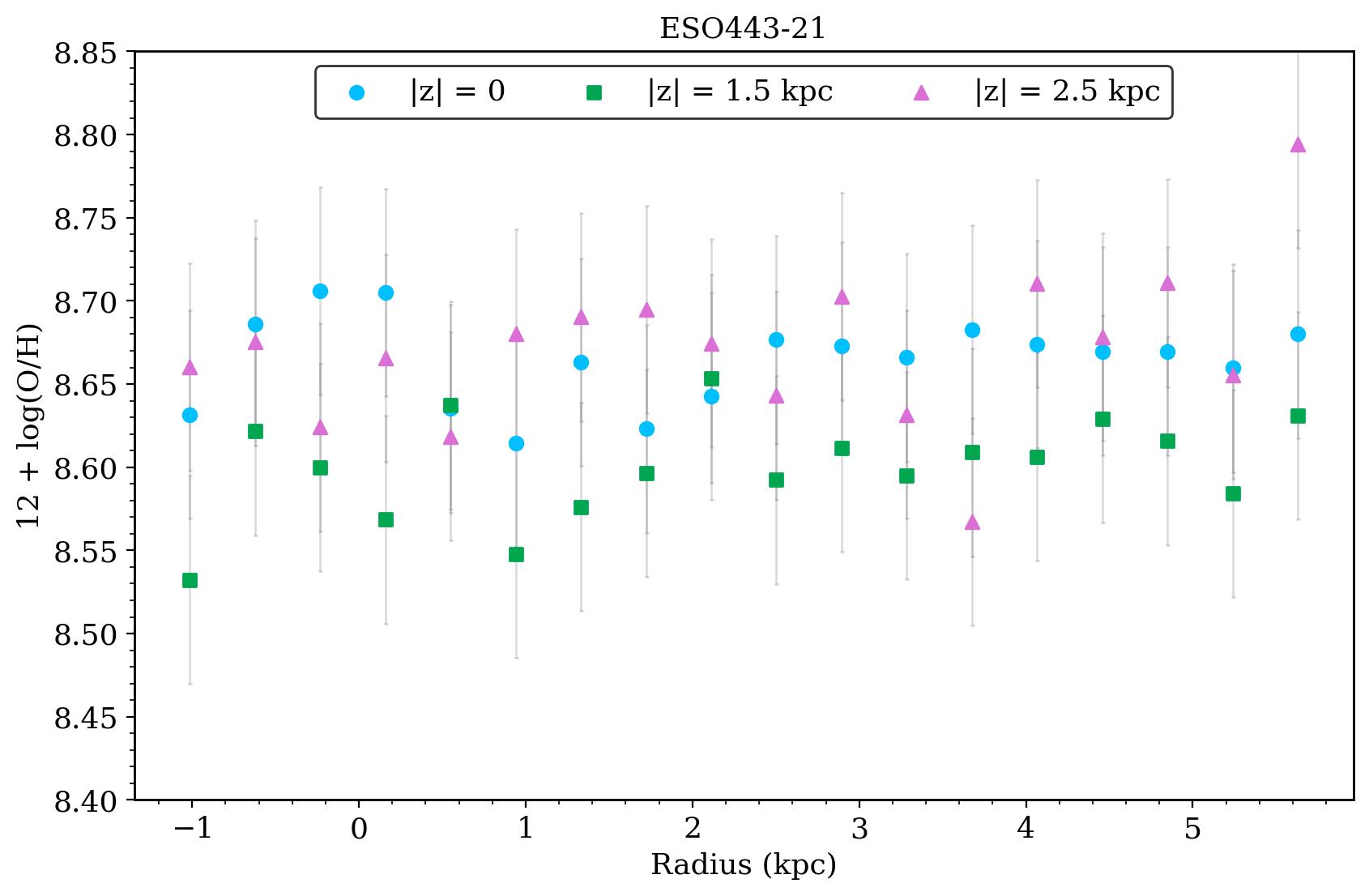}
    \includegraphics[width=0.49\textwidth]{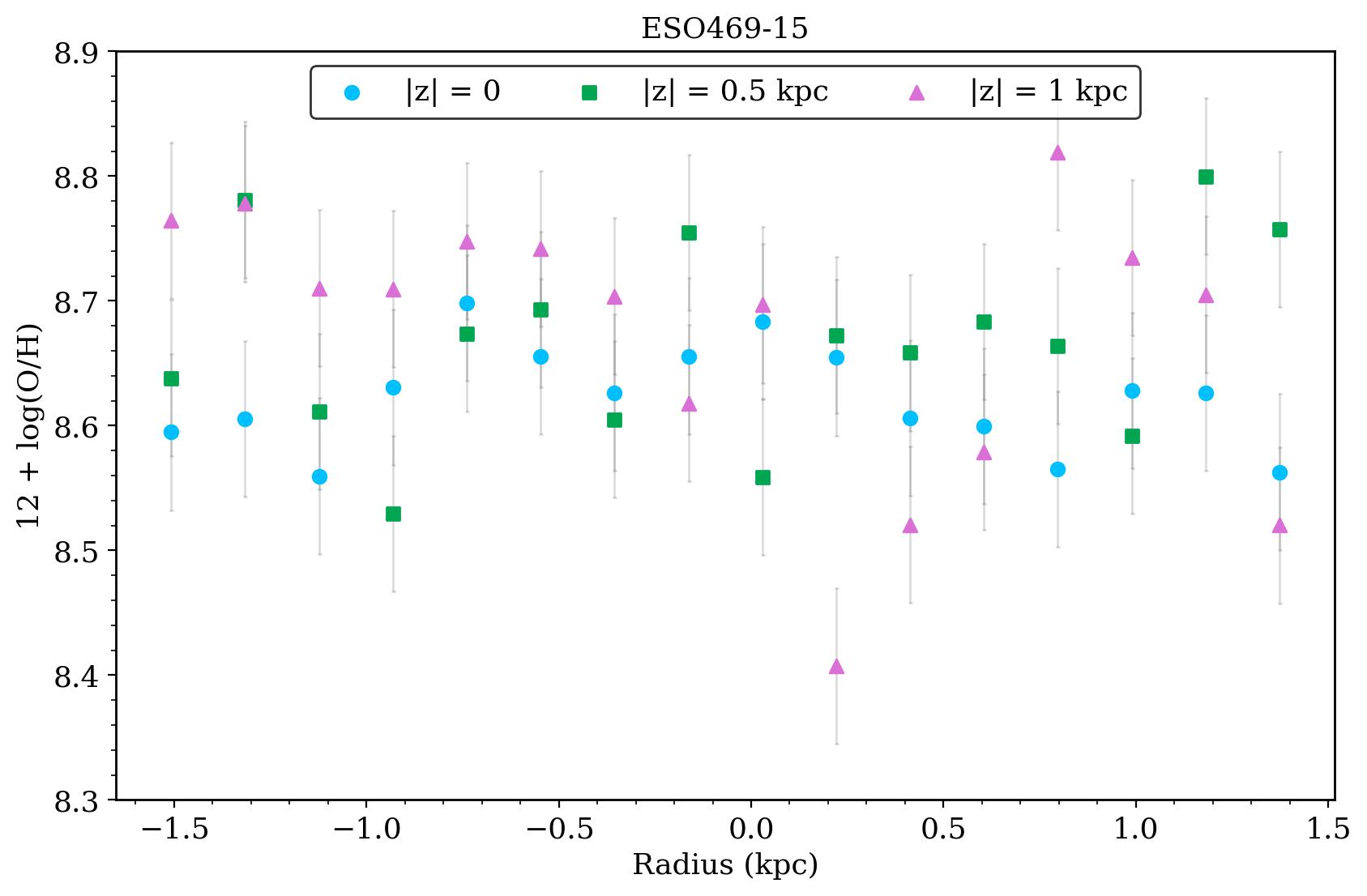}

\caption{Radial 12 + log(O/H) gradient at three different distances from the midplane: 0 (midplane, i.e., galactic metallicity gradient, with blue dots), 1 kpc (green squares) and 2.5 kpc (magenta triangles) for IC1553, 0, 1 kpc and 2 kpc for ESO157-49, 0, 1.5 kpc and 2.5 kpc for ESO443-21 and 0, 0.5 kpc and 1.0 kpc for ESO469-15.}
\label{fig:H_O_grad}     
\end{figure*}

\begin{figure*}[ht!]
\centering
    \includegraphics[width=0.5\textwidth]{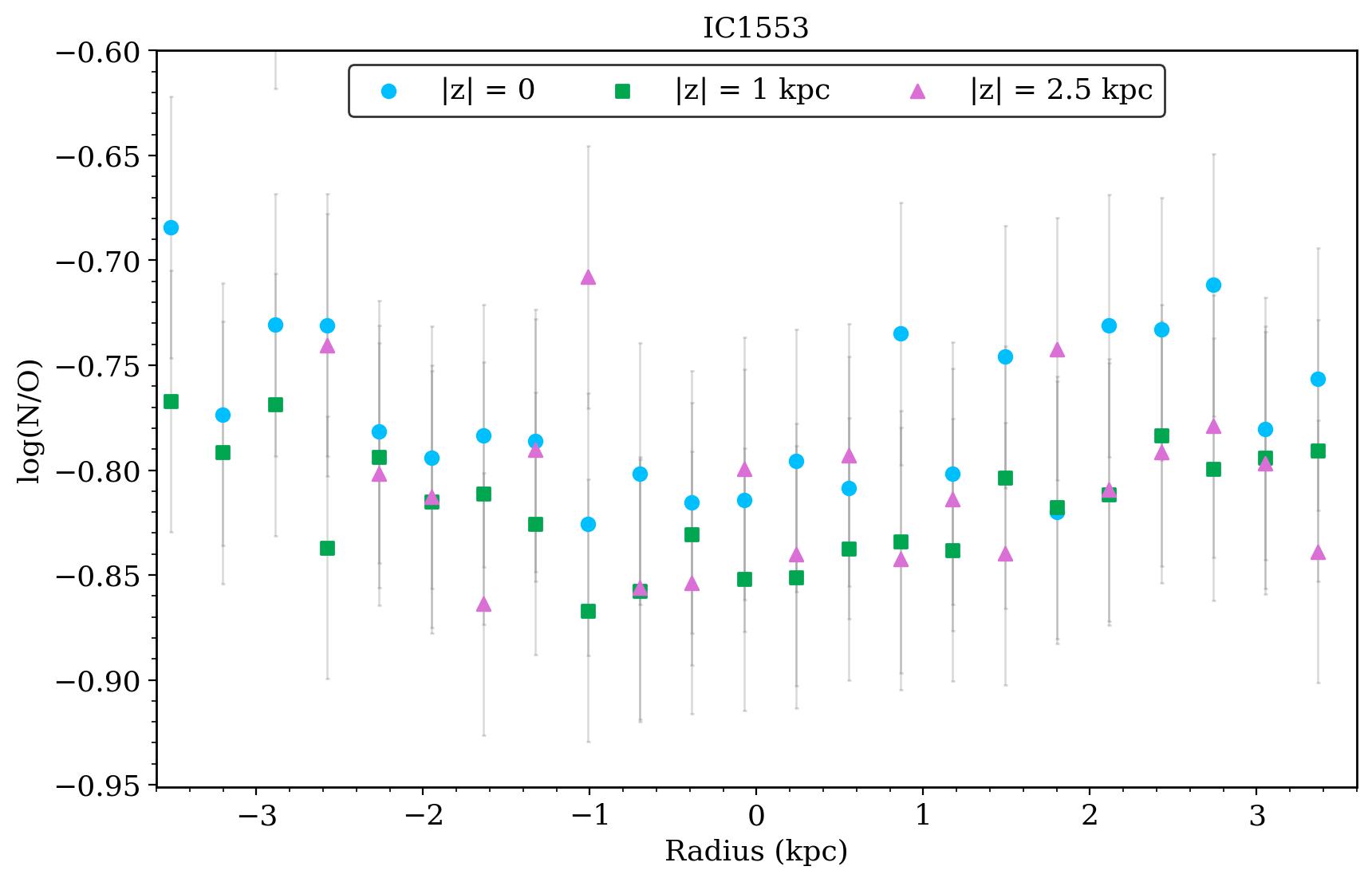}
    \includegraphics[width=0.49\textwidth]{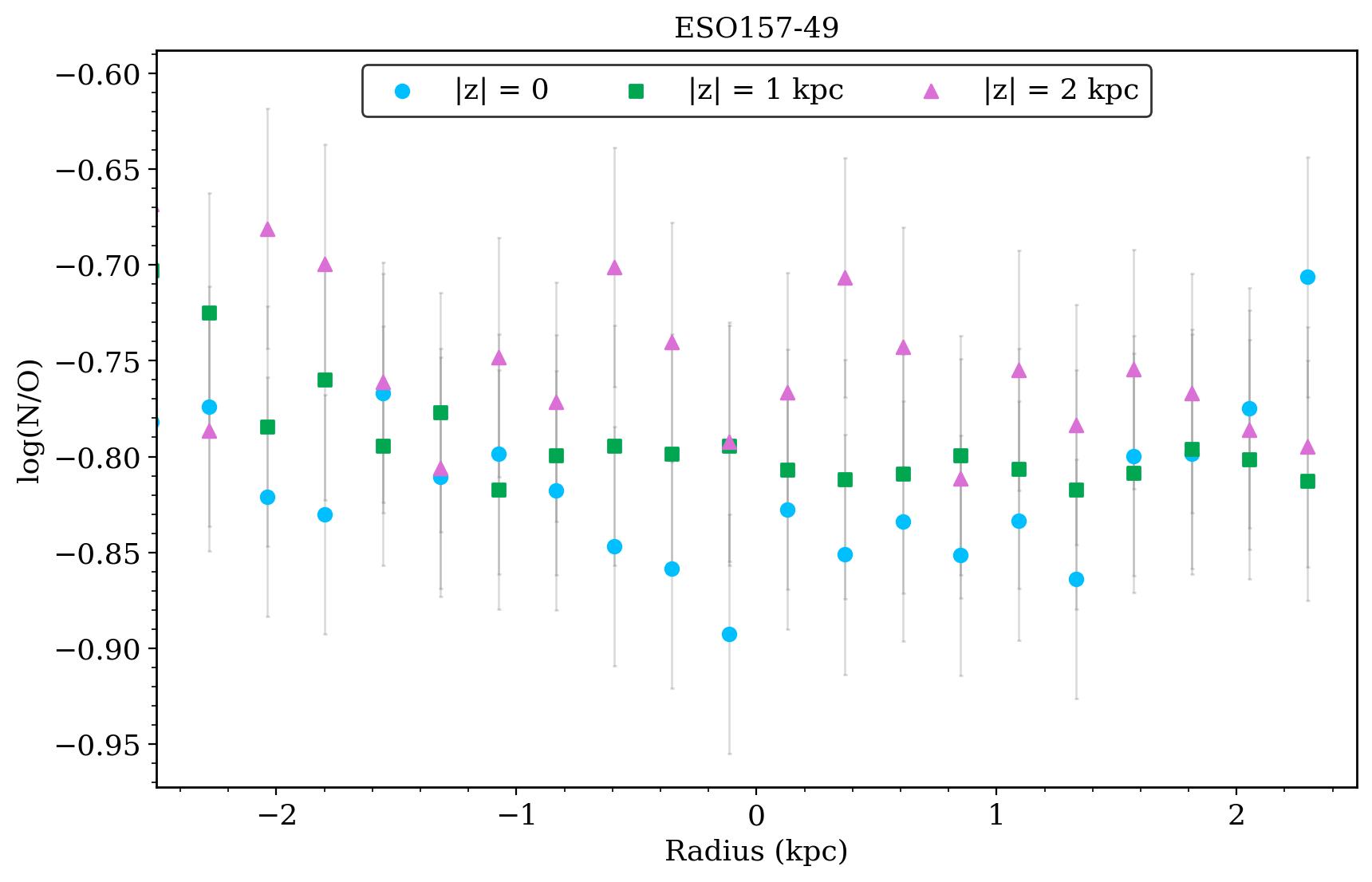}
    \includegraphics[width=0.5\textwidth]{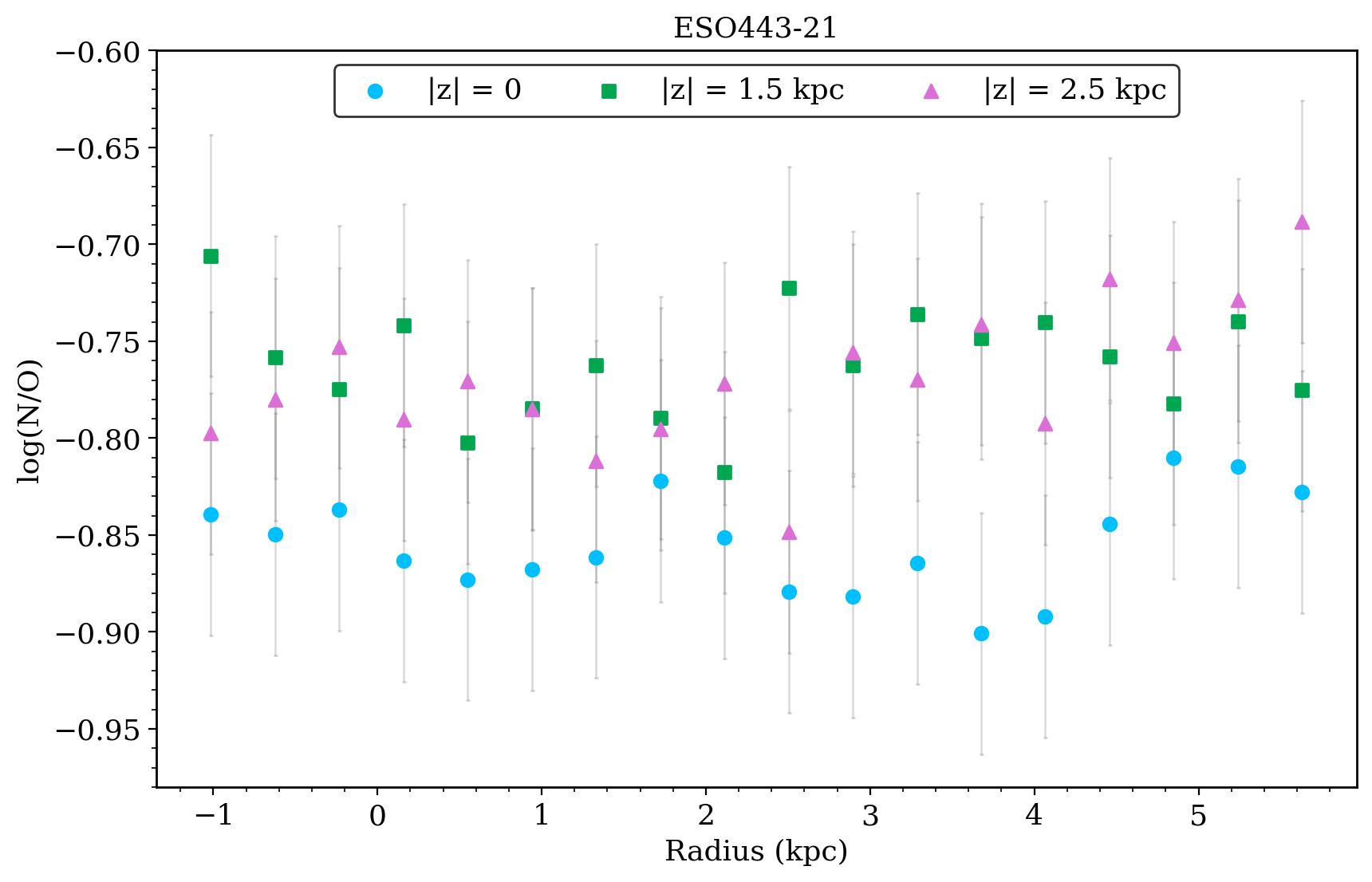}
    \includegraphics[width=0.49\textwidth]{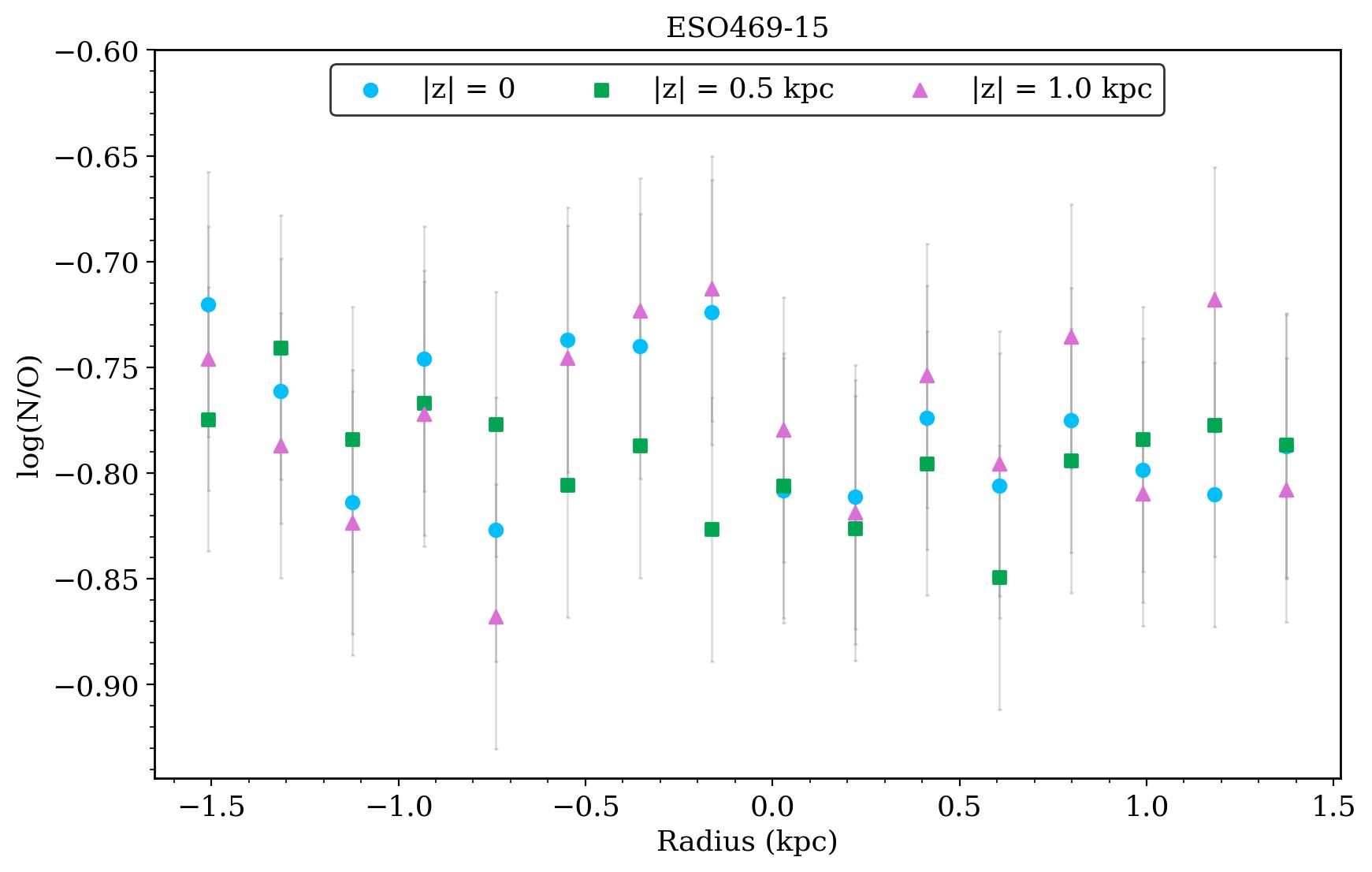}

\caption{Similar to Fig. \ref{fig:H_O_grad} for log(N/O).}
\label{fig:N_O_grad}     
\end{figure*}

\subsection{ESO469-15}
In ESO469-15, the influence of the extraplanar \hii regions on the abundance determination is clearly evident. The 12+log(O/H) values tend to be more tightly clustered around solar values in the SF-dominated regions, with a smaller dispersion compared to ESO157-49 and ESO443-21. In contrast, the inferred log(N/O) values exhibit a notably broader dispersion, reaching significantly higher values than those observed in the other galaxies, spanning from log(N/O) = -1.0 to -0.5. The ionisation parameter also reaches higher values for the SF-dominated distribution, spanning from log(U) = -3.2 to -2.8. The contribution of shocks in this galaxy does not exceed 30\%, and the models that best reproduce the ionisation budget require assuming an average stellar metallicity of 2Z$_{\odot}$. This strong contribution from extraplanar \hii regions is also well reflected in the line ratios, with the low-ionisation species in the SF-dominated distribution exhibiting typical values found in SF regions ($\sim$0.2 and 0.3 for \sii/\ha and \nii/\ha, respectively). Moreover, these ratios do not differ substantially between the two regimes, particularly for \sii/\ha and \oiii/\hb, where the shift is almost negligible. The relatively low contribution of shocks compared to SF is further reflected in the small dispersion of the shock-contaminated distributions, which reach slightly sub-solar values for 12+log(O/H) and log(N/O) ($\sim$8.65 and -0.8, respectively).

\section{Discussion}\label{sec:discuss}
Overall, the metallicity values inferred with \hcm do not change significantly across the sample, suggesting that the partial contribution of fast shocks into the ionisation budget does not strongly bias the results given by the photoionisation models. The oxygen abundance, 12+log(O/H), derived by \hcm remains relatively similar between SF-dominated and shock-contaminated regions, with differences typically within $\sim$0.1-0.15 dex and therefore comparable to the shape of the models grid sampling. In contrast, log(N/O) tends to be systematically lower in the shock-contaminated gas. These values are calculated using \nii and \sii lines, so the differences measured can be due to the larger variation found in the \sii line in comparison with the \nii line in shock-contaminated regions, but the differences in N/O are still within the sampling associated with the model grids. The ionisation parameter is generally similar or only slightly lower than in the SF-dominated regime. The enhanced low-ionisation emission-line ratios observed in shock-contaminated regions reflect the different excitation conditions rather than significant changes in the derived oxygen abundance, supporting the use of \hcm to estimate chemical abundances when fast shocks partially contribute to the ionisation budget alongside photoionisation, as the bias potentially introduced is small, and comparable to the models grid sampling (0.1-0.15 dex).

The differences between the loci of the two types of distributions are never higher than 0.1-0.15 dex, which lies within the \hcm model grids sampling. This suggests that adopting the same SED in the photoionisation model grids used by \hcm in this analysis does not significantly affect the results produced by the code, introducing potentially a bias in the results comparable to the model grid sampling. Given the reliability of the code in this context, together with the small contribution of shocks to the metallicity estimation, we can explore the dependence of these derived values on height. Figures \ref{fig:H_O_grad} and \ref{fig:N_O_grad} show the 12+log(O/H) and log(N/O) radial distributions respectively.

The inferred O/H abundance distributions presented in Fig. \ref{fig:H_O_grad} show no clear dependence on the distance from the midplane. In most galaxies, the halo and midplane distributions span a similar range of oxygen abundances, typically between 8.4 and 8.7. The only exception is IC1553, with variations of up to 0.2 dex between the disc and the halo. In this galaxy, the oxygen abundance measured in the disc ranges from $\sim$ 8.4 in the outer regions to $\sim$ 8.6 in the central parts, whereas the halo shows systematically higher values, ranging between $\sim 8.6$ and $\sim 8.8$. This behaviour strongly contrasts with the log(N/O) distributions.  ESO443-21 and IC1553 exhibit clear differences in their respective N/O distributions. Although both galaxies show a similar radial trend, with log(N/O) marginally increasing with radius in all distributions, they differ significantly in the values of log(N/O) measured in the halo compared to those in the midplane at |z| = 0. IC1553 presents intense SF activity in the disc at R > 1 kpc, leading to an SF-feedback-dominated regime in the halo and a bi-conical structure (also seen in Fig. \ref{fig:maps} top central panel), as discussed in \citetalias{BETISII}. This SF-powered regime is consistent with the decrease in log(N/O) as a function of distance from the midplane, with log(N/O) changing from $\sim$ -0.75 in the disc plane to $\sim$ -0.85 in the halo. The opposite behaviour is observed in ESO443-21, where such intense SF activity is absent. In this galaxy, log(N/O) reaches values as low as $\sim$ -0.9 in the midplane, while increasing to values of up to $\sim$ -0.7 in the halo and at larger galactocentric radii. A similar trend is observed in ESO157-49, although in this case it is more strongly concentrated towards the central regions of the galaxy. In the case of ESO469-15, the sampled distance from the midplane is relatively small (up to 1 kpc), and therefore no clear trend is observed. In comparison with other studies, \citet{2023A&A...678A..84D} measure the oxygen abundances of two \hii regions in IC1553 located at $z = 0$ using the S32 diagnostic, obtaining values of $12 + \log(\mathrm{O/H}) = 8.47-8.48$. These measurements are consistent with our results at $z = 0$ and galactocentric distances of $\rm R \approx -2$ to $-3$ kpc, corresponding to the locations of the two \hii regions analysed by \citet{2023A&A...678A..84D} (see their Fig. 3). Furthermore, the O/H abundances obtained here are consistent with those inferred by \citet{Rautio2024}, who performs CLOUDY modelling of both the eDIG and the disc for the same sample, finding maximum abundance offsets of $\approx$ 0.2 dex between the eDIG and the disc at $z = 0$. This is in good agreement with our results in Fig. \ref{fig:H_O_grad}, where the maximum offsets are also found to be approximately 0.15–0.2 dex. Furthermore, these values are also consistent with our $12 + \log(\mathrm {O/H})$ measurements for IC1553 at $z = 0$, as well as with those reported by \citet{2023A&A...678A..84D}.

Multiple hydrodynamic simulations modelling OB-associations \citep[e.g. ][]{Springel03,MelioliFountains09} show that while fountains and galactic outflows eject metals to kiloparsec heights, the freshly ejected metals tend to fall back close to the same Galactocentric region from which they originated, preventing significant radial abundance profile changes. That means that galactic fountains and SF feedback significantly enrich galactic halos with metals including oxygen, though the metals tend to circulate locally rather than spread radially across the disc. \citet{MelioliFountains2015} found that 25-80\% of metal-rich supernova ejecta is vented into the intergalactic medium depending on galaxy mass, and more especially in dwarf galaxies, with about half of retained metals located far (> 500 pc) from the galactic plane. For oxygen, \citet{VasilievOHalos15} showed that feedback-driven ionisation can produce high column densities of ionised oxygen ($10^{14.5-15.3}cm^{-2}$) in extended halos, demonstrating that hot outflows are capable of generating metal-enriched gaseous atmospheres. The similarity of the HCm-derived O/H values between disc and halo, as long as the small effect of fast-shocks contamination, could reflect genuine chemical homogeneity produced by galactic fountains, as predicted by hydrodynamical simulations and the circulation of metal-rich gas between these two components driven by feedback processes, as discussed in \citetalias{BETISII}. However, we cannot rule out that this similarity partly arises because the \hcm strong-line method is anchored, via its assumed empirical O/H-U relation, to the values characteristic of the \hii regions on which it is calibrated. Distinguishing between these two scenarios would require direct-method abundances, which are not accessible with our current data.

\noindent
Regarding the inferred N/O values, \hcm does not assume any empirical relation between O/H and N/O, deriving the N/O abundance ratio independently from the \nii and \sii lines, that is, from the N2S2 index (\nii/\sii). Both lines arise in the same low-ionisation gas, so the ratio should depend only weakly on the ionisation parameter. However, it traces N/O only if the S/O ratio is approximately constant, and any variation in S/O would appear as an apparent variation in N/O. The lower N/O values in the shock-contaminated regions may therefore partly reflect the behaviour of the \sii line rather than a genuine abundance difference, although these differences remain within the sampling of the model grid. We therefore regard the radial and vertical N/O trends within each regime as more reliable than the absolute N/O values, and the N/O trends reported here are likely to be more robust against the ionisation-regime biases that can affect O/H determinations.
Another possible scenario to explain the high N/O abundance ratios observed in galactic halos is that they may be associated with primary nitrogen production in very metal-poor massive stars \citep[e.g. ][]{ChiappiniNO05} and may become more apparent under conditions of weaker SF feedback (conditions observed in ESO443-21 and ESO157-49). Furthermore, the N/O abundance ratio is known to vary with galactocentric radius \citep[e.g. ][]{Pilyugin2004,PerezMontero2016}, reflecting the different evolutionary histories of different regions within a galaxy. In particular, shorter gas infall timescales are expected in the inner regions, whereas longer infall timescales combined with higher star formation efficiencies are characteristic of the outer regions \citep[e.g. ][and references therein]{MagriniNO2018}. These scenarios may account for the O/H radial variations observed in our sample, which are more pronounced in regions of reduced SF activity, such as in ESO443-21 at (R > 2 kpc) and in the central regions of ESO157-49. In the case of IC1553, it presents a particularly high SF activity in comparison with the other galaxies of the sample (also reflected in the particularly high star formation rate (SFR) of the disc depicted in Table \ref{tab:sample}) which may be indicative of strong and rapid ongoing SF-driven feedback between the disc and the halo, among with an observed bi-cone where low ionisation dominates and shocks contribute most as discussed in \citetalias{BETISII}. This high SF-feedback outside the influenced area of the bi-cone where the SF strongly dominates can be associated with the decrease in the oxygen abundance, as reflected by the higher log(N/O) and lower values of 12 + log(O/H) observed in the halo, which can explain the small removal or redistribution of oxygen-enriched gas from the disc.

The HCm grids assume photoionisation equilibrium, whereas the low densities of the eDIG imply recombination timescales of up to several Myr, which can exceed the variability timescale of the ionising sources. Recent radiation-hydrodynamic simulations including time-dependent ionisation of hydrogen and metals show that, in this regime, the ionisation state of the gas depends on its ionisation history \citep[e.g.][]{McCallum2024a,McCallum2024b}. 
In these simulations, S$^{+}$ is hardly affected, N$^{+}$ and O$^{+}$ are moderately affected, and ions with high ionisation potentials, such as O$^{2+}$ and
Ne$^{2+}$, are over-ionised at high altitudes with respect to the equilibrium solution. Moreover,
collisional ionisation in supernova-heated gas can reduce the N$^{+}$ fraction. Under these
conditions, the line ratios predicted by equilibrium models may differ from those of the eDIG, with
a possible impact on our results mainly through [O\,\textsc{iii}]/H$\beta$, which constrains
$\log(U)$. We note, however, that these simulations refer to a Milky Way-like patch, do not include fast shocks as an ionisation source, and still present discrepancies with
observations at low altitudes, so a quantitative correction of our abundances is not possible at
this stage. Moreover, our results show no significant systematic difference in the oxygen abundance between the disc and the halo (up to |z| $\gtrsim$ 2 kpc) in most galaxies of the sample, being the differences mostly where the SF dominates, and hence, where the density is higher than the low-density regime in which the simulations predict significant departures from equilibrium. Dedicated modelling of the non-equilibrium ionisation of metals in the eDIG is therefore required to assess this effect.

\section{Summary and conclusions}

In this work, we present the first spatially resolved estimation of the O/H and N/O abundances in the extraplanar diffuse ionised gas as a function of its ionisation conditions using integral field spectroscopy. We investigate the potential biases introduced by the partial contribution of fast shocks to the ionisation conditions of the gas, when deriving abundances with a methodology based on strong-line photoionisation models. We used four edge-on galaxies observed with MUSE, IC1553, ESO157-49, ESO443-21, and ESO469-15, selected from the \citetalias{BETISII} study. We selected the galaxies with the highest spatial resolution and the highest contrast in the relative contributions of fast shocks and star formation to the ionisation budget of the eDIG due to the particular characteristics of their galactic halos. To perform this estimation, we used the model-based code {\sc HII-CHI-Mistry} v5.5 from \citet{PerezMonteroHCm14}, which derives O/H, N/O, and U from measured strong emission-line ratios by comparing the observations with an extensive, user-selectable library of photoionisation models. Using as input the fluxes from the binned strong \oiii, \sii, and \nii de-reddened relative-to-\hb emission line maps, and assuming the POPSTAR stellar population models \citep{POPSTAR2009} together with a \citet{2003PASP..115..763C} IMF, we derived the binned 12+log(O/H), log(N/O) and log(U) maps. We divided the results for each galaxy into two subsets: one corresponding to the bins classified in \citetalias{BETISII} as eDIG regions where ionisation is dominated by star-formation feedback from the galactic plane (SF-dominated bins), and another corresponding to bins where fast shocks contribute to the ionisation budget (shock-contaminated bins). When applying this distinction, we find that galaxies with a higher contribution from shocks exhibit a small shift in the \hcm-derived metallicity values.

The individual galaxies exhibit distinct ionisation and \hcm-derived abundance patterns associated with the relative contribution of shocks and star formation. IC1553 and ESO157-49 show the strongest impact of shocks, with shock contributions reaching up to 40-50\% in shock-contaminated regions. In IC1553, shocks are mainly associated with enhanced low-ionisation emission, especially in the biconical structure of the halo, a lower ionisation parameter, and a decrease in log(N/O), while oxygen abundances remain close to solar values, manifesting a strong SF-feedback-dominated regime in the halo, that may be compatible with fountains and galactic outflows ejecting material to the halo.
In contrast, ESO157-49 displays lower values of both 12+log(O/H) and log(N/O) in shock-contaminated regions, together with only modest changes in the ionisation parameter and emission-line ratios.

ESO443-21 and ESO469-15 exhibit a smaller contribution from shocks, with star formation dominating the ionisation budget. In both galaxies, shock-contaminated regions are characterised by lower oxygen and nitrogen abundances, while the ionisation parameter remains broadly similar between regimes. Their emission-line ratios are generally closer to those expected for classical \hii regions, suggesting weaker shock signatures. A distinctive feature of ESO443-21 is the presence of a secondary sequence in the star-formation-dominated regions, characterised by lower metallicities, lower ionisation parameters, and enhanced line ratios, likely associated with the large ionised filamentary structure observed in the galaxy.

While the SF-dominated eDIG regions show approximately solar metallicities, the shock-contaminated regions display lower average O/H and N/O \hcm-derived, with offsets of $\sim$0.1-0.15 dex, on average. Moreover, when star formation dominates, the ionisation parameter exhibits a broad dispersion, spanning from low-ionisation regimes at log(U) $\approx$ -3.3 to higher ionisation values up to -2.8. In contrast, the shock-contaminated regions show a more uniform and consistently low ionisation regime, with log(U) $\approx$ -3.1.

Overall, our results show that the inference of chemical abundances in the eDIG through strong-line calibrations is closely linked to the ionisation conditions of the gas. Regions with a larger contribution from fast shocks systematically exhibit lower excitation, enhanced low-ionisation line ratios (\nii/\ha and \sii/\ha), lower ionisation parameters, and offsets of up to $\sim$0.1-0.15 dex in both O/H and N/O relative to SF-dominated regions. In contrast, regions whose ionisation is primarily driven by star formation feedback from the disc display properties more consistent with classical \hii regions, including higher excitation with lower \nii/\ha and \sii/\ha ratios as well as smaller abundance offsets and abundances closer to solar values. The observed dispersions and systematic shifts are consistent with previous findings for planar DIG \citep[e.g.][]{AbundancesPerez11,SandersDIGabundance17,KumariDIGabundance19,PoetrodjojoDIGabundance19,ValeDIGabundance19}.%, supporting the idea that the variations in electron temperature and density found in \citetalias{BETISII} for this sample, likely associated with shock-driven or diffuse low-density environments, can bias strong-line abundance determinations.

These results also suggest that the different physical regimes within the eDIG, which distinguish gas predominantly ionised by star formation from gas significantly influenced by shocks, do not substantially affect the abundances derived with \hcm. The differences found in O/H and N/O remain within the uncertainties of the solutions found by the code, suggesting that the results of the abundances determination do not strongly depend on the presence of fast-shocks in the eDIG ionisation regime, as these values can be biased as much as the intrinsic uncertainty associated to the shape of the model grid bins.

\noindent
Taking into account the low impact of the presence of fast-shocks as an ionisation regime to the \hcm metallicity determination, we study the variation of the radial distribution of the N/O and O/H ratios as a function of the distance from the midplane for our galaxies. Galaxies with reduced feedback activity, such as ESO443-21 and ESO157-49, show enhanced N/O values in their halos, whereas the intense feedback and outflow activity observed in IC1553, traced by its prominent bi-cone structure, can be associated with lower halo oxygen abundances and higher N/O ratios, suggesting the preferential redistribution or removal of oxygen-enriched gas by feedback-driven outflows. The small O/H abundance offsets between the eDIG and the galactic disc found in this work are consistent with previous studies based on theoretical CLOUDY models \citep{Rautio2024} and measurements of individual \hii regions in the galactic plane of IC1553 \citep{2023A&A...678A..84D}. Overall, our results support a scenario in which SF-driven feedback and galactic fountains efficiently transport metals from the disc into the halo, producing a chemically mixed disc-halo system with only weak radial variations in oxygen abundance. The similar O/H distributions observed in both components are consistent with the local circulation and recycling of oxygen-rich material predicted by hydrodynamical simulations. In contrast, the N/O ratio appears to retain information about the star formation history and chemical evolution of different galactic regions, exhibiting stronger radial and vertical variations, particularly in environments with weaker SF feedback.

The \hcm-derived O/H and N/O values in the eDIG correlate systematically with the ionisation regime of the gas, in a way that mirrors the abundance biases reported for the planar DIG. This confirms that strong-line methods calibrated on \hii regions are subject to small systematic offsets when applied to gas in  shock-influenced, low-density conditions, and provides the first spatially resolved characterisation of this effect in edge-on halos. In order to quantify the intrinsic metallicities in these regimes, future work should derive the chemical abundances using the direct method and assess whether the effects and biases identified in the model-based estimates are also present in the direct-method determinations. 

\begin{acknowledgements}

The authors acknowledge financial support from the State Agency for Research of the Spanish MCIU through Center of Excellence Severo Ochoa’ award to the Instituto de Astrofísica de Andalucía CEX2021-001131-S funded by MCIN/AEI/10.13039/501100011033, and from the project PID2022-136598NB-C32 “Estallidos8”. R.G.D. acknowledge support by the project ref. AST22\_00001\_Subp\_15 funded from the EU – NextGenerationEU. L.G. acknowledges financial support from MCIN and AEI 10.13039/501100011033 under projects PID2023-151307NB-I00, CEX2020-001058-M, and by the MaX-CSIC Excellence Award MaX4-SOMMA-ICE. EPM also acknowledges the assistance from his guide dog Rocko, without whose daily help this work would have been much more difficult.

\end{acknowledgements}

%\section*{Data availability}

\bibliographystyle{aa}
\bibliography{aa}

\appendix

\section{Model constraints}

\begin{figure*}[ht!]
\centering
    \includegraphics[width=0.5\textwidth]{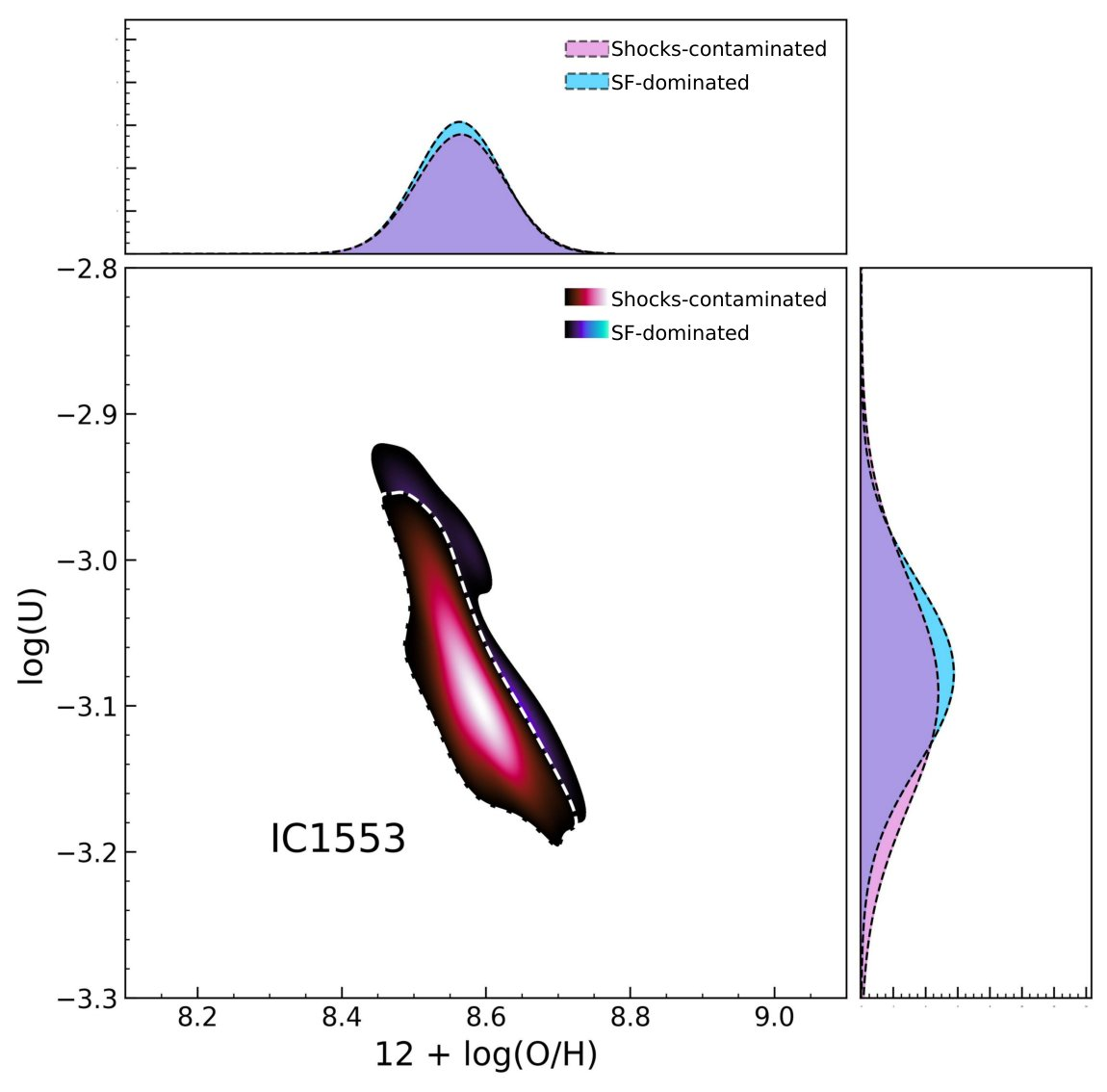}
    \includegraphics[width=0.49\textwidth]{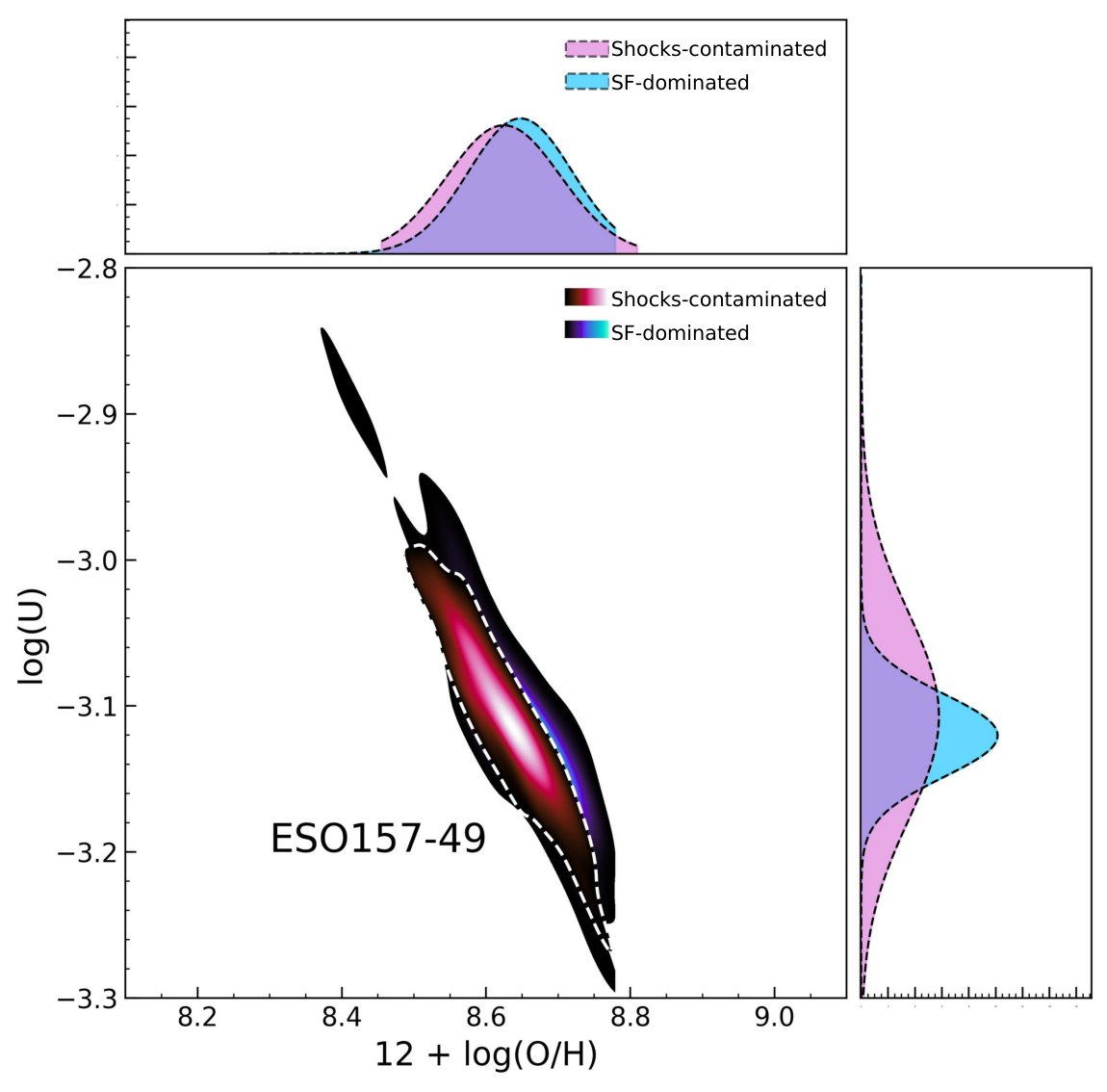}
    \includegraphics[width=0.5\textwidth]{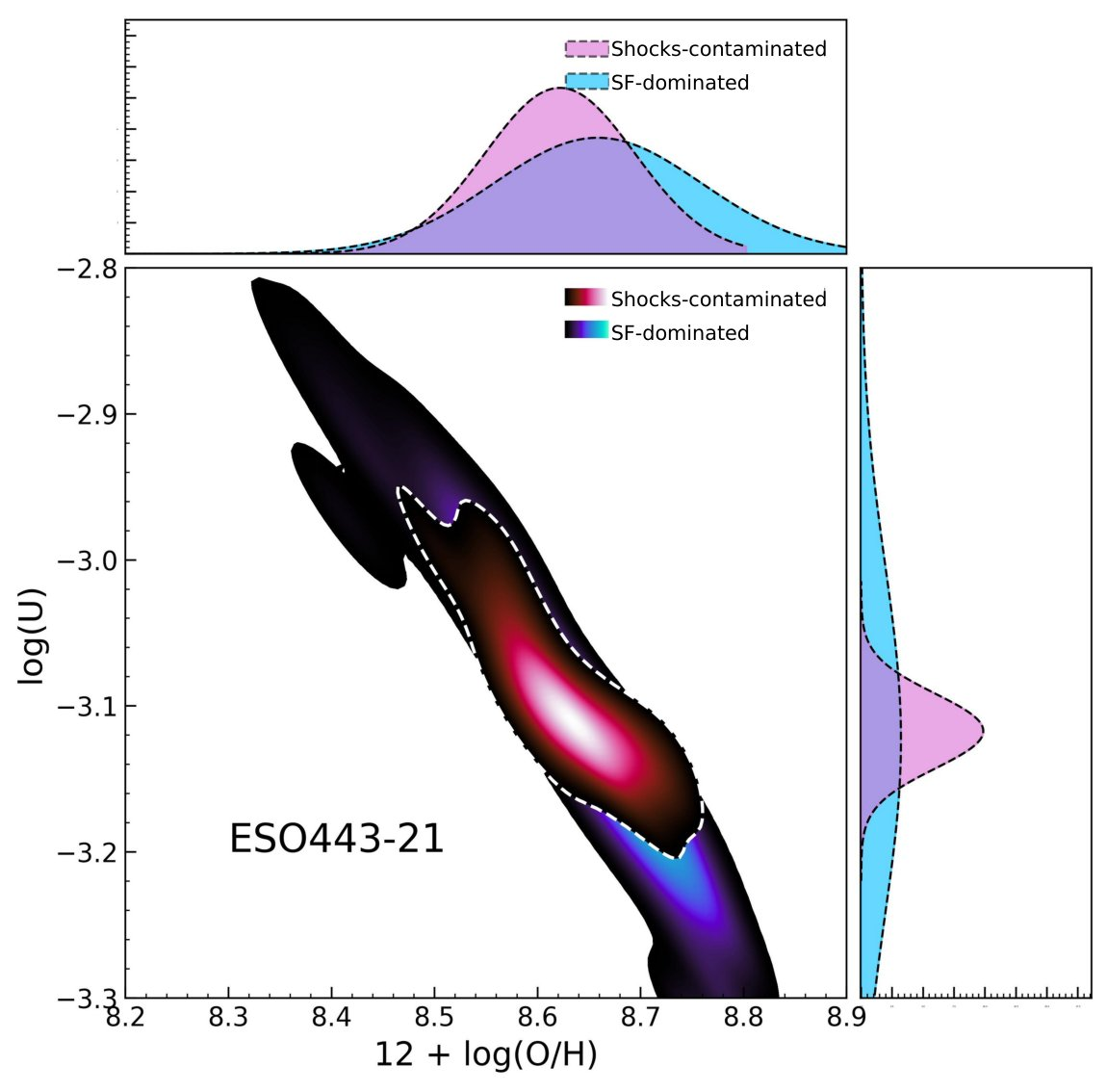}
    \includegraphics[width=0.49\textwidth]{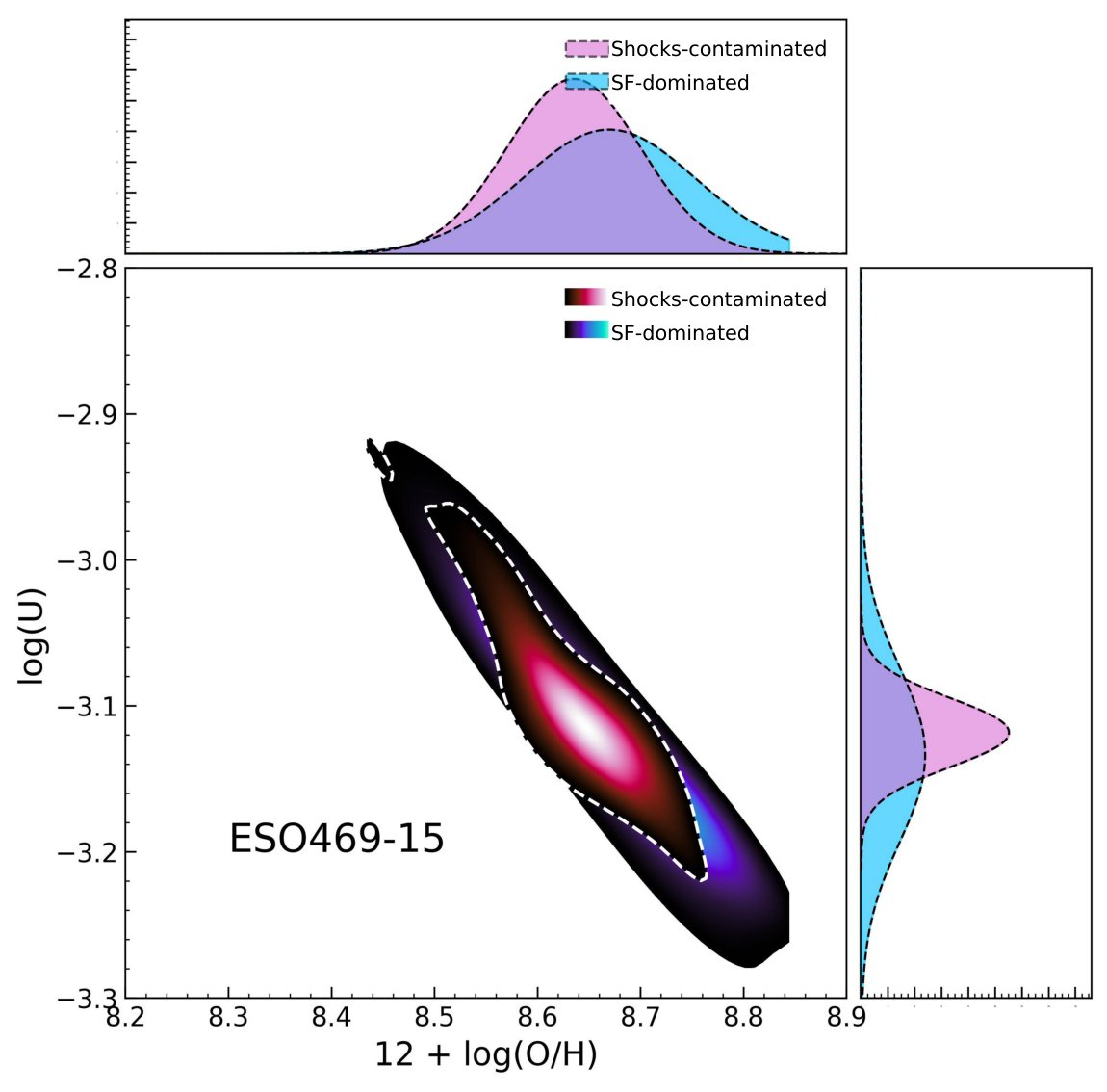}

\caption{12+log(O/H) vs log(U) for our results. We again divide the results into shocks and SF-dominated 2D dispersions as in previous figures.}
\label{fig:OHvsU}     
\end{figure*}

In this analysis, we need to take into account that the MUSE wavelength coverage does not allow the detection of \oii$\lambda$3727\A or \oiii$\lambda$4363\A at these low redshifts, some assumptions must be made. First, the \hcm models assume an empirical relation between O/H and U, showing a trend in which higher metallicities correspond to lower ionisation parameters, and vice versa. This restricts the possible values of O/H and U to the combinations presented in \citet{PerezMonteroHCm14}, which span from log(U) = -1.5 at the lowest metallicities (12+log(O/H)=7.1) and log(U)=-3.75 at the highest metallicities (12+log(O/H)=9.2) and follow a linear trend with higher dispersion in the low-metallicity regime. In Fig.\ref{fig:OHvsU} we show the 12+log(O/H)-log(U) relation for our results, again separating the data into two 2D distributions corresponding to shock-contaminated and SF-dominated regions. We can clearly see that, in the four galaxies, the trend is also linear, and the values are inside the theoretical limits presented in the models for both shocks and SF distributions in every galaxy.

Another assumption is required when \oii$\lambda$3727\A is not available. In this case, \hcm relies on the \nii, \sii, and \oiii emission lines to estimate the metallicity. The models described in \citet{PerezMonteroHCm14} present solutions that do not differ significantly from those obtained using strong-line calibrations such as O3N2 or N2S2. Moreover, the results of this work are consistent with previous studies that have investigated the metallicity of the DIG in the galactic plane using strong-line calibrators. For instance, \citet{SandersDIGabundance17} found that the DIG in the galactic plane tends to show lower levels of excitation than the surrounding ionised gas, and that subtracting this low-excitation DIG increases the inferred ionisation parameter and enhances the \oiii emission. This behaviour is similar to what we observe, where regions affected by shock-like conditions tend to exhibit lower excitation and higher \nii/\ha and \sii/\ha ratios. In addition, \citet{ValeDIGabundance19} reported a shift of $\sim$0.1 dex in O/H derived from strong-line calibrators in regions with low \ha equivalent width, where the DIG contribution is not subtracted. Furthermore, \citet{PoetrodjojoDIGabundance19}, using integral-field spectroscopy of M83, reported a strong negative metallicity gradient associated with the DIG, deviating from the trend typically observed in \hii regions. This is consistent with the $\sim$0.1 dex offset found in this work in regions less influenced by star formation in the disc, with the shift being less pronounced in regions whose ionisation conditions and line-ratio characteristics are consistent with those expected for \hii regions. Besides, the tendency of the gas-phase abundances to approach solar values in the SF-dominated regions is consistent with findings in other galactic halos, where galactic fountains powered by SF in the disc shift the metallicity of the halo towards solar values \citep[e.g.][]{BregmanMetalHalo13}. Moreover, the broad dispersions found in our results are also consistent with previous studies of the galactic plane. \citet{KumariDIGabundance19}, using the O3N2, O3S2, and N2S2H$\alpha$ calibrators, reported metallicity offsets of 0.11–0.13 dex between star-forming and DIG/low-ionisation regions, with associated dispersions of $\sim$0.09 dex. These values are fully compatible with the offsets and scatter measured in this work.

\end{document}